\documentclass[a4paper,12pt]{article}
\usepackage{graphicx} 

\usepackage{graphicx} 
\usepackage{latexsym}	
\usepackage{amsmath}
\usepackage{amsthm}
\usepackage{amssymb}
\usepackage{amsmath}
\usepackage{amsthm}

\usepackage{graphicx}
\def\ba{\begin{eqnarray}}
\def\ea{\end{eqnarray}}

\def\ba{\begin{eqnarray}}
\def\ea{\end{eqnarray}}

\def\be{\begin{equation}}
\def\ee{\end{equation}}

\theoremstyle{plain}

\begin{document}

\title{Scattering amplitudes in Quadratic Gravity in a general formalism}

\author{ Osvaldo P. Santill\'an$^1$
\thanks{Electronic addresses:firenzecita@hotmail.com}\\
\textit{\small{$^1$Instituto de Matem\'atica Luis Santaló (IMAS) and CONICET,}}\\
\textit{\small{Ciudad Universitaria, 1428 Buenos Aires, Argentina.}}}
\date{}

\maketitle

\begin{abstract}
In \cite{salvio}, inspired by the works \cite{pauli}-\cite{donogue}, a prescription for calculating the correlation functions in Quadratic Gravity \cite{stelle1}-\cite{stelle2} was presented and further exploited in \cite{salvio2}-\cite{salve}.  A covariant formalism ensuring positive definite probabilities is worked out, which is the main drawback of Quadratic Gravity. The Gauss-Ostrogradsky method for Quadratic Gravity defines two momentum densities $P_1$ and $P_2$ and two coordinate densities $Q_1$ and $Q_2$, one pair is standard, the other ghost like. The approach in \cite{salvio} involves the continuation $P_2\to i P_2$ and $Q_2\to i Q_2$ of the ghost variables acting on kets $|>$ after taking mean values. The notable result is that, in the euclidean setting, this procedure leads to Quadratic Gravity path integral $Z(J)$, thus renormalizability is not spoiled. In view of these findings, it is natural to ask how the LSZ rules of the model have to be formulated, and this is the topic of the present work. A formalism adapted to full quartic or higher order theories is worked out, extending the results of \cite{yomismo}. The main technical point is to determine the creation annihilation  algebra for the  graviton  modes, adapted to the present prescription, and properly dealing with gauge symmetry.  This makes the problem harder than the quantization of the Pais-Uhlenbeck model. Two possible quantization schemes are discussed, they depend on whether the above prescriptions are applied at the beginning or at the end of the LSZ calculation.
\end{abstract}

\section{Introduction}

Quadratic Gravity, or Stelle gravity \cite{stelle1}-\cite{stelle2} is a renormalizable gravity model, which contains a massless mode and two massive ones. In the original reference, the renormalization program for this model was achieved in terms the Slavnov-Taylor quantization method \cite{Slavnov}-\cite{Taylor}. However, one of the massive modes contributes to the kinetic energy with a wrong sign, and was interpreted as a ghost which spoils unitarity. The theory is of higher order, its equations of motion are of fourth order and its Cauchy problem is well posed \cite{yo2}-\cite{yo3}. This  unitarity problem is considered the major obstruction of the model, at least for small perturbations around  a flat background.

On the other hand, starting from the seminal works of Dirac and Pauli \cite{dirac}-\cite{pauli}, several authors \cite{wick}-\cite{donogue} and more recently \cite{kleefeld} started different programs for quantization of higher order theories, with the intention of avoiding ghosts. These works are not only related to Quantum Gravity, instead they present different motivations, some of them \cite{wick}-\cite{wick2} are even related to Statistical Physics. All of them bring several features which are not evident at first sight. An interesting approach was due to Boulware and Gross \cite{gross}, who made a formal description of the non positive definite metric in the Hilbert space. These authors consider states such that their inner product $<n|m>=\eta_{m n}$ is not positive definite. The reference \cite{strumia3} describes a covariant $\eta_{nm}$, contra-variant $\eta^{nm}$ and mixed metric $\delta_n^m$. The mixed one is positive definite and is the one to be employed for calculating probabilities. The Hilbert space itself has covariant and contra variant components. The mean values have to be adapted to the covariant or contra variant Hilbert space, the result is that negative and non bounded from below quantities become bounded by this covariant and contra-variant distinction. These authors \cite{gross} also present several useful formulas for the path integral for these theories, together with some interesting insights. In particular, that depending on the use of the covariant, contra-variant or mixed version, a self adjoint observable may be presented in terms of non hermitian matrices. Despite these  findings, those authors open the possibility that the path integral they define may not exist \cite{gross}. The main point is the possible presence of real and unbounded exponentials in the path integral, which make the full result divergent.

On the other hand, the references \cite{strumia1}-\cite{strumia3} made several efforts to physically interpret the results of the above results. In particular, in \cite{salvio} the author made the remarkable observation that the quantization scheme of Boulware-Gross applied to the Quadratic Gravity leads, in the euclidean setting, to a perfectly defined path integral. This was  further exploited in \cite{salvio2}-\cite{salve}. The potentially dangerous real exponentials  in the path integral become imaginary and the full result is well defined (in a path integral sense). The main observation is that, in a Hamiltonian setting, there are variables which behave normally under time reversal, and there are ones that not. It is not possible to state in brief the results of those detailed works, but it is worthy to mention that the quantization scheme with these variables, denoted as $P_2$ and $Q_2$ is consistent after the continuation to imaginary values $P_2\to i P_2$ and $Q_2\to i Q_2$. Stated roughly, if the commutator $[a,a^\dag]$ of an state of the mode has a wrong sign, this continuation removes the unwanted minus sign. 

Furthermore, both the results of \cite{gross} and \cite{salvio}, strongly point out that this quantization does not enter in any conflict with unitarity. The final result of the whole continuation is the euclidean version of Quadratic Gravity. Therefore, the physical quantities may be calculated in the euclidean setting, and continued to the Lorenzian one by going to real times. The results will be renormalizable, and it is possible that unitarity will be preserved.

The results of all the references mentioned above are of significant relevance. However, a full presentation requires an analysis of the LSZ formulas which, in  this context, have to be taken with care. The point is that, in this quantization method, the creation and annihilation algebra may have some particularities not present in an standard quantum theory. A program about this topic was initiated in \cite{yomismo}. The LSZ rules require a second quantization of the free model consistent with the principles employed in all these references. In the present work, the idea is to present an LSZ deduction fully adapted to a fourth order theory, regardless how the modes are decomposed.  A proper hamiltonian treatment, missing in \cite{yomismo} is included here.  There are found two possible rules, depending on the step where the above prescription is applied.

In authors opinions, the steps given here are in concordance with the prescription in \cite{salvio}-\cite{salvio2}.

The present work is organized as follows. Section 2 reviews the Gauss-Ostrogradsky quantization for the Pais-Uhlenbeck oscillator, and how the ghost emerges. Section 3 contains a description of what the authors understand about these non ordinary quantizations, written in pretty much detail. In Section 4 the LSZ rules for the Pais-Uhlenbeck oscillator are derived, in a formalism fully adapted for higher order theories. It also reviews the notable result of reference \cite{salvio} about the finiteness of the path integral when the above described quantization is presented, which presents a quantization of the model which does not allow  divergences due to exponential integrations in the path integral and, at the end, results in the euclidean version of Stelle gravity.  Section 5 is  a review of the results of \cite{salvio} about the quantization of Quadratic Gravity with the above formalis.  In Section 6 and 7 the LSZ formulas adapted to Stelle gravity fourth order model is derived, the universal form of the effective action is specified and the effect of the non standard creation annihilation operators is clarified in the last part of this section. Section 8 contains a brief statement of the quantization prescription, for convenience and concreteness. Section 9 contains possible research programs related to these results.

\section{Quantization formulas  for ghosts}
It is known that Quadratic Gravity posses ghosts, which are by definition states with negative norm. The next sections are devoted to describe results in the literature dealing with this type of problems.
\subsection{The Pais-Uhlenbeck quartic oscillator}
Before studying Quadratic Gravity, consider the prototype of a ghost system, namely the Pais-Uhlenbeck lagrangian \cite{pais} for a real scalar field $\phi$
\begin{equation}\label{federica}
L=-\frac{3}{2}\partial_\mu \phi \partial^\mu \phi-\frac{9\gamma}{2}(\square\phi)^2-V(\phi).
\end{equation}
The potential $V(\phi)$ will be quite general in the present discussion. As the theory is of higher order, the procedure of doubling the number of variables can be applied, and there will be two coordinates and two momenta involved instead of one. The coordinates will be chosen as
\begin{equation}\label{coordenadas}
Q_1=\beta\phi, \qquad Q_2=\lambda\dot{\phi},
\end{equation}
with $\beta$ and $\lambda$ constant factors. In the following the values $\beta=\lambda=1$ can be chosen, but these factors are written in order to specify certain liberty on the choice of coordinates and momenta which do not alter the canonical commutation relations. In fact, the quantization to be employed now may allow complex values for these constants and still give real mean values for these operators \cite{strumia3}. This method, maybe awkward at first sight, has roots from  Dirac and Pauli times \cite{dirac}-\cite{pauli}.

The canonical momenta corresponding to \eqref{coordenadas} can be found by the following formula, valid for a higher order formalism
$$
 P_1=\frac{\partial L}{\partial \dot{Q_1}}-\frac{d}{dt}\bigg(\frac{\partial L}{\partial \ddot{Q_1}}\bigg)=\frac{1}{\beta}[3\dot{\phi}-9\gamma\square \dot{\phi}],
$$
\begin{equation}\label{momentum}
 P_2=\frac{\partial L}{\partial \dot{Q_2}}-\frac{d}{dt}\bigg(\frac{\partial L}{\partial \ddot{Q_2}}\bigg)=\frac{9\gamma}{\lambda} \square \phi,
\end{equation}
while the Hamiltonian density is found by the standard prescription
$$
H=P_1 \dot{Q}_1+P_2\dot{Q}_2-L.
$$
The above procedure ensures that the hamiltonian equations of motion are equivalent to the Euler-Lagnrange ones
$$
\frac{\partial L}{\partial \phi}=\partial_\mu\bigg(\frac{\partial L}{\partial \partial_\mu \phi}\bigg)-\partial_\mu\partial_\nu \bigg(\frac{\partial^2 L}{\partial \partial_\mu\; \partial \partial_\nu\phi}\bigg).
$$
It is useful  to deduce from the second \eqref{momentum} that
$$
\ddot{\phi}=\Delta \phi-\frac{\lambda}{9\gamma} P_2.
$$
With the help of this formula, the  Hamiltonian density can be calculated explicitly in terms of the canonical coordinates \eqref{coordenadas} and momentum \eqref{momentum}, the result is
\begin{equation}\label{canon}
H=\frac{\beta P_1 Q_2}{\lambda}-\frac{\lambda^2 P_2^2}{18\gamma}+\frac{\lambda P_2\Delta Q_1}{\beta}-\frac{3}{2}\frac{Q_2^2}{\lambda^2}+\frac{3}{2\beta^2}(\nabla Q_1)^2+V(Q_1).
\end{equation}
Here $\Delta$ is the three dimensional  Laplacian. 

The equation of motion of the model
are given by
\begin{equation}\label{moviliza}
-3\square \phi+9\gamma\square^2\phi=\frac{\partial L_i}{\partial \phi},
\end{equation}
where the interaction lagrangian $L_1$ is due to the potential term $V(\phi)$. In the free case, the last equation shows that
$$
\phi=\phi_1+\phi_2,
$$
the first mode is massless and the second has mass $m^2=\frac{1}{3\gamma}$. This means that the field can be Fourier expanded as
$$
\phi(x) =\int\frac{d^3k}{(2\pi)^{\frac{3}{2}}\omega(k)}[b(k) e^{i\omega(k)t-i k\cdot x}+b^\ast(k) e^{-i\omega(k)t+i k\cdot x}]
$$
$$
+\int\frac{d^3k}{(2\pi)^{\frac{3}{2}}\omega_2(k)}[c(k) e^{i\omega_2(k)t-i k\cdot x}+c^\ast(k) e^{-i\omega_2(k)t+i k\cdot x}],
$$
where the corresponding dispersion relation  is given by $\omega_2(k)=\sqrt{k^2+m^2}$. Also $\omega(k)=|k|$.
Note that, by this dispersion relations and by \eqref{momentum}
\begin{equation}\label{livin}
P_1=3\dot{\phi}_1.
\end{equation}
The quantization method requires to express the mode coefficients $b(k)$, $b^\ast(k)$, $c(k)$ and $c^\ast(k)$  in terms of  the canonical variables $P^i$ and $Q_i$ with $i=1,2$. Then, by assuming the standard equal time commutation relations given by
\begin{equation}\label{canon}
[Q_i(x,t), P_j(y, t)]=i\hbar \delta_{ij} \delta(x-y),
\end{equation}
the commutation relations for the quantum operators $b(k)$, $b^\dag(k)$, $c(k)$ and $c^\dag(k)$ can be deduced. This is a $4\times 4$ system of equations which, in principle, may be enough for determining those quantities.  

This naive expectation however, may have a problem.  It will be shown below, it is not possible to enforce the standard creation annihilation algebra  if the coordinates commute \cite{yomismo}\begin{equation}\label{luli}[Q_i(x,t), Q_j(y,t)]=0.\end{equation}
In fact, it will be seen that the solution of the $4\times 4$ system is such that $[Q_i(x,t), Q_j(y,t)]\neq 0$. 

The presence of negative norm states can be justified as follows. The second of the formulas \eqref{momentum} shows that this massless mode does not contribute to $P_2$ as $\square \phi_1=0$. Furthermore, from  the first \eqref{momentum} and the definition $\dot{\phi}=\lambda Q_2$,
it is seen that this massless mode does not contribute to $\beta P_1-\frac{3 Q_2}{\lambda}$ either.  Therefore, only the massive mode $\phi_2$ contributes to these two quantities and, since the massive mode involves only the coefficient $c(k)$ and $c^\ast(k)$, the system
\begin{equation}\label{ecua}
\lambda \beta P_1-3Q_2=-9\lambda\gamma\square \dot{\phi}_2,\qquad P_2=\frac{9\gamma}{\lambda}\square \phi_2,
\end{equation}
is a $2\times 2$ one allowing the calculation  $c(k)$ and $c^\ast(k)$ in terms of the modes of $\lambda \beta P_1-3 Q_2$ and $P_2$. This makes the problem more tractable. To solve this $2\times 2$ linear system explicitly, write 
$$
\phi_2(x,t)=\int\frac{d^3k}{(2\pi)^{\frac{3}{2}}\omega_2(k)}[c(k) e^{i\omega_2(k)t-i k\cdot x}+c^\ast(k) e^{-i\omega_2(k)t+i k\cdot x}].
$$
By taking into account that $\square \phi_2=m^2 \phi_2$ with $m^2=\frac{1}{3\gamma}$ the equations \eqref{ecua} are converted into
$$
\lambda P_2(x,t)=3 \phi_2(x,t)=3\int\frac{d^3k}{(2\pi)^{\frac{3}{2}}\omega_2(k)}[c(k) e^{i\omega_2(k)t-i k\cdot x}+c^\ast(k) e^{-i\omega_2(k)t+i k\cdot x}],
$$
\begin{equation}\label{ecuador}
3Q_2(x,t)-\lambda \beta P_1(x,t)=3\dot{\phi}_2=3i\int\frac{d^3k}{(2\pi)^{\frac{3}{2}}}[c(k) e^{i\omega_2(k)t-i k\cdot x}-c^\ast(k) e^{-i\omega_2(k)t+i k\cdot x}].
\end{equation}
This can be inverted to give
$$
c(k)=\frac{1}{3}\int d^3x e^{ikx}[\omega_2(k)\lambda P_2(x,0)-3iQ_2(x,0)+i\lambda\beta P_1(x,0) ],
$$
\begin{equation}\label{oscila}
c^\dag(k)=\frac{1}{3}\int d^3x e^{-ikx}[\omega_2(k)\lambda P_2(x,0)+3iQ_2(x,0)-i\lambda\beta P_1(x,0) ].
\end{equation}
If these expressions are prompted to operators $\widetilde{c}_k$ and $\widetilde{c}_k^\dag$, by assuming the canonical commutation relations \eqref{canon}, it is arrived to
\begin{equation}\label{lulu1}
[\widetilde{c}_{k^\prime}, \widetilde{c}^\dag_k]=\frac{2\lambda\omega_2(k)}{3}\delta(k-k^\prime).
\end{equation}
This is the standard commutation relation, up to a factor $3\lambda^{-1}$.
Consider now the remaining operators $b(k)$ and $b^\dag(k)$. Their commutation relation  can be determined from \eqref{coordenadas}, under the assumption that $[Q_1, Q_2]=0$.  This definition implies that
$$
Q_1 =\beta\int\frac{d^3k}{(2\pi)^{\frac{3}{2}}\omega(k)}[b(k) e^{i\omega(k)t-i k\cdot x}+b^\ast(k) e^{-i\omega(k)t+i k\cdot x}]
$$
$$
+\beta\int\frac{d^3k}{(2\pi)^{\frac{3}{2}}\omega_2(k)}[c(k) e^{i\omega_2(k)t-i k\cdot x}+c^\ast(k) e^{-i\omega_2(k)t+i k\cdot x}],
$$
$$
Q_2 =i\lambda \int\frac{d^3k}{(2\pi)^{\frac{3}{2}}}[b(k) e^{i\omega(k)t-i k\cdot x}-b^\ast(k) e^{-i\omega(k)t+i k\cdot x}]
$$
\begin{equation}\label{aha}
+i\lambda\int\frac{d^3k}{(2\pi)^{\frac{3}{2}}}[c(k) e^{i\omega_2(k)t-i k\cdot x}-c^\ast(k) e^{-i\omega_2(k)t+i k\cdot x}].
\end{equation}
As these two quantities commute, if  the unknowns $b_k$ and $b^\ast(k)$ are prompted to operators $\widetilde{b}_k$ and $\widetilde{b}_k^\dag$ the commutation relation $[\widetilde{b}_k,\widetilde{b}_k^\dag]$ has to cancel the terms coming from $[c_k,c_k^\dag]$. This leads to
\begin{equation}\label{commutacion2}
[b_k,b_{k^\prime}^\dag]=-\frac{2\lambda\omega_2(k)}{3}\delta(k-k^\prime).
\end{equation}
The minus sign in the last expression is the key result. It implies that one of the two pairs of creation/annihilation operators has a non standard $-1$ sign in the commutation relation in order to ensure that $[Q_1, Q_2]=0$.  This  non standard sign is unavoidable.  

Note that \eqref{aha} and \eqref{ecuador} imply that 
\begin{equation}\label{tito}
P_1=\frac{3}{\lambda\beta}i\lambda \int\frac{d^3k}{(2\pi)^{\frac{3}{2}}}[b(k) e^{i\omega(k)t-i k\cdot x}-b^\ast(k) e^{-i\omega(k)t+i k\cdot x}].
\end{equation}
As $P_2$ in \eqref{ecuador} only involves only $c(k)$ and $c^\ast(k)$, the wrong or correct sign with one of the commutators will not enter in conflict with the condition $[P_1, P_2]=0$, it only affects the commutator $[Q_1, Q_2]$.  The commutation relation 
$$
[Q_1(x,t), P_1(y,t)]=i\hbar\delta(x-y),$$
is the expected one if  the algebra of $c(k)$ and $c^\ast(k)$ is the wrong one. If  the standard oscillator algebra is imposed, then the last relation will be non standard
$$
[Q_1(x,t), P_1(y,t)]=-i\hbar\delta(x-y).$$
Every choice will lead to something non standard. It is not possible to impose \eqref{canon} for both pairs of canonical variables, at least if $[Q_1, Q_2]=0$. 

Note that \eqref{aha} and \eqref{ecuador} imply that 
\begin{equation}\label{tito}
P_1=\frac{3}{\lambda\beta}i\lambda \int\frac{d^3k}{(2\pi)^{\frac{3}{2}}}[b(k) e^{i\omega(k)t-i k\cdot x}-b^{\ast}(k) e^{-i\omega(k)t+i k\cdot x}].
\end{equation}
As $P_2$ in \eqref{ecuador} only involves $c(k)$ and $c^\ast(k)$, the incorrect or correct sign with one of the commutators will not enter into conflict with the condition $[P_1, P_2]=0$, it only affects the commutator $[Q_1, Q_2]$. 

At this point it is important to recognize the following subtle details. The above discussion implies that the good commutation relations may be simultaneously  imposed 
$$[P_1, Q_1]=i\hbar\delta(x-y),\qquad [P_2, Q_2]=i\hbar \delta(x-y),\qquad [Q_1, Q_2]=[P_1, P_2]=0.$$
However, this is at cost of declaring the wrong commutation oscillator algebra \eqref{commutacion2}. Instead, the switch of the sign of \eqref{commutacion2} leads to good oscillators 
\begin{equation}\label{commutacionheart}
[b_k,b_{k^\prime}^\dag]=\frac{2\lambda\omega_2(k)}{3}\delta(k-k^\prime).\qquad [c_k,c_{k^\prime}^\dag]=\frac{2\lambda\omega_2(k)}{3}\delta(k-k^\prime).
\end{equation}
but $[Q_2, P_2]=-i\hbar$, and the coordinates $Q_1$ and $Q_2$ do not commute anymore. This forces a choice between two unwanted situations. There are two algebras involved, one for the creation annihilation and the other is the quantum canonical commutations. They are not allowed to be healthy simultaneously.

Note already the following potential abuse of notation. The creation and annihilation operators are writen as $b(k)$
and $b^\dag(k)$. This notation, in the standard context, implies that one operator is the hermitian conjugate of the other. In the present case, this may be wrong. So, the notation will be changed in the next sections in order to do not generate confusion.

 The behavior of the energy levels of the system requires the analysis of its Hamiltonian. Its classical version is given explicitly in terms of $\phi$  by
\begin{equation}\label{fragila}
H=[3\dot{\phi}-9\gamma\square \dot{\phi}]\dot{\phi}+9\gamma \square \phi\ddot{\phi}+\frac{3}{2}\partial_\mu \phi \partial^\mu \phi+\frac{9\gamma}{2}(\square\phi)^2+V(\phi).
\end{equation}
This expression  is obtained after some algebra  from \eqref{canon} by employing the formulas \eqref{momentum} in order to express everything in terms of $\phi$. 

By decomposing $\phi=\phi_1+\phi_2$ and by taking into account their dispersion relations, it can be seen after some calculation, to be presented now, that the mixed terms involving $\phi_1$ and $\phi_2$ cancel out, and the resulting Hamiltonian density is simply the sum of two free Klein-Gordon ones corresponding to both modes.

The calculation leading to this conclusion goes as follows. After introducing the decomposition $\phi=\phi_1+\phi_2$ into \eqref{fragila} there appear terms that only involve $\phi_1$, other that involve $\phi_2$ and mixed terms. The mixed ones are
$$
H_{12}=[3\dot{\phi}_1-9\gamma\square \dot{\phi}_1]\dot{\phi}_2+9\gamma \square \phi_1\ddot{\phi}_2+3\partial_\mu \phi_1 \partial^\mu \phi_2+9\gamma\square\phi_1\square \phi_2
$$
$$
+[3\dot{\phi}_2-9\gamma\square \dot{\phi}_2]\dot{\phi}_1+9\gamma \square \phi_2\ddot{\phi}_1.
$$
From the dispersion relation of the modes, it is clear that  $\square\phi_1=0$ and $3\gamma\square\phi_2=\phi_2$. This leads to several simplifications, and the last expression becomes 
$$
H_{12}=3\dot{\phi}_1\dot{\phi}_2+3\partial_\mu \phi_1 \partial^\mu \phi_2
+3\phi_2\ddot{\phi}_1.
$$
Collecting the first and the last term as a total derivative, it follows that
$$
H_{12}=3\partial_t(\dot{\phi}_1\phi_2)-3\dot{\phi}_1\dot{\phi}_2+3\nabla \phi_1 \cdot\nabla \phi_2.
$$
This can be equivalently written as
$$
H_{12}=3\phi_2\ddot{\phi}_1+3\nabla \phi_1 \cdot\nabla \phi_2.
$$
This is the mixed term of the hamiltonian density.  The full hamiltonian is integrated over the spatial volume $\int_V d^3x$.  Inside the  spatial volume integral, the last term is equivalent by parts integration to 
$$
H_{12}=3\phi_2\ddot{\phi}_1-3\phi_2\Delta \phi_1.
$$
Since $\ddot{\phi}_1\sim -\omega^2 \phi_1$ while $\Delta \phi_1\sim -k^2\phi_1$ inside the mode expansion, the dispersion relation for $\phi_1$ namely, $\omega^2=k^2$  implies that the contribution of these terms cancel. This means that the mixing term can be set $H_{12}=0$.

The resulting hamiltonian density is therefore composed only by the decoupled terms involving only $\phi_1$ or only $\phi_2$. The insertion of the expression $\phi=\phi_1+\phi_2$ into \eqref{fragila} throws the following 
$$
H=\frac{3}{2}\dot{\phi}^2_1+\frac{3}{2}(\nabla\phi_1)^2+3\phi_2\ddot{\phi}_2-\frac{3}{2}\dot{\phi}^2_1+\frac{3}{2}(\nabla\phi_2)^2+\frac{3m^2}{2}(\phi_2)^2.
$$
Inside the Fourier expansion $\ddot{\phi}_2=\dot{\phi}_2^2\sim -\omega_2^2 \phi_2$, thus the last expression is equivalent to
$$
H=\frac{3}{2}\dot{\phi}^2_1+\frac{3}{2}(\nabla\phi_1)^2+\frac{3}{2}\dot{\phi}^2_2+\frac{3}{2}(\nabla\phi_2)^2+\frac{3m^2}{2}(\phi_2)^2+V(\phi_1,\phi_2).
$$
This clearly the sum of two free  Klein-Gordon hamiltonian densities, for a massless and a massive particle, with an interaction $V(\phi_1, \phi_2)$.  By omitting the zero point energy, the free part of the hamiltonian written in terms of the modes is given by
\begin{equation}\label{klip}
H_f=\int d^3k[ \omega(k)b^\dag(k)b(k)+ \omega_2(k)c^\dag(k)c(k)].
\end{equation}
This expression would not raise any alarm if the creation and annihilation operators were standard. However, it was shown above that one of them may deviate from the standard commutation relation. 
Therefore it is needed to understand if this Hamiltonian is bounded or instead unbounded from below, if these non standard commutators are employed.

The problems described above may be understood in terms of path integrals. The Hamiltonian \eqref{canon} depends only linearly on $P_1$. This raises a problem when trying to quantize the model by the method of path integrals.  If the coefficients $\beta$ and $\lambda$ are chosen equal to one, then the euclidean path version of this theory  wlll be divergent, since the integration on the variable $P_1$ yields an infinite result. This divergence is clear, since the integral in $P_1$ involves a real exponential and the integrand will be unbounded.
The fact that the Hamiltonian is linear in $P_1$ seems then problematic\footnote{Related work is the reference \cite{kleefeld}, where  apparently a self-consistent way to quantize these systems has been presented.}.

\section{The Dirac-Pauli quantization and its generalizations}

\subsection{Generalities}
The following is a personal, but step by step elaboration, of what the author understands about the quantization schemes developed in \cite{strumia1}, \cite{dirac}-\cite{wick2}, \cite{gross}. 
Consider a harmonic oscillator $H=b^c b$ with the algebra
\begin{equation}\label{todomal}
[b, b^{c}]=-1,
\end{equation}
which has a wrong $-1$ in front. Naively, the standard oscillator is
$$
a=\frac{\hat{x}+i\hat{p}}{2},\qquad a^\dag=\frac{\hat{x}-i\hat{p}}{2},
$$
and $b$ and $b^c$ correspond to $\hat{x}\to i \hat{x}$, $\hat{p}\to i \hat{p}$.  Note that $b^c$ is not necessary $b^\dag$.  Furthermore $i\hat{p}$
and $i\hat{x}$ are now anti hermitian.  Instead, one may introduce the replacement $\hat{x}\to \hat{x}$, $\hat{p}\to -\hat{p}$.  In this case the operators are still hermitian, but with the impulse sign flipped. 

Here there appears the following ambiguity. For $\hat{X}=\hat{x}$ and $\hat{P}=-\hat{p}$ the hamiltonian becomes
$$
H=b^c b=\hat{X}^2+\hat{P}^2+\textbf{constant}.
$$
For $\hat{X}=i\hat{x}$ and $\hat{P}=i\hat{p}$ instead
$$
H=-\hat{X}^2-\hat{P}^2+\textbf{constant}.
$$
This ambiguity means nothing at this point, since the norm of the states corresponding to these choices has not been yet determined. 

For all the above choices
$$
[H, b]=[b^c b,b]=b^c b b-b b^c b=[b^c, b]b=b.
$$
$$
[H, b^c]=-b^c.
$$
Therefore
$$
[H, b]|n>=H b|n>-E_nb|n>=b|n>,\qquad H b|n>=(E_n+1)b|n>.
$$
The energy rises instead of lowering. Similarly
$$
Hb^c|n>=(E_n-1)b^c|n>.
$$
The energy lowers instead of rising. Declare that
$$
b|0>=0,\qquad b^c|0>=|1>,
$$
then the spectrum is not bounded from below since
$$
 H b^c|n>=(E_n-1)b^c|n>,\qquad  H b^c|0>=(E_0-1)|1>=E_1 |1>.
 $$
 Then $E_1=E_0-1$. Since $E_0=0$, then $E_1=-1$. Higher orders will also have negative energy.
Also note that $b^c\neq b^\dag$.  If heuristically $b=ia$ and $b^c=ia^\dag$ then $(b^c)^\dag=-b$.
$$
< 1|1>=< 0|(b^c)^\dag b^c|0>=-< 0|bb^c|0>
$$
$$
=< 0|b^cb|0>-< 0|1|0>=-1.
$$
The first state has negative norm. Keep going to get
$$
<n|m>=(-1)^n\delta_{nm}=\eta_{nm}.
$$
Odd states have negative norm, even ones have positive ones. Here $<n|=(|n>)^\dag$.
Clearly, this algebra  corresponds to a norm that is not the standard one
$$
<\psi|\psi^{\prime}>\neq\int dx\psi(x)^\ast \psi^\prime(x)dx,
$$
as this last expreession is be positive definite.  In fact
$$
|\psi>=\sum c_n |n>,\qquad <\psi^\prime|=\sum d_m^\ast <m|,
$$
$$
<\psi|\psi^{\prime}>=\sum c_n d^\ast_m \eta_{nm}=\sum_{n=2k} c_n d^\ast_n-\sum_{n=2k+1} c_n d^\ast_n.
$$
Taking into account the parity of the harmonic oscillator, this norm is
\begin{equation}\label{say}
<\psi'|\psi> =  \int dx\, [ \psi^{\prime*}_{\rm even}(x) \psi_{\rm even}(x) -\psi^{\prime*}_{\rm odd}(x)\psi_{\rm odd}(x)]= \int dx \, \psi^{\prime*}(x)  \psi(-x).
\end{equation}
It looks like the quantity playing the role of the completness operator is
$$
1=\int dx |x><-x|,
$$
and therefore $<x|y>=\delta(x+y)$. This is different from QM where  $<x|y>=\delta(x-y)$. This point is of course strange, since the coordinate wave function has support not on the same point, but on oposite points along the origin $x\to -x$.

Consider the norm $\eta_{ij}$ as a infinite dimensional metric, not  positive definite. By making an analogy with the Minkowski metric $\eta_{\mu\nu}$, which also has positive and negative entries, it is clear 
that although metric $\eta_{\mu\nu}$ is not positive definite, the mixed metric  $\eta_\mu^\nu=\delta_\mu^\nu$ is. Now take into account  that
$$
\eta |n>=\eta_{kn}|k>.
$$
Then 
$$
|\psi>=\sum c_n |n>,\qquad <\psi^\prime|=\sum d_m^\ast <m|,
$$
$$
<\psi|\eta|\psi^{\prime}>=\sum c_n d^\ast_m <m|\eta|n>=\sum c_n d^\ast_m \eta_{kn}<m|k>,
$$
$$
=\sum c_n d^\ast_m \eta_{kn}\eta_{km}=\sum c_n d^\ast_m \delta_{kn}\delta_{km}(-1)^{n+m}
$$
$$
=\sum c_n d^\ast_m \delta_{mn}(-1)^{n+m}=\sum c_n d^\ast_n.
$$
The probabilities will be defined in these terms with a $\eta$ inserted
$$
<\psi^{\prime}|\eta|\psi>= \int dx\, [ \psi^{\prime*}_{\rm even}(x) \psi_{\rm even}(x) +\psi^{\prime*}_{\rm odd}(x)\psi_{\rm odd}(x)]
$$
$$
= \int dx \, \psi^{\prime*}(x)  \psi(x).
$$
This is of course positive definite. One may define covariant and contravariant states
$$
|_n>=|n>,\qquad |^n>=\eta^{nm}|_m>,
$$
where $\eta^{nm}=\eta_{nm}$ since  $\eta^2=1$.  Then it follows that while
$$
<_n|_m>=\eta_{nm},\qquad <^n|^m>=\eta^{nm},
$$
are not positive definite, the mixed inner product
$$
\qquad <_n|^m>= <^n|_m>=\delta^{m}_n,
$$
is positive definite.  The above can be defined as a probability.
The completeness relations would be
$$
1=\int dx |x><-x|,\qquad \eta=\int  dx |x><x|,
$$ consistent with the above prescription. 

Based on the above findings, consider the possibility of employing the usual operators for $x$ and $p$ in QM, for covariant states the operators \cite{dirac}-\cite{pauli}, \cite{gross}
$$ 
\hat{x} |x> = ix |x>,\qquad p |x>  = \frac{d}{dx} |x>,
$$
which satisfy $[\hat{x}, \hat{p}]=-i$.  From 
$$
<y|z>=\int <y|-x><x|z>dx\rightarrow f(y,z)=\int <y|-x>f(x,z)dx,
$$
it follows the expected relation
$$
<y|x>=\delta(x+y).
$$
This is exactly what was found above. Define  the adjoint $\hat{x}^\dag$ of $\hat{x}$ by
$$
<y|\hat{x}|x>=<x|\hat{x}^\dag|y>^\ast.
$$
Then, by postulating that $<x|\hat{x}^\dag=if(x)<x|$, it is clear that
$$
ix<y|x>=(if(x)<x|y>)^\ast\rightarrow ix\delta(x+y)=-if(x)\delta(x+y).
$$
Thus $f(x)=-x$ and therefore
$$
<x|\hat{x}^\dag=-ix<x|.
$$
It is a temptation to conclude that $\hat{x}$ is anti-hermitian. However
\begin{equation}\label{esono}
<x^\prime|\hat{x}^\dag|x>=<x|\hat{x}|x^{\prime}>^\ast=[ix^\prime\delta(x+x^\prime)]^\ast=ix\delta(x+x^\prime)=<x^\prime|\hat{x}|x>.
\end{equation}
Therefore the operator $\hat{x}$ is still self adjoint. This may sound strange, but the point is that the completeness relation employed are not standard, from there this rare result. More even strange, this definition of self adjointness, with the non standard Dirac delta, does not insure that the mean values are real. In fact, consider  a generic state
$$
|\alpha>=\int f(y)|y>dy,\qquad <\alpha|=\int f^\ast(y)<y|dy
$$
Then the normalization of this state is
$$
<\alpha|\eta|\alpha>=\int f^\ast(z) f(y) <z|\eta|y>dydz=\int f^\ast(y) f(y)dy.
$$
This is positive definite. Consider an anzatz for the mean value
$$
<\alpha|\eta \hat{x}|\alpha>=\int f^\ast(z)f(y)<z|\eta \hat{x}|y>dydz
$$
\begin{equation}\label{notame}
=\int f^\ast(z)f(y)<z|\hat{x}|-y>dydz=-i\int f^\ast(y)yf(y)dy.
\end{equation}
The mean values are pure imaginary. So, this definition of self adjoint does not insure real mean values.
Neither does the definition without the $\eta$, as it can be verified by direct calculation.

In the present context, every author has to specify what definition of the quantities analogous to standard QM is using, and which properties are preserved and which not. There are plenty of well sounded affirmations that can be false.

The analogous follows for $p$ defining $\hat{p}|p>=ip|p>$. In particular, 
$$
<p^\prime|p>=\delta(p+p^{\prime}).
$$
Since 
$$
<z|\hat{x}|y>=i y\delta(y+z)=-iz\delta(y+z),
$$
then $<x|\hat{x}=-ix<x|$. By writing
$$
\hat{x}=-\frac{d}{dp}, \qquad \hat{p}=i p,
$$
it is clear that $[\hat{x}, \hat{p}]=-i$. Therefore
$$
<x|\hat{x}|p>=-ix<x|p>=-\frac{d}{dp}<x|p>,
$$
and the first and the last term consitute a differential relation with solution
$$
<x|p>=\frac{1}{\sqrt{2\pi}}e^{ipx}.
$$
Define the covariant states
$$
|_x>=|x>,\qquad <_x|=<x|,\qquad |_p>=|p>,\qquad <_p|=<p|.
$$
Then the contravariant ones are
$$
|^x>=\eta|_x>=|-x>,\qquad <^x|=<-x|=<_x|\eta,
$$
$$
|^p>=\eta|_p>=|-p>,\qquad <^p|=<-p|=<p|\eta.
$$
Furthermore, with this notation
$$
I=\int |^x><_x|dx,\qquad <_y|^x>=\delta(x-y),
$$
\begin{equation}\label{noto}
<_x|_p>=\frac{1}{\sqrt{2\pi}}e^{ipx},\qquad <^x|_p>=\frac{1}{\sqrt{2\pi}}e^{-ipx}.
\end{equation}
Endowed with these relations, note that $<_x|^y>$ looks like the correct product for studying probabilities, since it corresponds to the correct Dirac delta and the correct normalization.

In order to finish this section, some more comments are in order. Given a Hilbert space with a covariant or contravariant distinction,  an arbitrary  state $|\psi>$ can be expanded in the covariant or contravariant basis
$$
|\psi>=\psi^n|_n>=\psi_n|^n>.
$$
Here $\psi^n=<^n|\psi>$ and  $\psi_n=<_n|\psi>$. The  mean values of an observable $O$ have four different presentations 
$$
O=O_{mn}|^m><^n|=O^{mn}|_m><_n|=O_{m}^{\;\;n}|^m><_n|=O^m_{\;\;n}|_m><^n|,
$$
where 
$$O_{nm}=<_n|O|_m>,\qquad O_n^{\;m}=<n|O|^m>,
$$
$$
O^n_{\;m}=<^n|O|_m>,\qquad O^{nm}=<^n|O|^m>.$$
 In particular, this applies  for the operator $1$, which is related directly to the metric $\eta_{nm}$ or to its upper components by the following formula
$$
1=\eta_{nm}|^n><^m|= \eta^{nm}|_n><_m|=|^n><_n|.
$$
This is perfectly consistent with the completeness relations for $|_x>$ derived above. 

Given an operator $O$ whose action on a state is given by $|\psi^\prime>=O|\psi>$ then its adjoint operator $O^\dag$ is defined by the standard relation  $$<\psi^\prime|=<\psi| O^\dag.$$
In particular this means that 
\begin{equation}\label{conjug}
<\psi_1|O|\psi_2>=<\psi_2|O^\dag|\psi_1>^\ast.
\end{equation}
The adjoint of a matrix in all the presentations are defined by  $$O_{nm}^\dag=O_{mn}^\ast,\qquad O^{\dag mn}=O^{nm\ast}, \qquad O_{n}^{\dag m}=O^{m\ast}_{\;n}.
$$ 
 A self adjoint operator  is defined by $O=O^\dag$.  A self adjoint operator looks hermitian for the presentations $O_{nm}$ or $O^{nm}$ but for mixed indices the components are related by the isospectral transformation
$$
O_{n}^m=(\eta O^{\ast T} \eta^{-1})_n^m.
$$
It is a remarkable fact that the  last matrix can be anti hermitian and still have real eigenvalues. It is convenient to employ the symbol $\dag$  to denote the adjoint of an operator, but not for a matrix, in order to do not generate confusion. Note that the discussion between \eqref{esono} and \eqref{notame} shows that some intuitive properties of operators with this propery may fail.

The references \cite{pauli}-\cite{donogue},\cite{strumia1}-\cite{strumia3}, \cite{salvio} are the pioner works about the topic. In the authors understanding, the above is the main content of these references.

\subsection{The transition amplitudes as a path integral}
Consider the generic transition amplitude \cite{gross}, which in the covariant and contravariant formalism may be written \cite{strumia3}
$$
<_{x_f, X_f, t_f}|^{x_i, X_i, t_i}>=\prod_{l=1}^N \int [dx_l][dX_l]<_{x_f,X_f, t_f}|^{x_N, X_N, t_N}>...
$$
$$
<_{x_1, X_1, t_1}|^{x_i, X_i, t_i}>.
$$
Here the capital letter $X$ represent a ghost like variable, while the usual one $x$ represent a standard variable.
The reason for the choice  $<_x|^y>$ in the expressions above is that this is $\delta(x-y)$, which means that the states overlap only in the same point. For $<_x|_y>$, which corresponds to $\delta(x+y)$ it overlaps on separated points on the real axis. This is unwanted.  Alternatively, this results follows by successively inserting identities 
$$
1=\int |^{x,X, t}><_{x, X, t}| dx dX.
$$
Some factors in the last expression may be worked out by the identity
$$
<_{x_{l+1}, X_{l+1}, t_{l+1}}|^{x_l, X_l, t_l}>=\int <_{x_{l+1}, X_{l+1}, t_{l+1}}|^{p_{l+1}, P_{l+1}, t_{l+1}}>
$$
$$
<_{p_{l+1}, P_{l+1}, t_{l+1}}|^{x_l, X_l, t_l}>dp_l dP_l.
$$
From the expression of the evolution at small $\delta t_l=t_{l+1}-t_l$ given in terms of the system`s Hamiltonian by
$$
<_{p_{l+1}, P_{l+1}, t_{l+1}}|^{x_l, X_l, t_l}>=<_{p_{l}, P_{l}, t_{l}}|e^{-iH\delta t}|^{x_l, X_l, t_l}>
$$
$$
=\frac{1}{(2\pi)^m} \exp\{-iH(p_l,x_l, P_l, X_l) \delta t_l-i(p_l x_l-P_l X_l)\},
$$
where \eqref{noto} has been taken into account,  $m$ is the number of degrees of freedom and $H(p_l,x_l, P_l, X_l)$ is the mean value
$$
H_a(p_l,x_l, P_l, X_l)=\frac{<_{p_{l}, P_{l}, t_{l}}|H(\hat{p}_l,\hat{x}_l, \hat{P}_l, \hat{X}_l)|^{x_l, X_l, t_l}>}{<_{p_{l}, P_{l}, t_{l}}|^{x_l, X_l, t_l}>},
$$
An important point is that the above mean value has to be constructed taking into account that the eigenvalues of $P_i$ and $X_i$ are imaginary. Therefore, at the classical level it corresponds to put imaginary values $-iP_l$ and $-iX_l$.  The minus in $-iP_i$ is due to the fact it acts on the left. The minus in $-i X$ is due to the fact that it acts on the right, but with the upper index state. Thse hamiltonian is then 
\begin{equation}\label{aprescribir}
H_a(p_l,x_l, P_l, X_l)=H_c(p_l,x_l,-i P_l, -iX_l).
\end{equation}
where $H_c(p_l,x_l, P_l, X_l)$ is the classical hamiltonian. The ghost variables have been continued to 
imaginary values. 

Collecting all these results and by redefining $X\to-X$ in the integration variable, it follows that
$$
<x_f, X_f, t_f|x_i, X_i, t_i>=\int [Dx][DX][Dp][DP]
$$
$$
\exp\bigg\{i\int_{t_i}^{t_f}dt[p\dot{x}+P\dot{X}-H_c(p,x, -iP,iX)]\bigg\}.
$$
The last expression imitates usual path integral in QM, however with the subtle detail that the Hamiltonian is continued to imaginary values of the ghost variables. This may generate real divergent exponentials, and the integral may not exist. This was one of the unwanted features that were anticipated by the authors \cite{gross}. Fortunately, this does not happen for the Pais-Uhlenbeck model. This notable result  was found, for instance, in \cite{salvio}.

Another possible drawback of the above quantization is the appearence of modified hamiltonian $H_c(p,x, -iP,iX)$ which may spoil desired features of the original model. For Quadratic Gravity, it may be the renormalization property. Or symmetries which are considered fundamental can be spoiled as well. This has to be analyzed carefully model by model.

\subsection{The path integral corresponding to the Pais-Uhlenbeck model}

The above arguments can be applied to the Pais-Uhlenbeck model \eqref{federica}, but partially. The word "partially" is due to the following subtlety. First of all, in the exposition of the Dirac-Pauli quantization given above  some ficticious $\hat{p}$ and $\hat{x}$ were introduced. A wrong algebra $[\hat{x}, \hat{p}]=-1$ for them implied a wrong algebra for the creation annihilation operators $[c, c^c]=-1$. Instead, for the Pauli-Uhlenbeck model \eqref{federica}, there were two canonical coordinates $Q_1$ and $Q_2$ in \eqref{coordenadas} and two momenta $P_1$ and $P_2$ in \eqref{momentum}. These are defined in terms of two creation and annihilation pair $b(k)$, $b^c(k)$ and $c(k)$, $c^c(k)$. The correct algebra for the canonical operators $Q_1$, $Q_2$, $P_1$, $P_2$ gave a wrong oscillator algebra for one pair of $b(k)$, $b^c(k)$ and $c(k)$, $c^c(k)$, as shown in \eqref{lulu1} and \eqref{commutacion2}.  Also, the opposite is true namely, a correct oscillator algebra makes a wrong canonical algebra. This is a subtle difference to keep in mind in the following.
In one case wrong implies wrong, in the other wrong implies right.

By taking this into account, the  Hamiltonian \eqref{canon}, which it written again here by convenience after choosing $\beta=\lambda=1$
\begin{equation}\label{canonj}
H= P_1 Q_2-\frac{P_2^2}{18\gamma}+P_2\Delta Q_1-\frac{3}{2}Q_2^2+\frac{3}{2}(\nabla Q_1)^2+V(Q_1),
\end{equation}
only depends on $P_1$ by a linear term, which in the path integral may lead to a divergent exponential. However, it is too early to say so, since the path integral may be sandwiched by non standard states.
Sometimes,  it is preferable to employ the euclidean path integral quantization. This leads to
an imaginary time $t\to i\tau$. The definition \eqref{livin} and \eqref{ecuador} leads to
$$
Q_1=\phi_1+\phi_2,\qquad Q_2=\dot{\phi}_1+\dot{\phi}_2, \qquad P_1=3\dot{\phi_1},\qquad P_2=3\phi_2.
$$
Under a change of time $Q_2$  and $P_1$ pick an imaginary value. This does not change the behaviour of the hamiltonian since $P_1 Q_2\to -P_1 Q_2$, still a real exponential. However, this statement changes if one of the operators is not standard, such as the Pauli-Dirac ones $\hat{p}\to i\hat{p}$ and the same for $\hat{x}$.  In this case, the prescription of the previous section \eqref{canon1} of continuing the values of $Q_2\to i Q_2$ and  $P_2\to -iP_2$  has to be employed leading to
\begin{equation}\label{canonjk}
H= iP_1 Q_2+\frac{P_2^2}{18\gamma}-iP_2\Delta Q_1+\frac{3}{2}Q_2^2+\frac{3}{2}(\nabla Q_1)^2+V(Q_1).
\end{equation}
The dangerous term now is converted in a harmless imaginary term $iP_1Q_2$ whose exponentiation and integration does not lead  to divergent results.  To see this note that the euclidean path integral is 
$$
Z_e(0)=\int DP^{1} D\phi_1 DP^{2} D\phi_2 \exp\bigg\{-\int [iP_1 \dot Q_1 + iP_{2} \dot Q_2-H]dVd\tau\bigg\}
$$
$$
=\int DP^{1} D\phi_1 DP^{2} D\phi_2
$$
$$
\exp\bigg\{-\int [iP_1 \dot \phi_1 + iP_{2} \dot \phi_2- iP_1 Q_2-\frac{P_2^2}{18\gamma}+iP_2\Delta Q_1-\frac{3}{2}Q_2^2-\frac{3}{2}(\nabla Q_1)^2]dVd\tau\bigg\}.
$$
The integration over $P_1$ leads to a Dirac delta that enforces $Q_2 = -Q^{\prime}_1$, where the prime indicates derivative with respect to the euclidean time. The path integral now is related to $D\dot{Q}_1$ due to this identification, but a change  to the variable $DQ_1$ may be done by multiplying the integration measure by $\prod_{x_{E^\prime}} \Delta \tau^{-1}$.  The integral over $P_{2}$ is a Gaussian integral after setting  $Q_2 = -Q^{\prime}_1$. The general formula 
$$
\int d^N x e^{x^TAx+b x}=\frac{\pi^{\frac{N}{2}}}{\sqrt{\det A}}e^{-\frac{1}{4}b^TAb},
$$
can be employed to make this integral. After some calculation, it is found that the result with currents $J$ turned on given by
\begin{equation}\label{euclides}
Z_e(J)=\int D\phi \exp\bigg\{-\int L_edVd\tau+\int J \phi dVd\tau \bigg\}.
\end{equation}
Here $L_e$ is the euclidean version of the model corresponding to the action
$$
S=\int d\tau \int d^3x[\frac{3}{2}\partial_\mu \phi \partial_\mu \phi+\frac{9\gamma}{2}(\partial_\mu\partial_\mu \phi)^2+V(\phi)].
$$
Therefore, the apparent divergent exponential which can invalidate the model is avoided with this approach. The correlation functions can be calculated in the euclidean formalism and analitically continued to the  Lorenzian setting, by going from imaginary $\tau$ to real $t=i\tau$ times.   This result is presented in detail in \cite{salvio}. Note that the prescription $Q_2\to i Q_2$ and  $P_2\to -iP_2$  corresponding to the standard oscillator algebra, as mentioned in \eqref{commutacionheart}. This leads us to a discussion as how to define mean values of observables.

\subsection{The question about observables and their mean values}

Consider again the wrong  harmonic oscillator \eqref{todomal}. If the mean value of the hamiltonian is chosen  by the formula
$$
<\psi|H|\psi>=\sum c_n c^\ast_m<n|H|m>=\sum c_n c^\ast_m\eta_{nm}E_n,
$$
then, as the energy levels $E_{n}$ are always negative, for odd states the contribution is positive, for even states is negative. The result is not bounded from below. Instead the ansatz
 $$
 <\psi|H\eta|\psi>=\sum c_n c^\ast_m<n|H\eta|m>=\sum c_n c^\ast_m\eta_{km}\eta_{kn}E_n=\sum |c_n|^2 E_n,
 $$
is always negative since $E_n<0$.  However, the map  $\hat{p}\to -i\hat{p}$ and  $\hat{x}\to -i\hat{x}$ makes the new operators hermitian again. Therefore, for a harmonic oscillator
$$
H=\hat{p}^2+\hat{x}^2,
 $$
it may be reasonable to posulate that $H\to -H$ and this prescription solves the energy negative problem. 

In fact, note that the prescription \eqref{aprescribir} indicates something analogous.  It takes the observable, which in this case is the Hamiltonian $H_c$, and takes the analytical continuation to imaginary values of the ghost variables, resulting in
$$
H_a(p_l,x_l, P_l, X_l)=H_c(p_l,x_l,-i P_l, -iX_l).
$$
As the mean value for $\hat{x}$ in \eqref{notame} is imaginary., this continuation will make its value real.

These situations suggest, at least for the author, that the rule for taking mean values is the following. For the energy, coordinates and impulses, the mean values are
$$
\overline{p}=-i<\hat{P}\eta>,\qquad \overline{x}=-i<\hat{X}\eta>, \qquad \overline{E}=(-i)^2<H\eta>.
$$
Also
$$
\overline{p}^n=(-i)^n<\hat{P}^n\eta>,\qquad \overline{x}^n=(-i)^n<\hat{X}^n\eta>,
$$
$$
\overline{p^n x^m}=(-i)^{n+m}<\hat{P}^n\hat{X}^m\eta>.
$$
For a generic observable $O(P, X)$ constructed in powers of $\hat{X}$ and $\hat{P}$ these rules give the corresponding mean values. Calculate the mean value $<O(P,X)\eta >$ and analitically continue to values $P=\to-iP$ and $ X\to-iX$. This is in harmony with the all above findings.

Note that the continuation $i P_2$, $-iX_2$ confesses the use of the good oscillator algebra for the Pauli-Uhlenbeck model. Then the hamiltonian \eqref{klip} of the model becomes standard. Its mean values are positive and the problem of unbounded energies is solved. This gives the author confidence about the above prescription.

Some further general comments  would be convenient. As stated at the beginning of this section,  two different situations were analyzed. The first is constituted  by simple oscillators, where wrong canonical algebra implies wrong oscillators.  In the other case, wrong canonical algebra implies good oscillator algebra and viceversa. This is the case for the Pauli-Uhlenbeck oscillator. If an observable is written in terms of creation and annihilation operators
$$
O(b, b^c, c, c^c),
$$
then the standard algebra for both creation/annihilation pairs should be considered.  For example, one may employ \eqref{livin}, \eqref{ecuador} and the definition $Q_1=\phi$ to invert and express the creation and annihilation operator in terms of the the canonical variables $Q_i$ and $P_i$. Then employ the above paragraph methods of continuation to imaginary values, and the prescription given in \eqref{aprescribir} will be found.  This procedure may be general, but below it will be applied only to the Pais-Uhlenbeck model and to Quadratic Gravity.

\section{The S matrix and the LSZ formula}
The last point is to understand how to write the $S$ matrix LSZ reduction formula for the Pais-Uhlenbeck model. The LSZ reduction formula may be derived by lines similar to  the standard textbooks. Since non standard creation annihilation algebra is involved, and insertions of $\eta$ terms may appear, one has to be sure how these features affect the calculation. Therefore, it is advisable to follow the step by step procedure of the textbooks while taking into account these new features. This is the purpose of the present section.

\subsection{The S-matrix in a generic context}

Recall the definition of the  matrix $\hat{S}$ in terms of the scattering amplitude
$$
A_i^{\;f}=<_{f, \text{out}}|^{i, \text{in}}>=<_{f, \text{in}}|\hat{S}|^{i, \text{in}}>.
$$
If there are no ghosts or the algebra is standard, the distinction between covariant and contravariant states in innocuous since, in this case, $|_i>=|^i>$. Nevertheless, it will be convenient to work with the general formalism, covering both the standard and non standard cases.  It follows that
$$
\hat{S}=\sum_{h}|^{h, \text{in}}><_{h, \text{out}}|,
$$
since
$$
<_{f, \text{in}}|\hat{S}|^{i, \text{in}}>=<_{f, \text{in}}|\sum_{h}|^{h, \text{in}}><_{h, \text{out}}|^{i, \text{in}}>
$$
$$
=<_{f, \text{out}}|^{i, \text{in}}>=A_i^{\;f}.
$$
On the other hand, $\hat{S}^\dag$ is defined by
$$
<_{f, \text{in}}|\hat{S}|^{i, \text{in}}>^\ast=<^{i, \text{in}}|\hat{S}^\dag|_{f, \text{in}}>.
$$
Take the definition
$$
\hat{S}^\dag=\sum_{h}|_{h, \text{out}}><^{h, \text{in}}|.
$$
This definition works fine since
$$
<^{i, \text{in}}|\hat{S}^\dag|_{f, \text{in}}>=\sum_{h}<^{i, \text{in}}|_{h, \text{out}}><^{h, \text{in}}|_{f, \text{in}}>
$$
$$
=<^{i, \text{in}}|_{f, \text{out}}>=<_{f, \text{out}}|^{i, \text{in}}>^\ast=<_{f, \text{in}}|\hat{S}|^{i, \text{in}}>^\ast.
$$
This is the desired quantum mechanical property defining $S^\dag$.  In addition
$$
S^\dag |_{h,\text{in}}>=|_{h, \text{out}}>,\qquad <_{h, in}|S=<_{h, out}|,
$$
and since
$$
1=<_{f,\text{out}}|^{f,\text{out}}>
=<_{f,\text{in}}|S S^\dag|^{f,\text{in}}>,
$$
it is deduced that
\begin{equation}\label{nofederal}
\hat{S}\hat{S}^\dag=\eta,
\end{equation}
the last statement follows from the fact that for covariant and contra variant states $\eta_{\mu}^{\nu}=\delta^{\nu}_{\mu}$. 

In the above formulas, if a non standard algebra is employed, a generic state is 
$$
|_{f,\text{in}}>=|..,[n^{-}_i, -k_i],..., [n^{+}_i, k_i],.., \text{in}>,
$$
this notation $[n_i^+, k_i]$ means that there are $n_i^+$ particles with impulses $k_i$ and so on.
Then
$$
|^{f,\text{in}}>=|..,[n^{+}_i, -k_i],..., [n^{-}_i, k_i],.., \text{in}>.
$$
Note the switch $n_i^{+}\leftrightarrow n_i^{-}$ due to the action of $\eta$, which changes the sign of $k_i$.
The identity can be written as
$$
I=\sum_{\text{all states}}|..,[n^{-}_i, -k_i],..., [n^{+}_i, k_i],.., \text{in}>
$$
$$
<..,[n^{+}_i, -k_i],..., [n^{-}_i, k_i],.., \text{in}|.
$$
 In addition the creation operator is
$$
a_{in}^\dag(k_i)|..,[n^{-}_i, -k_i],..., [n^{+}_i, k_i],.., \text{in}>
$$
$$
=c(k_i)\sqrt{n^+_i+1}|..,[n^{-}_i, -k_i],..., [n^{+}_i+1, k_i],.., \text{in}>.
$$
The constant $c(k_i)$ may be anything, since it is not assumed that these operators are standard.
Write it as
$$
a_{in}^\dag(k_i)=\sum_{\text{all states}}|..,[n^{-}_i, -k_i],..., [n^{+}_i+1, k_i],.., \text{in}>
$$
$$
<..,[n^{+}_i, -k_i],..., [n^{-}_i, k_i],.., \text{in}|c(k_i)\sqrt{n^+_i+1}.
$$
Given the previous discussion,  consider  the $S$ matrix for a given process involving the Pais-Uhlenbeck scalar $\phi$. The action on  the creation operator on the ghost state state $b_{in}^\dag(k)$ of momentum $k$ is defined by
$$
S^\dag b^\dag_{in}(k_i)S=(\sum_{\text{all states r, h}}|r,\text{out}><r, \text{in}|)
 b^\dag_{in}(k_i)(|h,\text{in}><h, \text{out}|)
$$
$$
=(\sum_{\text{all states r}}|r,\text{out}><r, \text{in}|)
$$
$$
(\sum_{\text{all states } }c(k_i)\sqrt{n^+_i+1}|..,[n^{-}_i, -k_i],..., [n^{+}_i+1, k_i],.., \text{in}>
$$
$$
<..,[n^{-}_i, -k_i],..., [n^{+}_i, k_i],.., \text{out}|
$$
\begin{equation}\label{unitarios}
=\sum_{\text{all states h}}c(k_i)\sqrt{n^+_i+1}|..,[n^{+}_i+1, -k_i],..., [n^{-}_i, k_i],.., \text{out}>
\end{equation}
$$
<..,[n^{-}_i, -k_i],..., [n^{+}_i, k_i],.., \text{out}|.
$$
Clearly, the last expression represents $b^\dag_{out}(-k)$, which leads to the conclusion that
$$
S^\dag b^\dag_{in}(k)S= b^\dag_{out}(-k),\qquad S^\dag b_{in}(k)S= b_{out}(-k),
$$
$$
S^\dag b^\dag_{in}(k)S= b^\dag_{out}(k)\eta,\qquad S^\dag b_{in}(k)S= b_{out}(k)\eta,
$$
the second relation is completely analogous to the first one. The value of the constant $c_k$ parameterizing the non standard nature of the operator is irrelevant in order to obtain this conclusion. The same relation follows for $c(k)$
and $c^\dag(k)$. Therefore, it is deduced that
\begin{equation}\label{boca}
S^\dag \phi_{in}(k)S= \phi_{out}(-k),\qquad S^\dag \phi_{in}(k)S= \phi_{out}(k)\eta,
\end{equation}
regardless the non standard commutation of the creation and annihilation operators.
At this point, the matrix $S$ in \eqref{unitarios} is unitary.

If the operators are standard then the last formulas reduce to the usual ones. The above formalism is unified for both cases, and will be useful in the following.

\subsection{Classical aspects of in/out fields}
Before deducing the LSZ formula, it is mandatory to deduce some classical relations for the fields in the model. In fact, these considerations are usual in all the standard textbook of QFT. The technical point is that they have to be generalized to a context with equations of motion of fourth order.  The equations of motion  of the Pauli-Uhlenbeck field $\phi(x)$ can be found from the lagrangian \eqref{federica} and is given by
\begin{equation}\label{moushon}
\hat{O} \phi_1=c\frac{\partial L_{int}}{\partial \phi},
\end{equation}
where $L_{int}$ contains the infinite vertices involving the interaction of $\phi_1$ with the rest of the fields. 
The operator $\hat{O}$ is the quartic operator 
$$
\hat{O}= \square^2-\frac{1}{3\gamma}\square,
$$
up to a constant. That is the reason for adding a constant $c$ in the interaction term, these constants are not too relevant in the following discussion, which follows closely the methods of \cite{ryder}. The kernel of the last operator, that is, a Green function for $\hat{O}$, is given by
$$
\hat{O}_y G(y-x)=\delta(y-x).
$$
It is clear that 
$$
G(x-y)=\frac{1}{\hat{O}_y}=\frac{1}{\square^2-\frac{1}{3\gamma}\square}=-\bigg(\frac{1}{\square}-\frac{1}{\square-\frac{1}{3\gamma}}\bigg)3\gamma.
$$
Therefore the Green function $G(x-y)$ is given by
$$
G(x-y)=-3\gamma[G^{m=0}(x-y)-G^{m}(x-y)],
$$
being $G^{m}(x-y)$ any of the standard Green functions for a massive scalar field. There are  plenty of such Green functions, as the addition of a homogeneous solution of the wave operator to any of those leads to a new Green function. In terms of  any of these Green kernels, the following integral relation for $\phi_1$
\begin{equation}\label{aintegrar}
\int[\phi_1(y)\hat{O}_y G(y-x)-G(y-x)\hat{O}_y \phi_1(y)] d^4 y =\phi_1(x)-c\int  G(y-x)\frac{\partial L_{int}}{\partial \phi(y)} d^4 y,
\end{equation}
is found. By assuming that $\phi_1$ vanishes fast enough at the spatial infinite, it is concluded that the spatial part of the left hand side is related to two types of integrals. The first is related to the laplacian $\Delta_y$ in three dimensions, appearing in $\square_y$,. This contribution is
$$
\int[\phi_1(y)\Delta_y G(y-x)-G(y-x)\Delta_y \phi_1(y)] d^3 y dy_0
$$
\begin{equation}\label{argument}
=\int[\phi_1(y)\nabla_y G(y-x)-G(y-x)\nabla_y \phi_1(y)] dS_y dy_0.
\end{equation}
The right hand side is obtained from the left by noticing that the left integrand is the gradient of  $\phi_1(y)\nabla_y G(y-x)-G(y-x)\nabla_y \phi_1(y)$.  If everything vanish in the infinite surface $S_y$ this gives no contributions. The other spatial contribution, due to the terms with $\square_y^2$, are given by
$$
\int[\phi_1(y)\Delta_y \Delta_y G(y-x)-G(y-x)\Delta_y \Delta_y\phi_1(y)] d^3 y dy_0=0,
$$
after integration by parts. There are also terms in $\square^2_y$ mixing spatial and time derivatives. These are of the form
$$
\int[\phi_1(y)\Delta_y \partial^2_{y_0} G(y-x)-G(y-x)\Delta_y \partial^2_{y_0}\phi_1(y)] d^3 y dy_0
$$
$$
=\int\partial_{y_0}[\phi_1(y)\Delta_y \partial_{y_0} G(y-x)
-G(y-x)\Delta_y \partial_{y_0}\phi_1(y)] d^3 y dy_0
$$
$$
-\int[\partial_{y_0}\phi_1(y)\Delta_y \partial_{y_0} G(y-x)
-\partial_{y_0}G(y-x)\Delta_y \partial_{y_0}\phi_1(y)] d^3 y dy_0.
$$
The last two terms cancel by an argument analogous to \eqref{argument}. Therefore the last equality can be worked out as follows
$$
\int[\phi_1(y)\Delta_y \partial^2_{y_0} G(y-x)-G(y-x)\Delta_y \partial^2_{y_0}\phi_1(y)] d^3 y dy_0
$$
$$
=\int\partial_{y_0}[\phi_1(y)\Delta_y \partial_{y_0} G(y-x)
-G(y-x)\Delta_y \partial_{y_0}\phi_1(y)] d^3 y dy_0
$$
\begin{equation}\label{argument2}
=\int\partial_{y_0}[\Delta_y \phi_1(y)\partial_{y_0} G(y-x)
-G(y-x)\Delta_y \partial_{y_0}\phi_1(y)] d^3 y dy_0.
\end{equation}
The unique spatial contribution is then \eqref{argument2}. The pure time components that come from $\square_y$
are of the form 
$$
\int[\phi_1(y)\partial^2_{y_0} G(y-x)-G(y-x)\partial^2_{y_0} \phi_1(y)] d^4 y 
$$
$$
=\int\partial_{y_0}[\phi_1(y)\partial_{y_0} G(y-x)-G(y-x)\partial_{y_0} \phi_1(y)] d^4 y
$$
\begin{equation}\label{argument3}
=\bigg(\int_{y^-_0} -\int_{y^+_0} \bigg)G(y-x)\overleftrightarrow{\partial}_{y_0} \phi_1(y) d^3 y.
\end{equation}
The first two integrals are three dimensional, the first is made in a spatial surface corresponding to the time $y_0^-$at the past of  $x_0$ and the second at another time $y_0^+$ at the future of $x_0$. The last integral is four dimensional, and is performed on the $4$-volume delimited by these two three dimensional surfaces.
The last pure time contribution is due to the time part of $\square_y^2$. It can be worked out as
$$
\int[\phi_1(y)\partial^4_{y_0} G(y-x)-G(y-x)\partial^4_{y_0} \phi_1(y)] d^4 y 
$$
$$
=\int\partial_{y_0}[G(y-x)\partial^3_{y_0} \phi_1(y)-\phi_1(y)\partial^3_{y_0} G(y-x)] d^4 y
$$
$$
+\int\partial_{y_0}[\partial^2_{y_0} G(y-x)\partial_{y_0}\phi_1(y)-\partial^2_{y_0} \phi_1(y)\partial_{y_0}G(y-x)] d^4 y
$$
\begin{equation}\label{argumento6}
=-\bigg(\int_{y^-_0} -\int_{y^+_0} \bigg)[G(y-x)\overleftrightarrow{\partial}_{y_0} \partial_{y_0}^2\phi_1(y) +\partial_{y_0}^2G(y-x)\overleftrightarrow{\partial}_{y_0} \phi_1(y)]d^3 y.
\end{equation}
By collecting all the above formulas \eqref{argument}-\eqref{argumento6}, the expression \eqref{aintegrar}t becomes 
$$
\phi_1(x)=-\bigg(\int_{y^-_0} -\int_{y^+_0} \bigg)G(y-x)\overleftrightarrow{\partial}_{y_0} \phi_1(y) d^3 y 
$$
$$
+\frac{1}{3\gamma}\bigg(\int_{y^-_0} -\int_{y^+_0} \bigg)[G(y-x)\overleftrightarrow{\partial}_{y_0} \partial_{y_0}^2\phi_1(y) +\partial_{y_0}^2G(y-x)\overleftrightarrow{\partial}_{y_0} \phi_1(y)]d^3 y
$$
$$
+c\int_{y_0^-}^{y_0^+}  G(y-x)\frac{\partial L_{int}}{\partial \phi(y)} d^4 y.
$$
At the present point $G(y-x)$ can be any Green function, defined by any possible boundary condition. Particularly  useful examples are the advanced  $\Delta_a(x)$ or retarded $\Delta_r(x)$ Green functions, which vanish for $x_0>0$ and $x_0<0$, respectively, and vanish outside the light cone.  If these functions are employed, then the two spatial integrals in the last formulas are reduced into one.  For instance
$$
\phi_1(x)=\int_{y^-_0}\Delta_r(y-x)\overleftrightarrow{\partial}_{y_0} \phi_1(y) d^3 y 
$$
$$
+c\int_{y^-_0}[\Delta_r(y-x)\overleftrightarrow{\partial}_{y_0} \partial_{y_0}^2\phi_1(y) +\partial_{y_0}^2\Delta_r(y-x)\overleftrightarrow{\partial}_{y_0} \phi_1(y)]d^3y
+c\int  \Delta_r(y-x)\frac{\partial L_{int}}{\partial \phi(y)} d^4 y.
$$
The first term in the last formula has a simple interpretation. This term is such that
$$
\hat{O}_x \int_{y^-_0}\Delta_r(y-x)\overleftrightarrow{\partial}_{y_0} \phi_1(y) d^3 y
-\hat{O}_x\int_{y^-_0}[\Delta_r(y-x)\overleftrightarrow{\partial}_{y_0} \partial_{y_0}^2\phi_1(y) +\partial_{y_0}^2\Delta_r(y-x)\overleftrightarrow{\partial}_{y_0} \phi_1(y)]d^3y
$$
$$
=\int_{y^-_0}\delta(y-x)\overleftrightarrow{\partial}_{y_0} \phi_1(y) d^3 y
-\int_{y^-_0}[\delta(y-x)\overleftrightarrow{\partial}_{y_0} \partial_{y_0}^2\phi_1(y) +\partial_{y_0}^2\delta(y-x)\overleftrightarrow{\partial}_{y_0} \phi_1(y)]d^3y.
$$
In the limit  $y_0^-\to -\infty$ the last integral vanishes if $x_0$ is fixed, since this point will be always at the future of $y_0$. Therefore this particular quantity is killed by the action of $\hat{O}_x$ and may be identified as a free field, the incoming field. The previous formula, by taking into account the equations of motion \eqref{moushon}, may be expressed after taking this limit as
$$
\phi_1(x)=\phi_{1 in}(x) +c\int  \Delta_r(y-x)\frac{\partial L_{int}}{\partial \phi(y)} d^4y,
$$
or by use of the equations of motion
\begin{equation}\label{inner}
\phi_1(x)=\phi_{1 in}(x) +c\int  \Delta_r(x-y)\hat{O}_y\phi_1(y) d^4y.
\end{equation}
Here, the second integral can be done in the full space time volume, as the past time is taken to $-\infty$ and the retarded Green function vanishes at the future. 

The last two  formulas make sense, as this field is determined by time derivatives up to order three, consistent with a fourth order equation of motion. 

By applying similar procedures such as the previous paragraphs, it can be deduced as well that
\begin{equation}\label{innere}
\phi_1(x)=\phi_{1 iut}(x) +c\int  \Delta_a(y-x)\frac{\partial L_{int}}{\partial \phi(y)} d^4y,
\end{equation}
or
\begin{equation}\label{outer}
\phi_1(x)=\phi_{1 out}(x) +c\int  \Delta_a(x-y)\hat{O}_y\phi_1(y) d^4y.
\end{equation}
All the above discussion is purely classical, at the moment this solution has not been prompted to operators.

\subsection{A first attempt to write the S-matrix}

The description of this first attempt will be brief by two reasons. First, it is expanded in \cite{yomismo}. Second, the next attempts constitute an improvement, for these reasons, they will be described in much more detail.

From the expansion in modes of the fields
$$
\phi_1(x,t)=\int\frac{d^3k}{(2\pi)^{\frac{3}{2}}\omega(k)}[b(k) e^{i\omega_2(k)t-i k\cdot x}+b^c(k) e^{-i\omega_2(k)t+i k\cdot x}],
$$
$$
\phi_2(x,t)=\int\frac{d^3k}{(2\pi)^{\frac{3}{2}}\omega_2(k)}[c(k) e^{i\omega_2(k)t-i k\cdot x}+c^c(k) e^{-i\omega_2(k)t+i k\cdot x}],
$$
and by taking into account the expression of their derivatives $\dot{\phi}_1$ and $\dot{\phi}_2$
it is an elementary Fourier  exercise to determine $b(k)$, $b^c(k)$, $c(k)$ and $c^c(k)$ in terms of the values of these quantities at $t=0$. The result is
$$
b(k)=\int d^3x e^{ikx}[\omega(k)\phi(x,0)-i\dot{\phi}(x,0)] \qquad b^c(k)=\int d^3x e^{-ikx}[\omega(k)\phi(x,0)+i\dot{\phi}(x,0)],
$$
and analogous formulas are valid for $c(k)$ and $c^c(k)$. These formulas may be expressed in terms of the operation $u\overleftrightarrow{\partial}v=u(\partial v)-(\partial u) v$ as follows
$$
b(k)=-i\int d^3x [e^{-i(\omega(k)t-kx)}\overleftrightarrow{\partial}_0\phi ]\bigg|_{t=0},\qquad b^\dag(k)=i\int d^3x [e^{i(\omega(k)t-kx)}\overleftrightarrow{\partial}_0\phi ]\bigg|_{t=0}.
$$
A small detail is in order. The algebra of these operators has for $\lambda=1$ a non standard $3$ factor, as shown in formula \eqref{commutacion2}. Therefore it is convenient to redefine $\phi\to \sqrt{3}\phi$. In these terms the desired amplitude may be written as
$$
<p_1,.., p_l,\text{out}|r_1, .., r_n,  \text{in}>=i<p_1,.., p_l,\text{out}|b^c_{in}(r_1)|r_2, .., r_n,  \text{in}>
$$
\begin{equation}\label{aplique}
=-\lim_{t\to -\infty}\int d^3x [e^{i(\omega(r_1)t-r_1x)}\overleftrightarrow{\partial}_0<p_1,.., p_l,\text{out}|\phi_{1 in}(x,t)|r_2, .., r_n,  \text{in}> ]\bigg|_{t},
\end{equation}
where in the first step the fact that $b^c(k)|n_k>=i\sqrt{n_k+1}|n_k+1>$.
The standard trick of QFT of converting this expression in a four dimensional integral only requires the use of the mode equation $\square \phi_1=0$. The result, which is standard and follows by repeating the formulas of practically every QFT textbook is \cite{yomismo}
$$
<p_1,.., p_l,\text{out}|r_1, .., r_n,  \text{in}>=(-i)^{n+l}\text{disconnected terms}
$$
\begin{equation}\label{lszalamia}
+Z^{-\frac{n+l}{2}}\int d^4x_1...d^4y_n e^{i\sum_{k=1}^l(\omega(r_k)t_k-r_kx_k)} e^{-i\sum_{m=1}^n(\omega(p_m)t_m-p_mx_m)}
\end{equation}
$$
\square_{y_1}..\square_{y_a}..(\square_{y_b}+M^2)...\square_{x_c}..\square_{x_n}<0|T\phi_1(y_1)..\phi_2(y_a)...|0>.
$$
As expected, everything is found in terms of correlation functions.  

The above approach is safe, but it may be difficult to be implemented. The point is that involves the correlation point of $\phi_1$ and $\phi_2$. The distinction between these modes is clear only asymptotically. For the fully interacting theory, the last expression may be misleading, since it may be non trivial to find a technique to distinguish between these two components, if possible. The correlation function 
$$
<\phi(x_1)..\phi(x_n)>,
$$
may be expressed perturbatively, in terms of free fields $\phi^f(x)$. After that, one try to infer the desired correlation function for mixing between $\phi_1$ and $\phi_2$. These type of procedures may be complicated.  For this reason, it may be interesting to look for alternatives.

In the following two more attempts will be described, which differ each other in the moment the prescription \eqref{aprescribir} is applied.

\subsection{A second attempt for an S-matrix}
Consider now the operator version of these classical identities. Assume now the mode expansion
$$
\phi_{in} =\int\frac{d^3k}{(2\pi)^{\frac{3}{2}}\omega(k)}[b(k) e^{i\omega(k)t-i k\cdot x}+b^\ast(k) e^{-i\omega(k)t+i k\cdot x}]
$$
\begin{equation}\label{modista}
+\int\frac{d^3k}{(2\pi)^{\frac{3}{2}}\omega_2(k)}[c(k) e^{i\omega_2(k)t-i k\cdot x}+c^\ast(k) e^{-i\omega_2(k)t+i k\cdot x}].
\end{equation}
 At this point, it will be considered that the algebra of commutators is the wrong one. The mapping $c(k)\to ic(k)$  and $c^c(k)\to i c^c(k)$ making it healthy will be considered at the end of the calculation. Due to the non standard oscillator algebra of $b(k)$ and $b^\dag(k)$
 $$
[\phi_{in}(x), \phi_{in}(y)]=\int \int\frac{d^3k}{(2\pi)^{\frac{3}{2}}\omega_1(k)}\frac{d^3l}{(2\pi)^{\frac{3}{2}}\omega_1(l)}\bigg[[b(k), b^\dag(l) ] e^{-i\omega(k)x_0+i k\cdot x} e^{i\omega(l)y_0-i l\cdot y}
$$
$$
-[b(l), b^\dag(k) ] e^{i\omega(k)x_0-i k\cdot x} e^{-i\omega(l)y_0+i l\cdot y}\bigg]
 $$
 $$
 +\int \int\frac{d^3k}{(2\pi)^{\frac{3}{2}}\omega_2(k)}\frac{d^3l}{(2\pi)^{\frac{3}{2}}\omega_2(l)}\bigg[[c(k), c^\dag(l) ] e^{-i\omega(k)x_0+i k\cdot x} e^{i\omega(l)y_0-i l\cdot y}
$$
$$
-[c(l), c^\dag(k) ] e^{i\omega(k)x_0-i k\cdot x} e^{-i\omega(l)y_0+i l\cdot y}\bigg]
 $$
$$
=i \int\frac{d^3k}{(\pi)^{\frac{3}{2}}\omega_1(k)}e^{i k\cdot (x-y)}\sin(\omega_1(k)(x_0-y_0))
-i \int\frac{d^3k}{(\pi)^{\frac{3}{2}}\omega_2(k)}e^{i k\cdot (x-y)}\sin(\omega_2(k)(x_0-y_0)).
$$
This is a linear combination of two propagators for a massless and massive particle, but one of them has a non standard minus sign in front. Each of these propagators are known to be written as a difference of a retarded and advanced Green functions
\begin{equation}\label{retame}
\Delta^m(x-y)=\Delta^m_{a}(x-y)-\Delta^m_{ r}(x-y),\end{equation}
and the analogous formula holds for the massless case, with dispersion $\omega_1(k)$. In other words, by denoting the last commutator as $i\Delta(x-y)$ it is clear that
$$
[\phi_{in}(x), \phi_{in}(y)]=<0|[\phi_{1in}(x), \phi_{1in}(y)]|0>=i\Delta(x-y)=i\Delta_a(x-y)-i\Delta_r(x-y)
$$
$$
=i\Delta^{m=0}_a(x-y)-i\Delta^{m=0}_r(x-y)-i\Delta^m_a(x-y)+i\Delta^m_r(x-y),
$$
two propagators are due to the massless mode and two due to the massive mode. The first identity is due to the fact that the commutator is a c-number.  The last expression allow us to infer that
\begin{equation}\label{comuta1}
<0|[\phi_{in}(x), \phi_{in}(y)]|0>=<0|[\phi(x), \phi(y)]|0>.
\end{equation}
This is seen as follows. The last identity, deduced from \eqref{inner} is true  up to a term proportional to 
$$<0|[\phi_{in}(x),\hat{O}_y \phi(y)]|0>.$$ Therefore, the identity will be valid if this term is zero. It is, since 
$$
<0|\hat{O}_y \phi(y)|p>=\hat{O}_y e^{ipy} <0|\phi(0)|p>=p^2(1-(3\gamma)^{-1}p^2)e^{ipy} <0|\phi(0)|p>=0.$$
The last identity is due to the fact that the modes of \eqref{federica} satisfy the dispersion relation $p^2(1-(3\gamma)^{-1}p^2)=0$.
Due to this property the insertion of the identity $\int dp |p><p|$ in this extra piece gives vanishing result and the desired identity is proved. In addition,  the value of $[\phi_{1}(x), \phi_{1}(y)]$ is known to be 
 a c-number \cite{ryder} and therefore \eqref{comuta1} may be replaced by
\begin{equation}\label{comuta3}
[\phi_{in}(x), \phi_{in}(y)]=[\phi(x), \phi(y)]=i\Delta(x-y),
\end{equation}
in other words, the vacuum expectation value may be deleted.

 By collecting all the above information, consider the fundamental operator for scattering in QFT namely the path integral
 $$
Z(J)=<0|I(J)|0>,\qquad I(J)=T e^{i\int \phi(x)J(x)dx},
 $$
 the last quantity satisfies
 $$
-i\frac{\delta I(J)}{\delta J(x)}=T(\phi(x)I(J)).
 $$
 The last formula combined with formulas \eqref{inner} and \eqref{outer} leads to the following formulas
 \begin{equation}\label{inner2}
-i\frac{\delta I(J)}{\delta J(x)}=I(x)\phi_{ in}(x)  +i\int  \Delta_r(x-y)\hat{O}_y\frac{\delta I(J)}{\delta J(y)} d^4y,
\end{equation}
\begin{equation}\label{outer2}
-i\frac{\delta I(J)}{\delta J(x)}=\phi_{ out}(x)I(x) +i\int  \Delta_a(x-y)\hat{O}_y\frac{\delta I(J)}{\delta J(y)} d^4y.
\end{equation}
 From the two equations \eqref{inner2} and \eqref{outer2} it is arrived to
 $$
\phi_{out}(x) I(x)-\phi_{in}(x) I(x)= -i\int  \Delta(x-y)\hat{O}_y\frac{\delta I(J)}{\delta J(y)} d^4y.
 $$
Taking into account the action of the $S$ matrix defined in \eqref{boca} and by multiplying by $S$ the resulting expression  leads to the following commutator
\begin{equation}\label{hair}
[\phi_{in}(x), SI(J)]=i\int  \Delta(x-y)\hat{O}_y\frac{\delta SI(J)}{\delta J(y)} d^4y.
\end{equation}
Note that two factors $\eta$ appear in the last step, one due to \eqref{boca} and the other from the relation \eqref{nofederal}.  As $\eta^2=I$,  the effect of this factors is neglected and no $\eta$ factor appears.

In addition, it has been shown in \eqref{comuta3} that the commutator of the fields is a c-number. By applying the following consequence of the Bakker-Campbell formula, which is true only for commutators which are c-numbers, 
\begin{equation}\label{naomi}
[A, e^B]=[A, B] e^B, 
\end{equation}
it can be seen that the general solution of the last equation \eqref{hair} is 
\begin{equation}\label{horensie}
SI(J)=e^{\int \phi_{ in}(z)\hat{O}_z\frac{\delta}{\delta J(z)}d^4z} F(J),
\end{equation}
with $F(J)$ arbitrary.  The proof is as follows. The last  formula and the last trial function imply
$$
[\phi_{in}(x), SI(J)]=\bigg[\int [\phi_{in}(x), \phi_{in}(z)] \hat{O}_z\frac{\delta}{\delta J(z)}d^4z \bigg]e^{-\int \phi_{ in}(y)\hat{O}_y\frac{\delta}{\delta J(y)}d^4y} F(J)
$$
\begin{equation}\label{inlove}
=-ie^{-\int \phi_{ in}(y)\hat{O}_y\frac{\delta}{\delta J(y)}d^4y}\bigg[\int \Delta(x-z) \hat{O}_z\frac{\delta F(J)}{\delta J(z)}d^4z \bigg].
\end{equation}
This characterizes the left hand side of the equation \eqref{hair}  The right hand side is
$$
-i\int  \Delta(x-y)\hat{O}_y\frac{\delta SI(J)}{\delta J(y)} d^4y=-i\int  \Delta(x-y)\hat{O}_ye^{-\int \phi_{1 in}(z)\hat{O}_z\frac{\delta}{\delta J(z)}d^4z}\frac {\delta F(J)}{\delta J(y)} d^4y
$$
\begin{equation}\label{inlove2}
=-ie^{-\int \phi_{1 in}(z)\hat{O}_z\frac{\delta}{\delta J(z)}d^4z}\int  \Delta(x-y)\hat{O}_y\frac {\delta F(J)}{\delta J(y)} d^4y.
\end{equation}
  The last two formulas \eqref{inlove} and \eqref{inlove2} are equal, and this shows that \eqref{hair} is satisfied for every $F(J)$. 
  This unknown function is determined by taking into account that, for flat Minkowski space, $S|0>=|0>$, and that
  $$
<0|e^A|0>=<0|:e^A:|0>=1,
  $$
  for any operator $A$. In these terms
$$
<0|IS(J)|0>=<0|F(J)|0>=<0|I(J)|0>=Z(J).
$$
In the last step it was assumed that there are no external fields, and that this implies $S|0>=|0>$. In presence of external fields, this hypothesis has to be revised or even abandoned.

Therefore, it is concluded that the scattering operator is given by
\begin{equation}\label{lsz3}
S=:e^{\int \phi_{ in}(z)\hat{O}\frac{\delta}{\delta J(z)}d^4z} :Z(J)\bigg|_{J=0},
\end{equation}
The LSZ reduction formula is then as in standard QFT, but the wave operator is replaced by the quartic wave one $\hat{O}$.  However, at this point, the wrong oscillator algebra was employed. The path integral then does not exist, as shown in the previous section, since there appear real divergent integrals. Following the prescription of continuing $P_2$ and $Q_2$ to imaginary values, and going to the euclidean formulation, the quantity $Z(J)$ becomes the euclidean path integral for the Pais-Uhlenbeck model. This is given in \eqref{euclides}, and written here by convenience
$$
Z_e(J)=\int D\phi \exp\bigg\{-\int L_edVd\tau+\int J \phi dVd\tau \bigg\},
$$
$$
S=\int d\tau \int d^3x[\frac{3}{2}\partial_\mu \phi \partial_\mu \phi+\frac{9\gamma}{2}(\partial_\mu\partial_\mu \phi)^2+V(\phi)].
$$
This mapping is consistent with the use of the standard oscillator algebra 
$$
[b_{k^\prime}, b^c_k]=\frac{2\lambda\omega_1(k)}{3}\delta(k-k^\prime),\qquad [c_{k^\prime}, c^c_k]=\frac{2\lambda\omega_2(k)}{3}\delta(k-k^\prime),
$$
in the in operator of the path integral
$$
\phi_{in} =\int\frac{d^3k}{(2\pi)^{\frac{3}{2}}\omega(k)}[b(k) e^{i\omega(k)t-i k\cdot x}+b^\ast(k) e^{-i\omega(k)t+i k\cdot x}]
$$
$$
+\int\frac{d^3k}{(2\pi)^{\frac{3}{2}}\omega_2(k)}[c(k) e^{i\omega_2(k)t-i k\cdot x}+c^\ast(k) e^{-i\omega_2(k)t+i k\cdot x}],
$$
Therefore the prescription is to employ the last formulas for $Z(J)$ and the fields $\phi_{in}$ with its creation and annihilation operators in the LSZ formula given in  \eqref{lsz3}. After that, go to the Lorenzian setting  by analytically continuing the obtained results $\tau \to i t$. This will lead to the desired scattering rules.



\subsection{Still two more attempts for the S-matrix}

In the second  approach described above, the calculation started with the wrong oscillator algebra. Then, at the end in \eqref{lsz3}, the mapping $P_2$, $Q_2$ to imaginary values, making the oscillator algebra standard, was employed.  It is natural to study the consequences of doing this mapping from the very beginning instead. This is the topic to be discussed now.

There are two ways to impose the standard oscillator algebra right from the scratch. The free asymptotic field  $\phi_{in}=\phi_{1in}+\phi_{2in}$ in \eqref{modista} is composed by an standard component $\phi_{1in}$ and a ghost like $\phi_{2in}$. The mapping of this second component to a standard oscillator algebra $c(k)\to ic(k)$, $c^c(k)\to i c^{c}(k)$ is equivalent to the mapping $$\phi_{in}=\phi_{1in}+\phi_{2in}\to\phi^\prime_{in}= \phi_{1in}+i\phi_{2in}.$$
This can be the initial point for applying the analytic continuation procedure. 

Another possibility is instead to tackle the problem in terms of the propagator. In this approach,
if the standard oscillator is employed, the formula \eqref{retame} changes a sign due to the effect of the mapping $c(k)\to ic(k)$ and $c^c(k)\to i c^{c}(k)$ and becomes
\begin{equation}\label{retame2}
\Delta^m(x-y)=\Delta^m_{a}(x-y)+\Delta^m_{ r}(x-y).\end{equation}
As it will be shown below, both procedures will lead to the same result.

Starting with the second, note that \eqref{retame2} is not the inverse of the quartic wave operator of the model
$$
\hat{O}^{-1}=-\bigg(\frac{1}{\square}-\frac{1}{\square-\frac{1}{3\gamma}}\bigg)3\gamma,
$$
instead, it is the inverse of the modified operator
$$
\hat{O}^m=-[\bigg(\frac{1}{\square}+\frac{1}{\square-\frac{1}{3\gamma}}\bigg)3\gamma]^{-1}.
$$
This can be expressed as
\begin{equation}\label{inthearmy}
\hat{O}^m=\frac{1}{3\gamma}\frac{\square(\square -\frac{1}{3\gamma})}{2\square-\frac{1}{3\gamma}}=\frac{\hat{O}}{6\gamma\square-1}.
\end{equation}
This operator is clearly non local. By certain prescription, one may assume that the free fields are such that
$$
\hat{O}^m\phi_{in}=0,
$$
since the numerator of $\hat{O}^m$  in \eqref{inthearmy} still makes them vanish.  By repeating all the arguments of the previous section, assuming that the full fields are now given by 
$$
\hat{Q}^m \phi=\frac{\partial L_i}{\partial \phi},
$$
one may reach to the formula analogous to \eqref{lsz3}, that is
\begin{equation}\label{lszvarias3}
S=:e^{\int \phi_{ in}(z)\hat{O}^m\frac{\delta}{\delta J(z)}d^4z} :Z(J)\bigg|_{J=0},
\end{equation}
the only difference between the last formula and \eqref{lsz3} is the use of the modified operator $\hat{O}^m$. The fields $\phi_{in}$ correspond to the standard algebra.

Another possibility is to take the fields given by the integral equation \eqref{inner} namely
$$
\phi_1(x)=\phi_{1 iut}(x) +c\int  \Delta_a(y-x)\frac{\partial L_{int}}{\partial \phi(y)} d^4y,
$$
but to change $c(k)\to i c(k)$ and $c^c(k)\to i c^c(k)$ to make the algebra standard, that is, to map 
$$
\phi_{in}=\phi_1+\phi_2\to \phi^m_{in}=\phi_1+i\phi_2.
$$
This is a mapping in the initial conditions, while keeping the classical Euler-Lagrangian differential equation as before.  After a calculation analogous to the above, the following equation \eqref{hair} for the modified field
\begin{equation}\label{hairy2}
[\phi^m_{in}(x), SI(J)]=i\int  \Delta(x-y)\hat{O}_y\frac{\delta SI(J)}{\delta J(y)} d^4y,
\end{equation}
is found. However, since the commutator of 
$$
[\phi^m_{in}(x), \phi^m_{in}(y)]\neq \Delta(x-y),
$$
due to the change $c(k)\to i c(k)$ and $c^c(k)\to i c^c(k)$ , the solution of \eqref{hairy2} is unlikely to be \eqref{horensie}. That is
\begin{equation}\label{horensie2}
SI(J)\neq e^{\int \phi_{ in}(z)\hat{O}_z\frac{\delta}{\delta J(z)}d^4z} F(J).
\end{equation}
Let us try to solve \eqref{hairy2}.  Following \eqref{horensie} it  may be postulated
\begin{equation}\label{horensie}
SI(J)=e^{\int \phi_{ in}(z)\hat{W}_z\frac{\delta}{\delta J(z)}d^4z} F(J),
\end{equation}
with $F(J)$ arbitrary.   The operator $\hat{W}$ is now different than $\hat{O}$. The task is to determine it in order to solve the equation \eqref{hairy2}. The Campbell-Baker formula \eqref{naomi} leads to

$$
[\phi_{in}(x), SI(J)]=\bigg[\int [\phi_{in}(x), \phi_{in}(z)] \hat{W}_z\frac{\delta}{\delta J(z)}d^4z \bigg]e^{-\int \phi_{ in}(y)\hat{W}_y\frac{\delta}{\delta J(y)}d^4y} F(J)
$$
$$
=-ie^{-\int \phi_{ in}(y)\hat{W}_y\frac{\delta}{\delta J(y)}d^4y}\bigg[\int \Delta^m(x-z) \hat{W}_z\frac{\delta F(J)}{\delta J(z)}d^4z \bigg].
$$
Here $\Delta^m(x-y)$ is due to the commutator $[\phi_{in}(x), \phi_{in}(z)] $ and should by no means by confused with $ \Delta(x-z)$.  It has a term with a sign changed due to the change of the oscillator algebra.
The last calculation characterizes the left hand side of the equation \eqref{hairy2}  The right hand side is
$$
-i\int  \Delta(x-y)\hat{O}_y\frac{\delta SI(J)}{\delta J(y)} d^4y=-i\int  \Delta(x-y)\hat{O}_ye^{-\int \phi_{1 in}(z)\hat{W}_z\frac{\delta}{\delta J(z)}d^4z}\frac {\delta F(J)}{\delta J(y)} d^4y
$$
$$
=-ie^{-\int \phi_{1 in}(z)\hat{W}_z\frac{\delta}{\delta J(z)}d^4z}\int  \Delta(x-y)\hat{O}_y\frac {\delta F(J)}{\delta J(y)} d^4y.
$$
  The last two formulas are equal if 
  $$
\Delta^m(x-z) \hat{W}_z=\Delta(x-z) \hat{O}_z,
  $$
  leading to 
   $$
 \hat{W}_z=\frac{\Delta(x-z)}{\Delta^m(x-z)} \hat{O}_z.
  $$
  Since 
  $$
\Delta(x-z)=\bigg(\frac{1}{\square}-\frac{1}{\square-\frac{1}{3\gamma}}\bigg)\frac{1}{3\gamma},\qquad
\Delta^m(x-z)=\bigg(\frac{1}{\square}+\frac{1}{\square-\frac{1}{3\gamma}}\bigg)\frac{1}{3\gamma},
  $$
  Then 
  $$
\frac{\Delta(x-z)}{\Delta^m(x-z)}=\frac{1}{6\gamma\square -1},
  $$
  and therefore
  $$
 \hat{W}_z=\frac{1}{6\gamma\square-1} \hat{O}_z.
  $$
  By comparing the last expression with \eqref{inthearmy} it follows that 
  $$
\hat{W}_z=\hat{O}_z^m.
  $$
 In other words, the same scattering matrix as in \eqref{lszvarias3} is obtained  again! Therefore both methods of changing the initial conditions $\phi_{1 in}+\phi_{2in}\to \phi_{1 in}+i\phi_{2in}$  or to change the wave operator $\hat{O}\to \hat{O}^m$ leads to the same result. These two last quantization schemes are likely equivalent.

  Note however that the last two equivalent schemes are not the same as \eqref{lsz3}.  The path integral is the same, and so will be the correlation functions. Despite this, the scattering amplitudes for these two methods are different due to the different exponential operators acting on  $Z(J)$.

  Not less important,  the exponential in the LSZ formula in this case is non local due to the operator form \eqref{inthearmy}.

The next step is to apply these procedures to the Stelle Quadratic Gravity.

\section{Quadratic Gravity action in synchronous gauge }
Below, the LSZ rules for Quadratic Gravity will be described in detail.  But before, in  order to study the interacting theory,  the full Feynman integral $Z(J)$ should be calculated. This was done in the synchronous gauge in \cite{salvio}  and then generalized to arbitrary gauges. One of the most interesting results of  \cite{salvio} is that it calculates the  $Z(J)$ for this theory in the same fashion as the calculation done in \eqref{aprescribir}-\eqref{euclides}.  The resulting path integral the author finds is equivalent to Quadratic Gravity in an euclidean setting. This is a notable result since Quadratic Gravity is renormalizable and this fundamental property is not spoiled. In more precise terms \cite{stelle1}-\cite{stelle2}, the quantum effective action $\Gamma(h_{\mu\nu})$ can be found and  be renormalized with the help of finite counter terms.  It is natural to study of the LSZ rules for the model. This topic is not as simple is it may look, at least in authors opinion.  Below, an account of the main reasoing of \cite{salvio} is given. The reader may consult the original reference for further details or to take the formulas \eqref{steleuc} -\eqref{Stelleclid} as granted.
 
The action of the model  \cite{stelle1}-\cite{stelle2}, up to a derivative term, may be written as
\begin{equation}\label{cudra}
S=-\frac{1}{2\kappa}\int \sqrt{-g} d^4x\bigg[R+\frac{\gamma}{2} R^2-\frac{\alpha}{2}C_{\mu\nu\alpha\beta}C^{\mu\nu\alpha\beta}\bigg],
\end{equation}
with the first term is the Einstein-Hilbert lagrangian and
$$
C_{\mu\nu\alpha\beta}W^{\mu\nu\alpha\beta}=\frac{1}{2}R_{\mu\nu\rho\sigma}R^{\mu\nu\rho\sigma}-R_{\mu\nu}R^{\mu\nu}+\frac{R^2}{6}.
$$
The metric in the synchronous gauge is
 \begin{equation}\label{sincrono} ds^2 =-dt^2+ g_{ij}(x,t)dx^idx^j.  \end{equation}
In these coordinates the Christofell symbols read as follows
 \begin{equation} \label{Gammas}\Gamma_{\mu\nu}^{\rho}= \frac12 g^{\rho\sigma}(\partial_\mu g_{\sigma\nu}+\partial_\nu g_{\sigma\mu}-\partial_\sigma,g_{\mu\nu})\end{equation}
 \begin{equation} \Gamma^0_{ij}=\Gamma^0_{ji}= \frac12 \dot g_{ij}\equiv -K_{ij}, \quad \Gamma^l_{ij}, \quad \Gamma^l_{0j}=\Gamma^l_{j0}= \frac12 g^{lm}\dot g_{mj} \equiv -K^l_{~j},  \end{equation}
where a dot represents a derivative with respect to $t$. The curvature terms in the synchronous gauge is decomposed as \cite{salvio}
$$ 
R = R^3- 2 g^{ij} \dot K_{ij}-3K_{ij}K^{ij}+K^2, 
$$
$$
R_{\mu\nu}R^{\mu\nu} = \left[g^{ij} (\dot K_{ij}+K_{il}K^l_{~j})\right]^2-2(D_iK-D^jK_{ji})(D^iK-D^lK_l^{~i}) 
$$
$$
+(~R^3_{ij}-\dot K_{ij}-2K_{il}K^l_{~j}+KK_{ij})g^{il}g^{jm}(~R^3_{lm}-\dot K_{lm}-2K_{lp}K^p_{~m}+KK_{lm}),
$$
$$
R_{\mu\nu\rho\sigma}R^{\mu\nu\rho\sigma} = 4(\dot K_{ij}+K_{il}K^l_{~j})g^{il}g^{jm}(\dot K_{lm}+K_{lp}K^p_{~m})
$$
$$
-4(D^jK_{il}-D_lK^j_{~i})(D_jK^{il}-D^lK_j^{~i}) 
$$
\begin{equation}\label{lascurvaturas}
+(~R^3_{ijlm}+K_{il}K_{jm}-K_{jl}K_{im})(~R^{\hspace{-0.4cm}3 \hspace{0.3cm}ijlm}+K^{il}K^{jm}-K^{jl}K^{im}).
\end{equation}
Here \,$R^3_{ijlm}$, \,$R^3_{ij}$, \,$R^3$ and $D_i$ are, respectively, the three-dimensional Riemann tensor, Ricci tensor, Ricci scalar and covariant derivative built with the three-dimensional metric $g_{ij}$.the inverse three-dimensional metric is represented as usual by $g^{ij}$, and the following quantity $K\equiv K_i^{~i}$ was also introduced in those formulas. 

The quantity $\dot K_{ij}$ is time-reversal invariant and appears linearly, while $K_{ij}\equiv -\dot g_{ij}/2$ is odd but  appears quadratically. The full lagrangian is time reversal.

As done in previous section by following the Ostrogradsky method, the variables  $g_{ij}$, $\dot g_{ij}$ or, equivalently, of $g_{ij}$, $K_{ij}$ can be chosen. In this formalism $g_{ij}$ and $K_{ij}$ are independent canonical coordinates. The second choice will employed leading to the following conjugate momentum densities
\begin{equation}\label{momentas}\pi^{ij} \equiv \frac{\partial L}{\partial \dot g_{ij}} -\frac{d}{dt}\frac{\partial L}{\partial \ddot g_{ij}}, \quad P^{ij} \equiv \frac{\partial L}{\partial \dot K_{ij}} -\frac{d}{dt}\frac{\partial L}{\partial \ddot K_{ij}}. \end{equation}
Since $L_s$ is independent of $\ddot K_{ij}$, 
$$
P^{ij} = \frac{\partial L}{\partial \dot K_{ij}}.
$$
This leads to
$$ 
P^{ij} = -2 \sqrt{-g}[G^{ijlm}\dot{K}_{lm} + a K^{ij} K -a g^{ij} K_{lm} K^{lm}+\alpha R^{ij}
$$
\begin{equation}\label{dewo}
-g^{ij}[(\frac{\alpha}3+2 \gamma)(~R^3-3K_{lm}K^{lm}+K^2)], 
\end{equation}
where the  quantity
$$ G^{ijlm}= a\,\frac{g^{il}g^{jm}+g^{im}g^{jl}}2 +\left(4\gamma-\frac{\alpha}{3}\right) g^{ij}g^{lm}, 
$$
has been introduced. Note that, for non vanishing  $\alpha$ and $\beta$ the quantity
\begin{equation}\Gamma_{pqij} \equiv   \frac{1}{a} \frac{g_{pi}g_{qj}+g_{pj}g_{qi}}2 - \frac{4\gamma-\alpha/3}{12\gamma\alpha} g_{pq}g_{ij} \end{equation}
is the inverse of $G^{ijlm}$ , with the definition of inverse given by 
$$ \Gamma_{pqij}G^{ijlm} = \frac{1}{2} (\delta_p^l\delta_q^m+\delta_p^m\delta_q^l).$$
The crucial point is the following. The last expressions show that  $\dot K_{ij}$ can be expressed in terms of $P^{ij}$, $g_{ij}$ and $K_{ij}$. Inserting this expression in $L_s$  the resulting functional  $g_{ij}$, $K_{ij}$ and $P^{ij}$ with no dependence on $\pi^{ij}$. Therefore the Ostrogradsky Hamiltonian 
\begin{equation} H = \pi^{ij} \dot g_{ij} + P^{ij} \dot K_{ij} - L_s. \label{ostroha}\end{equation}
As the Stelle lagrangian $L_s$ does not depend on $\pi_{ij}$ it is clear that the only dependence is given by
\begin{equation} \pi^{ij} \dot g_{ij} = - 2 \pi^{ij} K_{ij}. \label{linearterm}\end{equation}
This is a linear term in $\pi_{ij}$ and, if the path integral formalism is applied, the resulting path integral $Z(J)$ will involve a linear exponential in $\pi_{ij}$ whose integration will diverge. Therefore the path integral would not exist.

The above non existence argument does not hold if the quantization presented in previous sections is applied. This remarkable observation can be found in  \cite{salvio}. The point is that $K_{ij}$ and its conjugate momentum $\pi_{ij}$ are ghost like variables, since they change the sign under time reversal, and in this quantization the Hamiltonian must be continued to imaginary values of these variables, as shown in \eqref{aprescribir}. This continued hamiltonian
$$
H_a(p_l,x_l, P_l, X_l)=H_c(p_l,x_l,-i P_l, -iX_l),
$$
will convert the divergent exponential into an imaginary one, which is harmless.

To see this statement more explicitly, consider the quantization of Stelle gravity in the synchronous gauge.  From \eqref{dewo} it is obtained that

$$
\dot K_{ab}=\Gamma_{abij}\bigg\{\frac{P^{ij}}{ 2 \sqrt{-g}} + \alpha K^{ij} K -\alpha g^{ij} K_{lm} K^{lm}+a_{ij}
$$
$$
-g^{ij}\left[\left(\frac{\alpha}3+2\gamma\right)(~R^{3} -3K_{lm}K^{lm}+K^2\right]\bigg\}, 
$$
By use of this formula replace $\dot{K}_{ij}$ in all the curvatures given in \eqref{lascurvaturas}. The result will depend on  $P^{ij}$, $g_{ij}$ and $K_{ij}$ but not on $\pi_{ij}$, as the last expression \eqref{dewo} does not depend on this quantity. The Hamiltoian is then an expression of the form
$$
H = \pi^{ij} \dot g_{ij} + P^{ij} \dot K_{ij} - L_s(g_{ij}, P^{ij},  K_{ij}).
$$
It is needed to analitically continue the classical Hamiltonian to imaginary values of the variables: $K_{ij} \to i K_{ij}$, $P^{ij} \to -i P^{ij}$ and remembering that $\dot{g}_{ij}=-2K_{ij}$
$$
H =-2i \pi^{ij} K_{ij} -i P^{ij} \dot{K}_{ij} - L_s(g_{ij},-iP^{ij}, i K_{ij}).
$$
In the last expression it is understood that $K_{ij}$ should be written as \eqref{dewo}.
The path integral, in its euclidean version, where the metric is
 $$ ds^2_E =d\tau^2 +g_{ij}(x)dx^idx^j. $$
is given by
$$
Z_e(0)=\int D\pi^{ij} DK_{ij} DP^{ij} Dg_{ij} \exp\bigg\{-\int [i\pi^{ij} \dot g_{ij} + P^{ij} \dot{K}_{ij}-H(g_{ij},\pi_{ij},-iP^{ij}, i K_{ij})]d\tau\bigg\}
$$
$$
=\int D\pi^{ij} DK_{ij} DP^{ij} Dg_{ij}
$$
$$
\exp\bigg\{-\int [i\pi^{ij} \dot g_{ij} + iP^{ij} \dot{K}_{ij}-2i \pi^{ij} K_{ij} -i P^{ij} \dot{K}_{ij} - L_s(g_{ij},-iP^{ij}, i K_{ij})]d\tau\bigg\}
$$
The integration over $\pi^{ij}$ leads to a Dirac delta that enforces $K_{ij} = -\frac{1}{2} g^{\prime}_{ij}$, where the sign $^{\prime}$ indicates derivative with respect to the euclidean time. The path integral now is related to $D\dot{g}_{ij}$ due to this identification, but a change of variable to $Dg_{ij}$  can be done by multiplying the integration measure by $\prod_{x_E} \Delta \tau^{-6}$. The  power of $6$   is due to the fact that $\pi^{ij}$, $g_{ij}$, $K_{ij}$ and $P^{ij}$ have a total of $6$ independent components at each spacetime point.  

The integral over $P^{ij}$ is a Gaussian integral after setting $K_{ij}=-\frac{1}{2} g_{ij}'$ , this follows after inspection of \eqref{lascurvaturas}. The general formula 
$$
\int d^N x e^{x^TAx+b x}=\frac{\pi^{\frac{N}{2}}}{\sqrt{\det A}}e^{-\frac{1}{4}b^TAb},
$$
can be employed to make this integral. The final result is remarkable \cite{salvio}-\cite{salvio2}. For finite initial and final euclidean times, after turning on a current $J^{ij}$, it is
\begin{equation} \label{steleuc}Z_e(J,\tau_i, \tau_f) = \int^{q(\tau_f) = q_f}_{q(\tau_i) = q_i}  \, C Dg_{ij} \, \exp\left(-\frac{S_E}{\hbar}  + \int_{\tau_i}^{\tau_f} d\tau\int d^3x \, J^{ij}g_{ij}\right), \end{equation}
where $S_E$ is the Euclidean version of the Stelle action
\begin{equation} S_E = \int_{\tau_i}^{\tau_f} d\tau\int d^3x\sqrt{g}\left(\frac{\alpha}{2} W_E^2+\gamma R_E^2 + \frac{2}{\kappa^2}R_E \right).\label{Stelleclid}\end{equation}
This result is important for the following reason. The theory that is constructed with this quantization is likely unitary. However, it was not warranted that it will coincide with the euclidean Stelle gravity version. The Stelle theory is known to be renormalizable, and this property would be broken if another lagrangian was to be found. The fact that, at the end, the euclidean version of Quadratic Gravity is obtained, means that all the good properties of the Stelle model will be preserved, in particular renormalizability, while keeping unitarity \cite{salvio}. 
The extra factor in the last path integral is
$$ C\sim \prod_{x_E'} \frac{1}{\sqrt{g(x_E)}}.  $$
This factor follows from the Gaussian integration formula directly. It is well known that it may introduce $\delta^4(0)$ singularities. However, if perturbation around flat spaces $\eta_{\mu\nu}$ are considered and dimensional regularization is employed, the well known Veltmann identities throw these terms to zero.
The quantity $J^{ij}$ is the external ``current" corresponding to $g_{ij}$ in the generating functional.

The above path integral is done in the synchronous gauge. Other gauges may be of interest, such as the De Donder one. The change to a general gauge leads to 
\begin{equation} Z(J) = N\int  \, {\cal D}g \,  \left(\det\frac{\delta f}{\delta\xi}\right)\, \delta(f) \exp\left(-\frac{S_E}{\hbar}  + \int d^4x_E \,J^{\mu\nu}g_{\mu\nu}\right), \label{gengauge}\end{equation}
where the normalization factor is 
$$
N=\frac{1}{\int  \, {\cal D}g \,  \left(\det\frac{\delta f}{\delta\xi}\right)\, \delta(f) \exp\left(-\frac{S_E}{\hbar}  \right)}
$$
The determinant $\det\frac{\delta f}{\delta\xi}$ is the Faddeev-Popov determinant that usually appears when fixing a gauge in such theories. 

After having the euclidean version of the model, the Lorentzian Green's functions may be found in the following way. Replace the euclidean currents $J^{\mu\nu}(\tau)$ with a Lorentzian currents $$J^{kl}(\tau)\to J^{kl}(it)/\hbar, \qquad J^{k4}(\tau)\to iJ^{k0}(it)/\hbar,\qquad J^{44}(\tau)\to -J^{00}(it)/\hbar,$$ and also make the replacement $$g_{kl}(\tau)\to g_{kl}(it),\qquad g_{k4}(\tau)\to -ig_{k0}(it),\qquad g_{44}(\tau)\to -g_{00}(it).$$ 
These substitutions lead to
\begin{equation} \label{lorenziana} Z(J) = N\int  \, Dg \,  \left(\det\frac{\delta f}{\delta\xi}\right)\, \delta(f) \exp\left(\frac{iS}{\hbar}  + i\int d^4x \,J^{\mu\nu}g_{\mu\nu}/\hbar\right), \end{equation}
where ${\cal Z}(J)$ is the generating functional of the Lorentzian Green's functions.
It is defined only as an analytic continuation of the Euclidean one. 

The next task is to characterize the second quantization and the LSZ rules of the model.
\section{Classical and  perturbative aspects of Quadratic Gravity}

In some slight sense, perturbative gravity is analogous to a gauge theory. The Pais-Uhlenbeck model \eqref{federica} does not contain gauge fields. The topic is now to understand how all the above features have to be adapted in presence of gauge like symmetries. This includes Gupta-Bleuler quantization and the determination of the oscillator algebra, taking into account the appearance of gauge dependent terms.
After presenting in the previous section the full path integral $Z(J)$ for the model, the next step is to study the perturbative aspects of the model, in particular the scattering rules.
\subsection{The free Stelle action in the De Donder gauge}
Consider again the Quadratic Gravity action \eqref{cudra} which, up to a total derivative, may be written as
$$
 S_s=\int \sqrt{-g}[-\frac{\gamma}{\kappa^2}R-\beta R^2+\alpha R_{\mu\nu}R^{\mu\nu}]d^4x.
$$
This is a renormalizable theory of gravity, although it contains states with negative norm. The present work describes, following the literature mentioned in the introduction, how to deal with this problem.

Here $\gamma=2$ and $\kappa^2=32 \pi G_N$.  The last action may be linearized in $h^{\mu\nu}=g^{\mu\nu}-\eta^{\mu\nu}$ by  taking into account that $\sqrt{-g}=1+\frac{h}{2}+O(h^2)$, together with the curvature expansions
$$
R_{\mu\nu}=\frac{1}{2}[\partial^\rho\partial_\mu h_{\nu\rho}+\partial^\rho\partial_\nu h_{\mu\rho}-\square h_{\mu\nu}-\partial_\mu \partial_\nu h],\qquad R=\partial_\mu\partial_\nu h^{\mu\nu}-\square h+O(h^2),
$$
These expansions imply that
$$
-\beta R^2=-\beta h\square^2h-\beta h^{\mu\nu}\partial_\mu\partial_\nu\partial_\alpha \partial_\beta h^{\alpha\beta}+2\beta h\square \partial_\mu\partial_\nu h^{\mu\nu}+\text{total derivatives},
$$
$$
\alpha R_{\mu\nu}R^{\mu\nu}=\frac{\alpha}{4}h\square^2 h+\frac{\alpha}{4}h_{\mu\nu}\square^2 h^{\mu\nu}-\frac{\alpha}{2}h_\mu^\sigma\square\partial_\sigma \partial_\nu h^{\nu\mu}
$$
$$
-\frac{\alpha}{2}h\square\partial_\mu \partial_\nu h^{\nu\mu}+\frac{\alpha}{2}h^{\mu\nu}\partial_\mu\partial_\nu\partial_\alpha \partial_\beta h^{\alpha\beta}
+\text{total derivatives},
$$
$$
-\beta R^2+\alpha R_{\mu\nu}R^{\mu\nu}=\frac{\alpha}{4}h^{\mu\nu}\square [\square h_{\mu\nu}-2\partial_\mu \partial^\sigma h_{\nu\sigma}]-(\beta -\frac{\alpha}{4})h\square [\square h-2\partial_\mu \partial_\nu h^{\nu\mu}]
$$
$$
-(\beta -\frac{\alpha}{2})h^{\mu\nu}\partial_\mu\partial_\nu\partial_\alpha \partial_\beta h^{\alpha\beta}+\text{total derivatives}.
$$
At this order, the GR lagrangian reduces to the Fierz-Pauli action
$$
-\frac{1}{16 \pi G }\int \sqrt{g}Rd^4x\simeq -\frac{2}{\kappa^2}\int \bigg[\frac{h^{\mu\nu}}{2}[\square h_{\mu\nu}-2\partial^\rho\partial_{(\mu} h_{\nu)\rho}]-\frac{h}{2}[\square h-2\partial_{\mu}\partial_\nu h^{\mu\nu}]\bigg]d^3x,
$$
this result follows only after expanding  the curvature $R$ up to second order in $h^{\mu\nu}$, and is valid up to total derivative terms. This last result of course has been known for years.

The free part of the Stelle action corresponds to the terms quadratic in $h^{\mu\nu}$. With the above formulas, it is straightforward to find them, the result will be lead to the  generalization of the Fierz-Pauli lagrangian adapted to the present context. The free quadratic Stelle action is
$$
S_s=-\frac{2}{\kappa^2}\int \bigg[\frac{1}{2}h^{\mu\nu}\bigg(1-\frac{\kappa^2 \alpha}{4}\Box\bigg)[\square  h_{\mu\nu}-2\partial^\gamma \partial_{(\nu}h_{\mu)\gamma}]
$$
\begin{equation}\label{reina}
+\frac{1}{2}h\bigg(1-\frac{\kappa^2(4\beta-\alpha)}{4}\square\bigg)[ 2\partial_\alpha \partial_\beta h^{\alpha\beta}-\square  h]
+\frac{\kappa^2}{2}(\beta-\frac{\alpha}{2}) h^{\mu\nu}\partial_\mu\partial_\nu\partial_\alpha \partial_\beta  h^{\alpha\beta}\bigg]d^4x.
\end{equation}
If the parameters $\alpha$ and $\beta$ vanish, the model reduces to   Fierz-Pauli.  This  non gauged Stelle action may be expressed in an alternative form
$$
S_s=-\frac{2}{\kappa^2}\int \bigg[\frac{1}{2}h^{\mu\nu}\bigg(1-\frac{\kappa^2 \alpha}{4}\Box\bigg)[\square  h_{\mu\nu}- 2\partial^\gamma \partial_{(\nu}h_{\mu)\gamma}]
$$
$$
+\frac{1}{2}h\bigg[\frac{2}{3}\bigg(1-\frac{\kappa^2(3\beta-\alpha)}{2}\square\bigg)+\frac{1}{3}\bigg(1-\frac{\kappa^2\alpha}{4}\square\bigg)\bigg][ 2\partial_\alpha \partial_\beta h^{\alpha\beta}-\square  h]
$$
\begin{equation}\label{reinaldo}
+\frac{\kappa^2}{2}(\beta-\frac{\alpha}{2}) h^{\mu\nu}\partial_\mu\partial_\nu\partial_\alpha \partial_\beta  h^{\alpha\beta}\bigg]d^4x.
\end{equation}
The motivation for writing this is to express the action in terms of the mass scales that will be found below in \eqref{anticipo} namely
\begin{equation}\label{travalie}
m_1^2=0,\qquad m_2^2=\frac{4}{\kappa^2\alpha},\qquad m_3^2=\frac{2}{\kappa^2(3\beta-\alpha)}.
\end{equation}
These mass scales appear after calculating the propagator and looking at its poles. However, at this point, the last is just an equivalent  and valid way to write the action. In any case, the equations of motion in vacuum without fixing the gauge are
$$
\bigg(1-\frac{\kappa^2 \alpha}{4}\Box\bigg)\square  h_{\mu\nu}-\eta_{\mu\nu}\bigg[\frac{2}{3}\bigg(1-\frac{\kappa^2(3\beta-\alpha)}{2}\square\bigg)+\frac{1}{3}\bigg(1-\frac{\kappa^2\alpha}{4}\square\bigg)\bigg]\square h
$$
$$
-2\bigg(1-\frac{\kappa^2 \alpha}{4}\Box\bigg)\partial^\gamma \partial_{(\nu}h_{\mu)\gamma}+\eta_{\mu\nu}\bigg[\frac{2}{3}\bigg(1-\frac{\kappa^2(3\beta-\alpha)}{2}\square\bigg)+\frac{1}{3}\bigg(1-\frac{\kappa^2\alpha}{4}\square\bigg)\bigg]\partial_\alpha \partial_\beta h^{\alpha\beta}
$$
$$
+\bigg[\frac{2}{3}\bigg(1-\frac{\kappa^2(3\beta-\alpha)}{2}\square\bigg)+\frac{1}{3}\bigg(1-\frac{\kappa^2\alpha}{4}\square\bigg)\bigg]\partial_\mu \partial_\nu h
$$
\begin{equation}\label{witold}
+\kappa^2(\beta-\frac{\alpha}{2}) \partial_\mu\partial_\nu\partial_\alpha \partial_\beta  h^{\alpha\beta}=0.
\end{equation}
The operator on the right is non invertible, since no gauge has been assumed. The divergence of these equations is
$$
\bigg(1-\frac{\kappa^2 \alpha}{4}\Box\bigg)\square  \partial^\mu h_{\mu\nu}-\bigg[\frac{2}{3}\bigg(1-\frac{\kappa^2(3\beta-\alpha)}{2}\square\bigg)+\frac{1}{3}\bigg(1-\frac{\kappa^2\alpha}{4}\square\bigg)\bigg]\square\partial_\nu h
$$
$$
-\bigg(1-\frac{\kappa^2 \alpha}{4}\Box\bigg)\square \partial^\gamma h_{\nu\gamma}+\bigg[\frac{2}{3}\bigg(1-\frac{\kappa^2(3\beta-\alpha)}{2}\square\bigg)+\frac{1}{3}\bigg(1-\frac{\kappa^2\alpha}{4}\square\bigg)\bigg]\partial_\nu\partial_\alpha \partial_\beta h^{\alpha\beta}
$$
$$
-\bigg(1-\frac{\kappa^2 \alpha}{4}\Box\bigg)\partial_\nu\partial^\gamma  \partial^\mu  h_{\mu\gamma}+\bigg[\frac{2}{3}\bigg(1-\frac{\kappa^2(3\beta-\alpha)}{2}\square\bigg)+\frac{1}{3}\bigg(1-\frac{\kappa^2\alpha}{4}\square\bigg)\bigg]\square \partial_\nu h
$$
\begin{equation}\label{witoldaa}
+\kappa^2(\beta-\frac{\alpha}{2}) \square\partial_\nu\partial_\alpha \partial_\beta  h^{\alpha\beta}=0.
\end{equation}
The first term cancels the third, the second cancels the sixth, and the third and fourth cancel the seventh, from there the result.

 The addition of the gauge fixing term \cite{stelle1}
 $$
L_g=-\frac{\kappa^2}{2\lambda}  h^{\mu\nu}\square  \partial_\mu\partial^\alpha h_{\alpha\nu},
 $$
adds to the equations of motion the term
 $$
-\frac{\kappa^2}{2\lambda}  \square  \partial_\mu\partial^\alpha h_{\alpha\nu}.
 $$
 The divergence of the equations of motion, as shown above, is zero. Therefore this term adds
  $$
-\frac{\kappa^2}{2\lambda}  \square^2\partial^\alpha h_{\alpha\nu}=0.
 $$
 This leads to $\square h_{\mu\nu}=0$ or $\partial^\alpha h_{\alpha\nu}=0$. If one chose
 $\square h_{\mu\nu}=0$, then all the terms with $\square$ in \eqref{witold} vanish and the resulting equation, when multiplied by $\eta^{\mu\nu}$, which leads to the trace, implies that $\partial^\nu \partial^\alpha h_{\alpha\nu}=0$.  This solution is very restrictive.  Instead, consider the possibility of having only  $\partial_\mu h^{\mu\nu}=0$. This is the Lorenz gauge and reduces the equations of motion into
$$
\bigg(1-\frac{\kappa^2 \alpha}{4}\Box\bigg)\square  h_{\mu\nu}-\eta_{\mu\nu}\bigg[\frac{2}{3}\bigg(1-\frac{\kappa^2(3\beta-\alpha)}{2}\square\bigg)+\frac{1}{3}\bigg(1-\frac{\kappa^2\alpha}{4}\square\bigg)\bigg]\square h
$$
$$
+\bigg[\frac{2}{3}\bigg(1-\frac{\kappa^2(3\beta-\alpha)}{2}\square\bigg)+\frac{1}{3}\bigg(1-\frac{\kappa^2\alpha}{4}\square\bigg)\bigg]\partial_\mu \partial_\nu h=0.
$$
Contracting the last differential equation with $\eta^{\mu\nu}$ gives
\begin{equation}\label{horale2}
\bigg(1-\frac{\kappa^2(3\beta-\alpha)}{2}\square\bigg)\square h=0.
\end{equation}
Therefore the trace of the metric $h$ is composed by masses $m^2_1=0$ and $m_3^2$.  The mode with mass $m_2^2$ does not give contribution $h_2$ to the trace. The equations of motion reduce to
$$
\bigg(1-\frac{\kappa^2 \alpha}{4}\Box\bigg)\square  h_{\mu\nu}-\frac{1}{3}\eta_{\mu\nu}\bigg(1-\frac{\kappa^2\alpha}{4}\square\bigg)\square h
$$
\begin{equation}\label{agu}
+\bigg[\frac{2}{3}\bigg(1-\frac{\kappa^2(3\beta-\alpha)}{2}\square\bigg)+\frac{1}{3}\bigg(1-\frac{\kappa^2\alpha}{4}\square\bigg)\bigg]\partial_\mu \partial_\nu h=0.
\end{equation}
Decompose the trace $h$ into its components $h_1$ and $h_3$. The last equation is then
$$
-\bigg(1-\frac{\kappa^2 \alpha}{4}m_3^2\bigg)m_3^2  h^3_{\mu\nu}+\frac{1}{3}\eta_{\mu\nu}m_3^2\bigg(1-\frac{\kappa^2\alpha}{4}m_3^2\bigg) h_3
$$
\begin{equation}\label{remando}
\partial_\mu\partial_\nu h_1+\frac{1}{3}\bigg(1-\frac{\kappa^2\alpha}{4}m_3^2\bigg)\partial_\mu \partial_\nu h_3=0.
\end{equation}
The task is now to constraints the polarizations arising from the chosen gauge $\partial_\mu h^{\mu\nu}=0$. For this, note that the Lorenz gauge condition implies that
$$
k^i_\mu \epsilon^{i\mu\nu}=0,
$$
where the index $i$ is indicating that the four momenta for every mode are different, since they satisfy a different mass shell relation. All three polarizations $\epsilon^i_{\mu\nu}$ with $i=1,2,3$ satisfy the last condition. Also, the fact that $h_2$ is zero is equivalent to
$$
\epsilon_2^{p\mu\nu}\eta_{\mu\nu}=0.
$$
The remaining condition follows from \eqref{remando} and is
$$
-\frac{1}{\omega_3(k)}\bigg(1-\frac{\kappa^2 \alpha}{4}m_3^2\bigg)m_3^2  \epsilon^3_{\mu\nu}+\frac{1}{3\omega_3(k)}\eta_{\mu\nu}m_3^2\bigg(1-\frac{\kappa^2\alpha}{4}m_3^2\bigg) \epsilon_3
$$
$$
-\frac{1}{3\omega_3(k)}\bigg(1-\frac{\kappa^2\alpha}{4}m_3^2\bigg)k^3_\mu k^3_\nu \epsilon_3=0,
$$
$$
k_\mu k_\nu \epsilon_1=0.
$$
The second one follows since $h_1$ and $h_3$ have different Fourier expansions and have to be considered as independent. But this implies that $\epsilon_1=0$. Contracting the first with $k^{3\mu}$ and taking into account that leads to nothing. Therefore $h_2=h_1=0$ and the polarization $\epsilon^i_{\mu\nu}$ is orthogonal to $k_i^\mu$ (this is a bit loosely speaking, as the definition of orthogonality in four dimensions is subtle).  If the direction of propagation is $\hat{z}$ it is known that the polarizations for the massless mode are
\begin{equation}\label{polariz}
\epsilon^+_{1\mu\nu}=\left(\begin{array}{cccc}
  0  & 0  & 0  & 0\\
  0  & 1 & 0  &  0\\
  0 & 0 & -1  & 0 \\ 
  0 & 0  &  0 & 0
\end{array}\right),\qquad 
\epsilon^x_{1\mu\nu}=\left(\begin{array}{cccc}
  0  & 0  & 0  & 0\\
  0  & 0 & 1  &  0\\
  0 & 1 & 0  & 0 \\ 
  0 & 0  &  0 & 0
\end{array}\right).
\end{equation}
For the massive mode there more possible polarizations. A massive graviton with mass $m$ is not relativistic and in the rest frame its 4-momentum is $k^\mu=(m,0,0,0)$.  A linearly independent set satisfying the transversality and trace zero conditions are
$$
\epsilon^{+}_{\mu\nu}=\left(\begin{array}{cccc}
  0  & 0  & 0  & 0\\
  0  & 1 & 0  &  0\\
  0 & 0 & -1  & 0 \\ 
  0 & 0  &  0 & 0
\end{array}\right),\qquad \epsilon^{1}_{\mu\nu}=\left(\begin{array}{cccc}
  0  & 0  & 0  & 0\\
  0  & 1 & 0  &  0\\
  0 & 0 & 0  & 0 \\ 
  0 & 0  &  0 & -1
\end{array}\right),
$$
\begin{equation}\label{polariza}
\epsilon^{x}_{\mu\nu}=\left(\begin{array}{cccc}
  0  & 0  & 0  & 0\\
  0  & 0 & 1  &  0\\
  0 & 1 & 0  & 0 \\ 
  0 & 0  &  0 & 0
\end{array}\right),\qquad \epsilon^{2}_{\mu\nu}=\left(\begin{array}{cccc}
  0  & 0  & 0  & 0\\
  0  & 0 & 0  &  1\\
  0 & 0 & 0  & 0 \\ 
  0 & 1  &  0 & 0
\end{array}\right),\qquad \epsilon^{3}_{\mu\nu}=\left(\begin{array}{cccc}
  0  & 0  & 0  & 0\\
  0  & 0 & 0  &  0\\
  0 & 0 & 0  & 1 \\ 
  0 & 0  &  1 & 0
\end{array}\right).
\end{equation}
Defining the sum over polarizations as
$$
P_{\mu\nu,\alpha\beta}=\sum_{\lambda} \epsilon^\lambda_{1\mu\nu} \epsilon^\lambda_{1\alpha\beta},
$$
it follows that the non zero components are
$$
P_{00,00}=0,\qquad P_{11,11}=2,\qquad P_{22,22}=P_{33,33}=1,
$$
$$
P_{12,12}=P_{21,21}=P_{13,13}=P_{31,31}=P_{23,23}=P_{32,32}=1,
$$
$$
P_{12,21}=P_{21,12}=P_{13,31}=P_{31,13}=P_{23,32}=P_{32,23}=1.
$$
By making a Lorenz transformation to a moving frame $k^\mu=(\omega(k),0,0,k)$ 
\begin{equation}\label{boost}
\Lambda^\mu_\nu=\frac{1}{m}\left(\begin{array}{cccc}
  \omega  & 0  & 0  & k\\
  0  & m & 0  &  0\\
  0 & 0 & m  & 0 \\ 
 k & 0  &  0 & \omega
\end{array}\right),
\end{equation}
the sum of polarizations can be calculated as
$$
P^\prime_{\mu\nu,\alpha\beta}=\Lambda^\gamma_\mu \Lambda^\delta_\nu \Lambda^\epsilon_\alpha\Lambda^\kappa_\beta P_{\gamma\delta,\epsilon\kappa}.
$$
The sum depends on the masses. The polarization tensors in the new frame can also be calculated by making Lorentz transformations to the above ones \eqref{polariza}.

The last thing to analyze is the third mode. After setting  $h_1=h_2=0$ it is reached to the equation 
$$
\bigg(1-\frac{\kappa^2 \alpha}{4} m_3^2\bigg)m_3^2  h_{\mu\nu}-\frac{1}{3}\eta_{\mu\nu}\bigg(1-\frac{\kappa^2\alpha}{4}m_3^2\bigg)m_3^2 h
+\frac{1}{3}\bigg(1-\frac{\kappa^2\alpha}{4}m_3^2\bigg)\partial_\mu \partial_\nu h=0.
$$
The polarization is then
\begin{equation}\label{horale}
\epsilon^3_{\mu\nu}=\frac{1}{3}\bigg(\eta_{\mu\nu}+\frac{k_\mu k_\nu}{m_3^2}\bigg) \epsilon.
\end{equation}
Therefore if the direction of propagation is $k_\mu=(\omega(k), 0, 0, k_3)$ the polarization matrices reduce to a single one given by
\begin{equation}\label{teplenie}
\epsilon^{3}_{\mu\nu}=\frac{1}{3}\left(\begin{array}{cccc}
  -1 +\frac{\omega_3^2}{m_3^2} & 0  & 0  & \frac{k\omega_3}{m_3^2}\\
  0  & 1 & 0  &  0\\
  0 & 0 & 1  & 0 \\ 
  \frac{k\omega_3}{m_3^2} & 0  &  0 & 1+\frac{k^2}{m_3^2}
\end{array}\right)\epsilon.
\end{equation}
The field is decomposed as
$$
h_{in}^{\mu\nu}=\sum_p\int\frac{d^3k}{(2\pi)^{\frac{3}{2}}\omega(k)}[\epsilon_{1\mu\nu} e^{i\omega(k)t-i k\cdot x}+\epsilon_{1\mu\nu}^\ast e^{-i\omega(k)t+i k\cdot x}]
$$
$$
+\sum_p\int\frac{d^3k}{(2\pi)^{\frac{3}{2}}\omega_2(k)}[\epsilon_{2\mu\nu} e^{i\omega_2(k)t-i k\cdot x}+\epsilon^\ast_{2\mu\nu} e^{-i\omega_2(k)t+i k\cdot x}].
$$
$$
+\sum_p\int\frac{d^3k}{(2\pi)^{\frac{3}{2}}\omega_3(k)}[\epsilon_{3\mu\nu} e^{i\omega_3(k)t-i k\cdot x}+\epsilon^\ast_{3\mu\nu} e^{-i\omega_3(k)t+i k\cdot x}].
$$
The sum is over all allowed polarizations, described above. Note that the third mode contains only one polarization, and it may look to be an scalar. However, for a further visualization of this fact, it may be interesting to study several gauges. This will be done in subsequent sections.

The next topic is how this gauge conditions are imposed after quantization on the physical states. The Gupta-Bleuler method is an effective way for doing that, as it is a successful method in gauge theories.

\subsubsection{The Gupta-Bleuler quantization scheme}
Consider again the Lorenz gauge $\partial_\mu h^{\mu\nu}=0$.  When making quantization of the model, one should be aware of imposing these classical considerations for the polarization operators. This awareness may be inspired by lessons about the Gupta-Bleuler quantization. Sometimes to impose a gauge at operator level, such as the Lorenz gauge $\partial_\mu A^\mu=0$ in QED, may be an over request. Instead, imposing a softer condition to the physical states is enough.  An example is  $\partial_\mu A^{+\mu}|\Psi>=0$, a condition heavily employed in this quantization method. The mean values of $\partial_\mu A^\mu$ as an operator is zero with this mild condition.

Consider the non gauged equations of motion without fixing given in \eqref{opes}. As before, the Lorenz gauge $\partial_\mu h^{\mu\nu}=0$ will be considered. However, both \eqref{opes} and the gauge has to be valid only when mean values between physical states $|\Psi>$ are taken. The field is expanded as
$$
h^{\mu\nu}=\sum_p\int\frac{d^3k}{(2\pi)^{\frac{3}{2}}\omega(k)}[a_{\mu\nu} a_p(k)e^{i\omega(k)t-i k\cdot x}+a_{\mu\nu}^\ast a_p^c(k)e^{-i\omega(k)t+i k\cdot x}]
$$
$$
+\sum_p\int\frac{d^3k}{(2\pi)^{\frac{3}{2}}\omega_2(k)}[b_{\mu\nu}b_p(k) e^{i\omega_2(k)t-i k\cdot x}+b^\ast_{\mu\nu} b_p^c(k)e^{-i\omega_2(k)t+i k\cdot x}].
$$
$$
+\sum_p\int\frac{d^3k}{(2\pi)^{\frac{3}{2}}\omega_3(k)}[c_{\mu\nu}c_p(k) e^{i\omega_3(k)t-i k\cdot x}+c^\ast_{\mu\nu} c_p^c(k)e^{-i\omega_3(k)t+i k\cdot x}].
$$
Here $a_p(k)$, $b_p(k)$ and $c_p(k)$ are annihilation operators, and the ones with the upperindex "c" are the creation ones. The notation "c" replaces $\dag$ in order to leave the possibility for a non standard creation-annihiliation algebra, which is a typical feature when ghost appear, as in the Stelle model. 

The sum above is over all the polarizations, not only the allowed by the gauge conditions. The physical states  will restrict these sums. Therefore the Gupta-Bleuler request is
$$
\partial^+_\mu h^{\mu\nu}|\Psi>=0.
$$
This implies that 
$$
\sum_{\lambda}k^\mu \epsilon^\lambda_{1\mu\nu} a_\lambda(k)|\Psi>=0,\qquad \sum_{\lambda}k^\mu \epsilon^\lambda_{2\mu\nu} b_\lambda(k)|\Psi>=0,
$$
\begin{equation}\label{condicion1}
\sum_{\lambda}k^\mu \epsilon^\lambda_{3\mu\nu} c_\lambda(k)|\Psi>=0.
\end{equation}
Here $\lambda$ included all the polarizations, not only \eqref{polariz} and \eqref{polariza}. The last gives a linear relation between the annihilation operators corresponding to  these additional polarizations, whose action destroys the physical states. The averaged equations of motion are now
$$
<\bigg(1+\frac{\kappa^2 \alpha}{4}\Box\bigg)\square  h_{\mu\nu}>-\eta_{\mu\nu}<\bigg[\frac{2}{3}\bigg(1+\frac{\kappa^2(3\beta-\alpha)}{2}\square\bigg)+\frac{1}{3}\bigg(1+\frac{\kappa^2\alpha}{4}\square\bigg)\bigg]\square h>
$$
$$
+<\bigg[\frac{2}{3}\bigg(1+\frac{\kappa^2(3\beta-\alpha)}{2}\square\bigg)+\frac{1}{3}\bigg(1+\frac{\kappa^2\alpha}{4}\square\bigg)\bigg]\partial_\mu \partial_\nu h>=0.
$$
Here the average indicates evaluation with a physical state $|\Psi>$.
Contracting the last differential equation with $\eta^{\mu\nu}$ gives
$$
<\bigg(1+\frac{\kappa^2(3\beta-\alpha)}{2}\square\bigg)\square h>=0.
$$
The fact that the  mean values do not contain the massive mode with $m_2^2$  imply that
\begin{equation}\label{condicion2}
\sum_{\lambda}\epsilon_2^\lambda b_\lambda(k)|\Psi>=0.
\end{equation}
The states \eqref{polariz} and \eqref{polariza} have zero traces.  The remaning are not traceless and have non zero $\epsilon^\lambda$. The condition \eqref{condicion2} gives a new linear relation for the annihilation operators.  Now, with this new feature, the averaged equations are
$$
<\bigg(1+\frac{\kappa^2 \alpha}{4}\Box\bigg)\square  h_{\mu\nu}>-<\frac{1}{3}\eta_{\mu\nu}\bigg(1+\frac{\kappa^2\alpha}{4}\square\bigg)\square h>
$$
$$
+<\bigg[\frac{2}{3}\bigg(1+\frac{\kappa^2(3\beta-\alpha)}{2}\square\bigg)+\frac{1}{3}\bigg(1+\frac{\kappa^2\alpha}{4}\square\bigg)\bigg]\partial_\mu \partial_\nu h>=0.
$$
The absence of the mode with $m_2^2$ in the average leads to 
$$
-\bigg(1-\frac{\kappa^2 \alpha}{4}m_3^2\bigg)m_3^2  <h^3_{\mu\nu}>+\frac{1}{3}\eta_{\mu\nu}m_3^2\bigg(1-\frac{\kappa^2\alpha}{4}m_3^2\bigg) <h_3>
$$
\begin{equation}\label{remandolas}
\partial_\mu\partial_\nu <h_1>+\frac{1}{3}\bigg(1-\frac{\kappa^2\alpha}{4}m_3^2\bigg)\partial_\mu \partial_\nu <h_3>=0.
\end{equation}
The independence between the modes $1$ and $3$ leads to 
$$
k_\mu k_\nu\sum_{\lambda}\epsilon_1^\lambda a_\lambda(k)|\Psi>=0.
$$
For the third mode, it is deduced from \eqref{horale} that 
$$
\sum_{\lambda}\epsilon^{3\lambda}_{\mu\nu}c_\lambda(k)|\Psi>=\sum_{\lambda}\frac{1}{3}\bigg(\eta_{\mu\nu}+\frac{k_\mu k_\nu}{m_3^2}\bigg) \epsilon^\lambda c_\lambda(k)|\Psi>.
$$
Quantization should be performed by considering general polarizaion and assuming that the physical states are constrained by the above conditions.

\subsection{The free action in the Stelle gauge}

The original reference \cite{stelle1} employs  the gauge fixing term described in the previous section, which is a generalization of the Lorenz gauge fixing term of GR, but with the particularity that induces terms which decay as $k^{-4}$. This behavior makes the renormalization analysis presented in that reference more accessible. It is not forbidden to employ other gauges, however, the renormalizability issues raised in that classic reference are much harder to be achieved. For this reason, it is important to discuss this gauge in some extent. 

In some contexts, it is sometimes more convenient to express the action \eqref{reina} and its equations of motion in terms of the Barnes- Rivers operators $P^a_{\mu\nu\alpha\beta}$, which  are given in terms of the elementary tensors,
\begin{equation}\label{ripol2}
\omega_{\mu\nu}=\frac{k_\mu k_\nu}{k^2}, \qquad \theta_{\mu\nu}=\eta_{\mu\nu}-\frac{k_\mu k_\nu}{k^2},
\end{equation}
in the following way
$$
P^2_{\alpha \beta,\mu\nu}=\frac{1}{2}(\theta_{\beta \mu}\theta_{\alpha\nu}+\theta_{\beta\nu}\theta_{\alpha\mu})-\frac{1}{3}\theta_{\beta\alpha}\theta_{\mu\nu},
$$
$$
P^1_{\alpha \beta,\mu\nu}=\frac{1}{2}(\theta_{\beta\mu}\omega_{\alpha\nu}+\theta_{\beta\nu}\omega_{\alpha\mu}+\theta_{\alpha\mu}\omega_{\beta\nu}+\theta_{\alpha\nu}\omega_{\beta\mu}),
$$
$$
P^{0-s}_{\alpha \beta,\mu\nu}=\frac{1}{3}\theta_{\beta\alpha}\theta_{\mu\nu},\qquad
P^{0-w}_{\alpha \beta,\mu\nu}=\omega_{\beta\alpha}\omega_{\mu\nu},
$$
\begin{equation}\label{ripol}
P^{0-sw}_{\alpha \beta,\mu\nu}=\frac{1}{\sqrt{3}}\theta_{\beta\alpha}\omega_{\mu\nu},\qquad
P^{0-ws}_{\alpha \beta,\mu\nu}=\frac{1}{\sqrt{3}}\omega_{\beta\alpha}\theta_{\mu\nu}.
\end{equation}
These formulas may be expressed in the coordinate space by making the change
$$
\omega_{\mu\nu}=\frac{\partial_\mu \partial_\nu}{\square}, \qquad \theta_{\mu\nu}=\eta_{\mu\nu}-\frac{\partial_\mu \partial_\nu}{\square},
$$
resulting in
$$
P_{\mu\nu\alpha\beta}^2=\frac{2}{3\square^2}\partial_\mu \partial_\nu \partial_\alpha \partial_\beta
+\frac{1}{3\square}(\partial_\mu \partial_\nu \eta_{\alpha\beta}+\partial_\alpha \partial_\beta \eta_{\mu\nu})
$$
$$
-\frac{1}{2\square}(\partial_\mu \partial_\alpha \eta_{\nu\beta}+\partial_\mu \partial_\beta \eta_{\nu\alpha}+\partial_\nu \partial_\beta \eta_{\mu\alpha}+\partial_\alpha \partial_\nu\eta_{\mu\beta})
$$
$$
+\frac{1}{2}(\eta_{\mu\alpha}\eta_{\nu\beta}+\eta_{\nu\alpha}\eta_{\mu\beta})-\frac{1}{3}\eta_{\mu\nu}\eta_{\alpha\beta},
$$
$$
P_{\mu\nu\alpha\beta}^1=-\frac{1}{\square^2}\partial_\mu \partial_\nu \partial_\alpha \partial_\beta+\frac{1}{2\square}(\partial_\mu \partial_\alpha \eta_{\nu\beta}+\partial_\mu \partial_\beta \eta_{\nu\alpha}+\partial_\nu \partial_\beta \eta_{\mu\alpha}+\partial_\alpha \partial_\nu\eta_{\mu\beta}),
$$
$$
P_{\mu\nu\alpha\beta}^{0-s}=-\frac{1}{3\square}(\partial_\alpha \partial_\beta \eta_{\mu\nu}+\partial_\mu \partial_\nu \eta_{\alpha\beta})+\frac{1}{3}\eta_{\mu\nu}\eta_{\alpha\beta},\qquad 
P_{\mu\nu\alpha\beta}^{0-w}=\frac{1}{\square^2}\partial_\mu \partial_\nu \partial_\alpha \partial_\beta,
$$
$$
P^{0-sw}_{\mu\nu\alpha\beta}=\frac{1}{\sqrt{3}\square}\eta_{\mu\nu}\partial_\alpha \partial_\beta-\frac{1}{\sqrt{3}\square^2}\partial_\mu \partial_\nu \partial_\alpha \partial_\beta,
$$
\begin{equation}\label{fortu}
P^{0-ws}_{\alpha \beta,\mu\nu}=\frac{1}{\sqrt{3}\square}\eta_{\alpha\beta}\partial_\mu \partial_\nu-\frac{1}{\sqrt{3}\square^2}\partial_\mu \partial_\nu \partial_\alpha \partial_\beta.
\end{equation}
The definition of the  Barnes-Rivers operators shows that
$$
(P^{2}+P^1+P^{0-2}+P^{0-w})_{\mu\nu}^{\;\;\;\alpha\beta}=I_{\alpha\beta,\mu\nu}=\frac{1}{2}(\delta^\alpha_\mu \delta^\beta_\nu+\delta^\alpha_\nu \delta^\beta_\mu).
$$
In other words, the last sum is an identity. Therefore 
\begin{equation}\label{proyectala1}
(P^{2}+P^1+P^{0-2}+P^{0-w})_{\mu\nu}^{\;\;\;\alpha\beta} h^{\mu\nu}=h^{\alpha\beta}.
\end{equation}
In addition  the set composed by these four operators are such that
\begin{equation}\label{proyectala2}
P^{i-a} P^{j-b}=\delta^{ij}\delta^{ab}P^{j-b},
\end{equation}
where $i, j$ takes values between $0$ and $2$ and $a, b$ takes values $s$ and $w$.  Furthermore, the mixed operators $P^{0-ws}$ and $P^{0-sw}$ are such
$$
P^{0-ws}P^{0-sw}=P^{0-w},\qquad P^{0-sw}P^{0-ws}=P^{0-s},
$$
and their non zero actions with the other is
$$
P^{i-ab}P^{j-c}=\delta^{ij} \delta^{bc}P^{j-ac},\qquad P^{i-a}P^{j-bc}=\delta^{ij} \delta^{ab}P^{j-ac}.
$$
Note that the first four operators are projection operators, due to \eqref{proyectala2}. 

The terms of the linearized Stelle  action \eqref{reina} can be worked out in terms of the Barnes-Rivers operators as follows. Consider a typical term of the linearized action, mapped to momentum space
$$
h^{\alpha\beta}\eta_{\alpha\mu}\eta_{\beta\nu} \square^2 h^{\mu\nu}\to k^4h^{\alpha\beta}\eta_{\alpha\mu}\eta_{\beta\nu}  h^{\mu\nu}.
$$
By employing the definition \eqref{ripol2}, it is evident that $\eta_{\mu\nu}=\theta_{\mu\nu}+\omega_{\mu\nu}$, and with this simple identity the last term in momentum space may be worked out as
$$
k^4h^{\alpha\beta}\eta_{\alpha\mu}\eta_{\beta\nu}  h^{\mu\nu}=k^4 h^{\alpha\beta}(\theta_{\alpha\mu}\theta_{\beta\nu}+\theta_{\alpha\mu}\omega_{\beta\nu}+\omega_{\alpha\mu}\theta_{\beta\nu}+\omega_{\alpha\mu}\omega_{\beta\nu})  h^{\mu\nu}
$$
$$
=k^4h^{\alpha\beta}(P^{2}_{\alpha\beta, \mu\nu}+P^{0}_{\alpha\beta, \mu\nu}+P^{1}_{\alpha\beta, \mu\nu}+P^{0-w}_{\alpha\beta, \mu\nu})  h^{\mu\nu}.
$$
In the last identity, the definition of the Barnes Rivers operators \eqref{ripol} was employed.  After going back to coordinate space, and doing the same with the other terms of the linearized action, the result is
$$
S=\frac{2}{\kappa^2}\int h^{\mu\nu}\square\bigg[\frac{1}{2}\bigg(1-\frac{\kappa^2 \alpha}{4}\square \bigg) P^2-\bigg(1-\frac{\kappa^2(\alpha-3\beta)}{2}\square\bigg) P^{0-s}
$$
\begin{equation}\label{opes}
+\bigg(\frac{1}{2}-\frac{\kappa^2(4\beta-3\alpha)\square}{8}\bigg) P^{0-w}\bigg]_{\alpha\beta,\mu\nu} h^{\alpha\beta}d^4x.
\end{equation}
 There are several identities satisfied for these operators, which have been worked out in the literature, for instance in \cite{barnes1}. One of the advantages of expressing the action in terms of the Barnes-Rivers operators is that, given an expression in terms of the Barnes-Rivers operators
$$
M=a_2 P^2+a_1 P^1+a_s P^{0-s}+a_w P^{0-w}+a_{sw}\sqrt{3}(P^{0-sw}+P^{0-ws}),
$$
its inverse is 
\begin{equation}\label{inverso}
M^{-1}=\frac{1}{a_2} P^2+\frac{1}{a_1} P^1+\frac{1}{a_s a_w-3 a_{sw}^2}\bigg[a_w P^{0-s}+a_s P^{0-w}-a_{sw}\sqrt{3}(P^{0-sw}+P^{0-ws})\bigg].
\end{equation}
Here the inverse means is related to the identity with four indices
$$
M_{\mu\nu \gamma\delta}(M^{-1})^{\gamma\delta \alpha\beta}=(M^{-1})^{\gamma\delta \alpha\beta}M_{\mu\nu \gamma\delta}=I_{\mu\nu}^{\alpha\beta}=\frac{1}{2}(\delta^\alpha_\mu \delta^\beta_\nu+\delta^\alpha_\nu \delta^\beta_\mu).
$$
The formula \eqref{inverso} shows  the advantage of employing these operators. Given an action written as 
$$
S=\int h^{\mu\nu}(x)\hat{O}_{\mu\nu\alpha\beta} h^{\alpha\beta}(x) d^4x,
$$with the operator $\hat{O}_{\mu\nu\alpha\beta}$ expressed in terms of the Barnes-Rivers operators, then \eqref{inverso} gives directly the propagator $\hat{D}_{\mu\nu\alpha\beta}=\hat{O}^{-1}_{\mu\nu\alpha\beta}$. Of course, there is no inverse for the operator in \eqref{opes}, since it is not gauge fixed and therefore non invertible. It is easy to see why it is not invertible. The coefficient for $P^1$ is zero and the inverse involves the inverse of the coefficient, which is infinite.

Now, a gauge has to be specified in order to obtain an invertible kinetic operator, leading to a non divergent propagtor.  The gauge fixing term will be the one  of the previous section namely
\begin{equation}\label{calibre}
L_{gf}=-\frac{\kappa^2}{2\lambda}F_\tau\square F^\tau.
\end{equation}
Here $F^\tau= \partial_\nu h^{\tau\nu}$.  The use of this gauge may be advantageous in some applications and non practical in others. The advantage of it is that it leads to terms in the gauge fixed propagators which behave like $k^{-4}$ and which facilitate the renormalization program of that reference. 

The gauge fixed action in the Stelle gauge follows by expressing the gauge fixing part in terms of the Barnes-Rivers operators, the result is 
$$
S=\frac{2}{\kappa^2}\int h^{\mu\nu}\square\bigg[\frac{1}{2}\bigg(1-\frac{\kappa^2 \alpha}{4}\square \bigg) P^2-\bigg(1-\frac{\kappa^2(\alpha-3\beta)}{2}\square\bigg) P^{0-s}
$$
$$
+\bigg(\frac{1}{2}-\frac{\kappa^2(4\beta-3\alpha)\square}{8}\bigg) P^{0-w}+\frac{\kappa^4}{8\lambda}\square^2[P^1-P^{0-w}]\bigg]_{\alpha\beta,\mu\nu} h^{\alpha\beta}d^4x.
$$
 The inverse of the kinetic operator that can be read from this expression is the gauge fixed propagator, by employing \eqref{inverso}
$$
D_{\alpha\beta, \mu\nu}=\frac{1}{\square}\bigg[\frac{2}{1-\frac{\kappa^2\alpha}{4}\square } P^2+\frac{8\lambda}{\kappa^4\square} P^1-\frac{1}{1-\frac{\kappa^2}{2}(3\beta-\alpha)\square } P^{0-s}
$$
$$
+\frac{2}{1-\frac{\kappa^4}{4\lambda}\square-\kappa^2(\beta-\frac{3}{4}\alpha)\square} P^{0-w}\bigg]_{\alpha\beta,\mu\nu}.
$$
It is seen that the gauge independent part namely, the one that corresponds to $\lambda\to 0$, involves three mass scales
\begin{equation}\label{anticipo}
m_1^2=0,\qquad m_2^2=\frac{4}{\kappa^2\alpha},\qquad m_3^2=\frac{2}{\kappa^2(3\beta-\alpha)}.
\end{equation}
These are the mass scales anticipated in the previous section in \eqref{travalie}. 

\section{The LSZ formula for Stelle gravity}

\subsection{The operator expression of the S matrix}
By collecting all the information given above,  the LSZ rules for the model may be written.  The analogy with \eqref{lsz3} will lead to
$$
\hat{S}=:\exp\bigg\{\frac{2}{\kappa^2}\int h^{\mu\nu}_{in}(x)\square\bigg[\frac{1}{2}\bigg(1-\frac{\kappa^2 \alpha}{4}\square \bigg) P^2-\bigg(1+\frac{\kappa^2(\alpha-3\beta)}{2}\square\bigg) P^{0-s}
$$
\begin{equation}\label{lsznose}
+\bigg(\frac{1}{2}+\frac{\kappa^2(4\beta-3\alpha)\square}{8}\bigg) P^{0-w}+\frac{\kappa^4}{8\lambda}\square[P^1-P^{0-w}]\bigg]_{\alpha\beta,\mu\nu} \frac{\delta}{\delta J_{\alpha\beta}(x)}d^4x.\bigg\}
:Z(J)\bigg|_{J=0}.
\end{equation}
Note that the operator on the left corresponds to the Stelle lagrangian with a replacement $h^{\alpha\beta}\to \frac{\delta}{\delta J_{\alpha\beta}}$.
 The expansion of the exponential in \eqref{lsznose} leads to terms related to correlation functions of the following form
\begin{equation}\label{elementi}
\prod_{i=1}^n \hat{h}^{in}_{\gamma\delta}(x^i)\hat{O}^{\gamma\delta\mu\nu}_{x^i}<\hat{T}h_{\mu\nu}(x^1)...h_{\alpha\beta}(x^n)>,
\end{equation}
where $\hat{O}$ is the operator defined in the exponential in \eqref{lsznose}.  The full element \eqref{elementi} is to be sandwiched between states which a definite momentum, considered as the standard ones.

However, the above result are not  complete without a proper prescription for the free fields $h^{in}_{\mu\nu}$. In particular, a characterization on how they act on the vacuum. The free fields involve six creation and annihilation operators $a(k)$, $a^c(k)$, $b(k)$, $b^c(k)$, $c(k)$ and $c^c(k)$, a pair for every mode with masses $m_1^2=0$ (the graviton), $m_2^2$ and $m_3^2$ respectively.
At this point, the algebra of these operators has not been defined. This is the topic of the next section. Note however that, different from the Pais-Uhlenbeck model, here there are six operators $a(k)$, $a^c(k)$, $b(k)$, $b^c(k)$, $c(k)$ and $c^c(k)$ and there are apparently four commutation relations $[Q_1, Q_2]$, $[P_1, Q_1]$, $[P_2, Q_2]$ and $[P_1, P_2]$ for determining them, which is undetermined.  It is be needed to clarify the meaning of this apparent indetermination, or freedom, for the creation destruction algebra.

\subsection{A rough attempt for an standard quantization}
Following the Gauss-Ostrogradsky method, the impulses of the gravitational field can be obtained from the action \eqref{reina}, which it is written here by convenience
$$
S_s=-\frac{2}{\kappa^2}\int \bigg[\frac{1}{2}h^{\mu\nu}\bigg(1-\frac{\kappa^2 \alpha}{4}\Box\bigg)[\square  h_{\mu\nu}-2\partial^\gamma \partial_{(\mu}h_{\nu)\gamma}]
$$
\begin{equation}\label{reinaldos}
+\frac{1}{2}h\bigg(1-\frac{\kappa^2(4\beta-\alpha)}{4}\square\bigg)[ 2\partial_\alpha \partial_\beta h^{\alpha\beta}-\square  h]
+\frac{\kappa^2}{2}(\beta-\frac{\alpha}{2}) h^{\mu\nu}\partial_\mu\partial_\nu\partial_\alpha \partial_\beta  h^{\alpha\beta}\bigg]d^4x.
\end{equation}
The coordinate variables corresponding  to this lagrangian can be chosen as
\begin{equation}\label{progreso1}
Q_{1\mu\mu}=h_{\mu\nu},\qquad Q_{2\mu\mu}=\dot{h}_{\mu\nu}.
\end{equation}
The momentum variables conjugated to the above are then calculated from formula \eqref{momentas}, the result is
$$
P_{1\mu\nu}=\frac{2}{\kappa^2}\bigg[-\dot{h}_{\mu\nu}+\dot{h}\eta_{\mu\nu}+\eta_{\mu\nu}\partial_\alpha h^{0\alpha}+\partial_{(\mu}h\delta_{\nu) 0}+ \partial_{(\mu}h_{\nu)0}\bigg]
$$
$$
+\frac{2}{\kappa^2}\bigg[\frac{\kappa^2 \alpha}{4}[ \square \dot{h}_{\mu\nu}- \partial^\gamma \partial_{(\nu}\dot{h}_{\mu)\gamma}]+\frac{\kappa^2(4\beta-\alpha)}{4}\eta_{\mu\nu}[ \partial_\alpha \partial_\beta \dot{h}^{\alpha\beta}-\square  \dot{h}]
$$
$$
-\frac{\kappa^2 \alpha}{4} \square \dot{h}_{0(\mu}\delta_{\nu) 0}-\frac{\kappa^2(4\beta-\alpha)}{4}\square \dot{h} \delta_{\mu 0}\delta_{\nu 0}+\kappa^2(\beta-\frac{\alpha}{2}) \delta_{\mu 0}\delta_{\nu 0}\partial_\alpha \partial_\beta  \dot{h}^{\alpha\beta}\bigg],
$$
$$
P_{2\mu\nu}=-\frac{2}{\kappa^2}\bigg[\frac{\kappa^2 \alpha}{4}[ \square h_{\mu\nu}- \partial^\gamma \partial_{(\nu}h_{\mu)\gamma}]+\frac{\kappa^2(4\beta-\alpha)}{4}\eta_{\mu\nu}[ \partial_\alpha \partial_\beta h^{\alpha\beta}-\square  h]
$$
\begin{equation}\label{progreso2}
-\frac{\kappa^2 \alpha}{4} \square h_{0(\mu}\delta_{\nu) 0}-\frac{\kappa^2(4\beta-\alpha)}{4}\square h \delta_{\mu 0}\delta_{\nu 0}+\kappa^2(\beta-\frac{\alpha}{2}) \delta_{\mu 0}\delta_{\nu 0}\partial_\alpha \partial_\beta  h^{\alpha\beta}\bigg].
\end{equation}
The gravitational field is decomposed as
$$
h^{\mu\nu}=\sum_p\int\frac{d^3k}{(2\pi)^{\frac{3}{2}}\omega(k)}[a_{\mu\nu} a_p(k)e^{i\omega(k)t-i k\cdot x}+a_{\mu\nu}^\ast a_p^c(k)e^{-i\omega(k)t+i k\cdot x}]
$$
$$
+\sum_p\int\frac{d^3k}{(2\pi)^{\frac{3}{2}}\omega_2(k)}[b_{\mu\nu}b_p(k) e^{i\omega_2(k)t-i k\cdot x}+b^\ast_{\mu\nu} b_p^c(k)e^{-i\omega_2(k)t+i k\cdot x}]
$$
$$
+\sum_p\int\frac{d^3k}{(2\pi)^{\frac{3}{2}}\omega_3(k)}[c_{\mu\nu}c_p(k) e^{i\omega_3(k)t-i k\cdot x}+c^\ast_{\mu\nu} c_p^c(k)e^{-i\omega_3(k)t+i k\cdot x}].
$$
In these expressions no gauge is assumed, and the sum involves all the possible polarizations. 
\\

\emph{A potentially wrong argument}
\\

The following is an argument which arguably contains a failure. It is written explicitly 
since, even potentially false, is instructive to  be reviewed. In fact, at some point it sounded reasonable and confused the present author.

If  standard quantization is employed, the  first request  is that $[Q_{1\mu\nu}, Q_{2\mu\nu}]=0$ for every choice of indices. However, different than  the Pais-Uhlenbeck oscillator,Quadratic Gravity model is a gauge theory. It may be expected, based on experience on spin 1 gauge fields, that this affirmation is true up either to momentum dependent terms, or terms not proportional to the identity.  For example the commutation relations of QED,  involve $[P_\mu, Q_\nu]\sim \eta_{\mu\nu}-k^{-2}k^\mu k^\nu$  when some gauge is chosen.  Based on these intuitions, it may be  assumed that the commutator will be given by 
$$
[Q_{1\mu\nu}, Q_{2\alpha\beta}]=[h_{\mu\nu}, \dot{h}_{\alpha\beta}]\equiv\bigg[\frac{1}{\omega_1(k)} [a(k), a^c(l)]+\frac{1}{\omega_2(k)} [b(k), b^c(l)]
$$
\begin{equation}\label{promi}
+\frac{1}{\omega_3(k)} [c(k), c^c(l)]\bigg]I_{\mu\nu,\alpha\beta}=0.
\end{equation}
Here the sum of polarizations is not restricted, and leads to the identity term. It is the action over the physical states $|\Psi>$ which, due to the Gupta-Bleuler method, will induce momentum dependent terms. In any case, the last condition  implies that one or two of the oscillators composing $h_{\mu\nu}$ will have a non standard minus sign, otherwise the last sum will not be zero. A ghost is insured with this condition. 

Consider the remaining canonical commutators. For simplicity, choose only spatial indices $i,j$ with $i\neq j$, which throw 
away several terms. The resulting expressions of the impulses are
$$
P_{1ij}=\frac{2}{\kappa^2}\bigg[-\dot{h}_{ij}+ \partial_{(i}h_{j)0}\bigg]
+\frac{2}{\kappa^2}\frac{\kappa^2 \alpha}{4}[ \square \dot{h}_{ij}- \partial^\gamma \partial_{(i}\dot{h}_{i)\gamma}],
$$
$$
P_{2ij}=-\frac{2}{\kappa^2}\frac{\kappa^2 \alpha}{4}[ \square h_{ij}- \partial^\gamma \partial_{(i}h_{j)\gamma}].
$$
Furthermore, the equations of motion for $h_{\mu\nu}$ imply that 
$$
P_{1ij}=\frac{2}{\kappa^2}\bigg[-\dot{h}^1_{ij}-\dot{h}^3_{ij}+ \partial_{(i}h_{j)0}\bigg]
+\frac{2}{\kappa^2}\frac{\kappa^2 \alpha}{4}[ \frac{2}{\kappa^2(3\beta-\alpha)} \dot{h}^3_{ij}- \partial^\gamma \partial_{(i}\dot{h}_{j)\gamma}],
$$
$$
P_{2ij}=-\frac{2}{\kappa^2}\frac{\kappa^2 \alpha}{4}[ \frac{4}{\kappa^2\alpha} h^2_{ij} +\frac{2}{\kappa^2(3\beta-\alpha)}  h^3_{ij}- \partial^\gamma \partial_{(i}h_{j)\gamma}].
$$
Here for instance $h^3_{ij}$ is the metric corresponding to the massive mode with $m^2_3=\frac{2}{\kappa^2(3\beta-\alpha)}$, and so on. In the expression for $P_{1ij}$ some terms corresponding to $h^2_{ij}$ are gone due to their equation of motion or, what is the same, due to its dispersion relation. The commutator between coordinates and impulses is 
$$
[P_{2ij}, \dot{h}_{kl}]\equiv-\frac{2}{\kappa^2}\bigg[ \frac{1}{\omega_2(k)}[b(k), b^c(l)]+\frac{m_3^2}{m_2^2}\frac{1}{\omega_3(k)}[c(k), c^c(l)]\bigg]I_{ijkl}
$$
$$
= iI_{ijkl}\delta(k-l)+\text{momentum dependent commutators},
$$
$$
[P_{1ij}, h_{kl}]\equiv-\frac{2}{\kappa^2}\bigg[- \frac{1}{\omega_1(k)}[a(k), a^c(l)]-\bigg(1-\frac{m_3^2}{m_2^2}\bigg)\frac{1}{\omega_3(k)}[c(k), c^c(l)]\bigg]I_{ijkl}
$$
$$
= iI_{ijkl}\delta(k-l)+\text{momentum dependent commutators}.
$$
The momentum dependent factors are due to terms such as $\partial_{(i}h_{j)0}$ or $\partial^\gamma\partial_{(i}h_{j)\gamma}$. The last request is that the terms proportional to the identity are multiplied by a unit factor. The momentum dependent operators are usual in gauge theories, and are unavoidable.

The subtraction ot the two last commutators leads to \eqref{promi}, which is a consistency check. Therefore it is possible in this case to insure canonical commutation relations for the impulses and coordinates, although \eqref{promi} lead to the conclusion that one state is a ghost.

The last condition to be imposed is that the commutator
$$
[P_{1ij}, P_{2kl}]\equiv [c(k), c^c(l)]=0+\text{momentum dependent terms}.
$$
The reason for which the commutator involves only the operators $c(k)$ and $c^c(k)$ is that the momentum independent terms of  $P_{1ij}$ depend on $h^1_{ij}$ and $h^{3}_{ij}$ while such  terms in $P_{2ij}$ only involve $h^{3}_{ij}$. The last equation then implies that the ghost is absent  from the physical states. It is, in some sense, something analogous to a Faddeev-Popov ghost. The analogy is partial, as it is not clear that this ghost is related to some gauge fixing. 
Taking into account the last conclusion, it follows that 
$$
\frac{1}{2\omega_1(k)}[a(k), a^c(l)]=i\delta(k-l),\qquad \frac{1}{2\omega_2(k)}[b(k), b^c(l)]=-i\delta(k-l),
$$
\begin{equation}\label{discofever}
 [c(k), c^c(l)]=0.
\end{equation}
The first mode, the standard graviton is standard. The second is a ghost. The third only participates as a internal line.

The above argument looks attractive,  since it presents the ghost as a kind of Faddeev-Popov ghost, which disappear from the asymptotic states.  However,  this observation may be not quite correct\footnote{Some comments from Diego Buscio are taken into account, and i am grateful to him since he found some errors in a first version of the present work.}. The state with mass $m_3^2$ has a polarization with one component, and it imitates an scalar. The expectation  that it leads to terms proportional to the identity has to be reviewed since, if it wrong, the state with mass $m_3^2$ is not absent.
Fortunately, there is a gauge which allows a better visualization of this problem. This is described next.

\subsection{A gauge that separate the scalar out from the graviton}

A convenient gauge choice for visualizing the scalar mode is the following. Consider a small perturbation around the Minkowski metric, but now written as
$$
g_{\mu\nu}=\eta_{\mu\nu}+(\widetilde{h}_{\mu\nu}-\frac{1}{2}\eta_{\mu\nu}\widetilde{h}).
$$
Inspired by the methods of  \cite{medeiros1}-\cite{medeiros2},  the perturbation may be decomposed as
$$
\widetilde{h}_{\mu\nu}=\Psi_{\mu\nu}+\frac{\eta_{\mu\nu}}{2}\bigg(\phi+\frac{\Psi}{3}\bigg).
$$
However, in the following, it may be more convenient to use the simple parametrization
$$
h_{\mu\nu}=\Psi_{\mu\nu}+\eta_{\mu\nu} B,\qquad h=\Psi+4B.
$$
The advantage of  defining a scalar field $B$  in the metric perturbation with a term proportional to $\eta_{\mu\nu}$ will be clear soon. The lagrangian \eqref{reina} is written in this gauge  as
$$
S_s=-\frac{2}{\kappa^2}\int \bigg[\frac{1}{2}\Psi^{\mu\nu}\bigg(1-\frac{\kappa^2 \alpha}{4}\Box\bigg)\square  \Psi_{\mu\nu}+2B\bigg(1-\frac{\kappa^2 \alpha}{4}\Box\bigg)\square B+B\bigg(1-\frac{\kappa^2 \alpha}{4}\Box\bigg)\square \Psi
$$
$$
-\Psi^{\mu\nu}\bigg(1-\frac{\kappa^2 \alpha}{4}\Box\bigg)\partial^\gamma \partial_{(\mu}\Psi_{\nu)\gamma}-2B\bigg(1-\frac{\kappa^2 \alpha}{4}\Box\bigg)\partial_\mu \partial_{\nu}\Psi^{\mu\nu}-B\bigg(1-\frac{\kappa^2 \alpha}{4}\Box\bigg)\square B
$$
$$
+\Psi\bigg(1-\frac{\kappa^2(4\beta-\alpha)}{4}\square\bigg)\partial_\mu\partial_\nu \Psi^{\mu\nu}+4B\bigg(1-\frac{\kappa^2(4\beta-\alpha)}{4}\square\bigg)\partial_\mu\partial_\nu \Psi^{\mu\nu}
$$
$$
+\Psi\bigg(1-\frac{\kappa^2(4\beta-\alpha)}{4}\square\bigg)\square B+4B\bigg(1-\frac{\kappa^2(4\beta-\alpha)}{4}\square\bigg)\square B
$$
$$
-\Psi\bigg(1-\frac{\kappa^2(4\beta-\alpha)}{4}\square\bigg) \square \Psi-8B\bigg(1-\frac{\kappa^2(4\beta-\alpha)}{4}\square\bigg) \square B-4B\bigg(1-\frac{\kappa^2(4\beta-\alpha)}{4}\square\bigg) \square \Psi
$$
$$
+\frac{\kappa^2}{2}(\beta-\frac{\alpha}{2}) \Psi^{\mu\nu}\partial_\mu\partial_\nu\partial_\alpha \partial_\beta  \Psi^{\alpha\beta}+\frac{\kappa^2}{2}(\beta-\frac{\alpha}{2}) B\square^2 B+\kappa^2(\beta-\frac{\alpha}{2})\square B \partial_\mu \partial_\nu \Psi^{\mu\nu} \bigg]d^4x.
$$
The equations of motion for $\Psi_{\mu\nu}$ are
$$
\bigg(1-\frac{\kappa^2 \alpha}{4}\Box\bigg)\square  \Psi_{\mu\nu}-2\eta_{\mu\nu}\bigg(1-\frac{\kappa^2(4\beta-\alpha)}{4}\square\bigg) \square \Psi-2\eta_{\mu\nu}\bigg(1-\frac{\kappa^2 (3\beta-\alpha)}{2}\Box\bigg)\square B
$$
$$
+2\bigg(1-\frac{\kappa^2 (3\beta-\alpha)}{2}\Box\bigg)\partial_\mu \partial_{\nu}B
-2\bigg(1-\frac{\kappa^2 \alpha}{4}\Box\bigg)\partial^\gamma \partial_{(\mu}\Psi_{\nu)\gamma}
$$
$$
+\bigg(1-\frac{\kappa^2(4\beta-\alpha)}{4}\square\bigg)\partial_\mu\partial_\nu \Psi+\eta_{\mu\nu}\bigg(1-\frac{\kappa^2(4\beta-\alpha)}{4}\square\bigg)\partial_\alpha\partial_\beta \Psi^{\alpha\beta}
$$
\begin{equation}\label{scandalous}
+\kappa^2(\beta-\frac{\alpha}{2}) \partial_\mu\partial_\nu\partial_\alpha \partial_\beta  \Psi^{\alpha\beta}=0.
\end{equation}
The equation of motion for $B$ is
$$
-6\bigg(1-\frac{\kappa^2( 3\beta-\alpha)}{2}\Box\bigg)\square B-2\bigg(1-\frac{\kappa^2( 3\beta-\alpha)}{2}\Box\bigg)\square \Psi
$$
$$
+2\bigg(1-\frac{\kappa^2 \alpha}{4}\Box\bigg)\partial_\mu \partial_{\nu}\Psi^{\mu\nu}+\kappa^2(\beta-\frac{\alpha}{2})\square \partial_\mu \partial_\nu \Psi^{\mu\nu}=0.
$$
Choose the gauge $\partial_{\nu}\Psi^{\mu\nu}=0$.  This is of course different than the one of the previous sections, which was $\partial_{\nu}h^{\mu\nu}=0$. Take the trace of the equation for $\Psi_{\mu\nu}$. It is
$$
\bigg(1-\frac{\kappa^2 \alpha}{4}\Box\bigg)\square  \Psi-8\bigg(1-\frac{\kappa^2(4\beta-\alpha)}{4}\square\bigg) \square \Psi-6\bigg(1-\frac{\kappa^2 (3\beta-\alpha)}{2}\Box\bigg)\square B=0.
$$
The equation for $B$ in this gauge becomes
$$
-6\bigg(1-\frac{\kappa^2( 3\beta-\alpha)}{2}\Box\bigg)\square B-2\bigg(1-\frac{\kappa^2( 3\beta-\alpha)}{2}\Box\bigg)\square \Psi=0.
$$
These equations are compatible if $\square \Psi=0$, leading to
$$
\bigg(1-\frac{\kappa^2( 3\beta-\alpha)}{2}\Box\bigg)\square B=0.
$$
Now, $\square \Psi=0$ implies that $\Psi_{\mu\nu}$ contains a massless mode. The last equation indicates that $B$ may contain a massless mode as well, plus a massive mode with mass $m_3^2=\frac{2}{\kappa^2( 3\beta-\alpha)}$. The equation for $\Psi_{\mu\nu}$ is then simplified
$$
\bigg(1-\frac{\kappa^2 \alpha}{4}\Box\bigg)\square  \Psi_{\mu\nu}
+2\bigg(1-\frac{\kappa^2 (3\beta-\alpha)}{2}\Box\bigg)\partial_\mu \partial_{\nu}B=0.
$$
The first operator $\square$ removes the massless mode from  $\Psi_{\mu\nu}$ and the operator
$(1-\frac{\kappa^2 (3\beta-\alpha)}{2}\Box)$ removes the massive mode in $B$. Therefore the last equation is completely equivalent to
$$
\bigg(1-\frac{\kappa^2 \alpha}{4}\Box\bigg)\square  \Psi^{\text{massive}}_{\mu\nu}
+2\partial_\mu \partial_{\nu}B^{\text{massless}}=0.
$$
This is an identity between a massive and a massless mode. It can be true only if every term is zero. Then  there is no massless mode in $B$ and $\Psi$ contains, besides a massless mode, a massive one with mass $m_3^2=\frac{4}{\kappa^2\alpha}$. Therefore, the quantum fields are
$$
\Psi_{\mu\nu}(k) =\sum_p\int\frac{d^3k}{(2\pi)^{\frac{3}{2}}\omega(k)}[a_{\mu\nu} a_p(k)e^{i\omega(k)t-i k\cdot x}+a_{\mu\nu}^\ast a_p^c(k)e^{-i\omega(k)t+i k\cdot x}]
$$
$$
+\sum_p\int\frac{d^3k}{(2\pi)^{\frac{3}{2}}\omega_2(k)}[b_{\mu\nu}(k)b_p(k) e^{i\omega_2(k)t-i k\cdot x}+b^\ast_{\mu\nu}(k) b_p^c(k)e^{-i\omega_2(k)t+i k\cdot x}].
$$
where  $\omega(k)=|k|$ and  $\omega_2(k)=\sqrt{k^2+m_2^2}$. Also the gauge $\partial_\mu \Psi^{\mu\nu}=0$ and the above found trace conditions lead to
$$
 k^i a_{\mu i}(k)=\omega a_{\mu 0}(k).\qquad k^i b_{\mu i}(k)=\omega_3 b_{\mu 0}(k).$$
In addition, since $\square \Psi=0$, the massive mode does not have a trace. Therefore
$$
b_{\mu\nu}(k)\eta^{\mu\nu}=0.
$$
The use of \eqref{scandalous} also shows that
$$
a_{\mu\nu}(k)\eta^{\mu\nu}=0.
$$

 For $B$, it contains only a massive mode 
$$
B(x) =\int\frac{d^3k}{(2\pi)^{\frac{3}{2}}\omega_3(k)}[c(k) e^{i\omega_3(k)t-i k\cdot x}+c^c(k) e^{-i\omega_3(k)t+i k\cdot x}],
$$
with $\omega_3^2-k^2=m_3^2$, and $m_3^2=\frac{2}{\kappa^2(3\beta-\alpha)}$.

The above gauge $\partial_\mu \Psi^{\mu\nu}=0$ is different from the one $\partial_\mu h^{\mu\nu}=0$ employed in the previous sections. It has the advantage that it represents the spin 2  excitation $\Psi^{\mu\nu}$ as a composition of two modes, while the third one is mapped entirely to $B$. 

Given this convenient decomposition, consider again the definitions of the canonical coordinates $Q_i$ and momentum $P_i$  given in \eqref{progreso1} and \eqref{progreso2}.  If  these coordinates and momentum are expressed in terms of the modes $\Psi_{\mu\nu}$ and $B$, it is clear that the contributions to the canonical commutation relations corresponding to $B$ involve terms such as
$$
\eta_{\mu\nu} B,\qquad \partial_\mu \partial_\nu B,
$$
and time derivatives of these expressions.  The first terms induce  factors  in the commutation relations such as 
$\eta_{\mu\nu}\eta_{\alpha\beta}$. These are not part of the identity $\eta_{\mu(\alpha}\eta_{\nu)\beta}$. In fact, $\eta_{11}\eta_{22}$ has different indices and is non zero. The second type of terms have derivatives and involve the momentum $k^\mu k^\nu$. Also, these factors do not contribute to the identity. So, the oscillator operators $c(k)$ and $c^c(k)$ are not fixed by the identity factors, and can be freely taken as standard ones. 

However, the problem again with the above argument is that these affirmations are valid taking into account a gauge namely, $\partial_\mu \Psi^{\mu\nu}=0$.  This argument may be a heuristic hint, but this is not enough. The mode with mass $m_3^2$ is included only in  field $B$, but this may change by a gauge transformation. Therefore, a more intrinsic description is needed.

\subsection{The use of generic gravitational perturbations}
Based on the above observations, perhaps it is more rigorous to decompose the graviton as in the theory of gravitational perturbations
$$
g=-(1+2A)dt^2+B_{0i}dt dx^i+[(1+2D)\delta_{ij}+E_{ij}]dx^i dx^j.
$$
The scalars are $A$ and $D$, $B_{i0}$ is the vector perturbation and $E_{ij}$ is a tensor part, which is traceless. It is important to remark that the denomination of being scalar, vector and tensors is related to the change of law under rotation of the three dimensional space, they are not true tensor quantities in Minkowski space. For instance, the scalar $D$ is invariant under rotations, but of course not  invariant under generic gauge transformations $x^\mu\to x^\mu+\xi^\mu$. Also, if $\xi^\mu$ depends on time then, for instance, $A\to A+\dot{\xi}^0$, etc.

It is instructive for instance to make the above characterization for the mode with $m_3^2$ in the gauge $\partial_{\mu}h^{\mu\nu}=0$, moving along the $z$ axis. This follows from \eqref{teplenie}, namely 
$$
\epsilon^{3}_{\mu\nu}=\frac{1}{3}\left(\begin{array}{cccc}
  -1 +\frac{\omega_3^2}{m_3^2} & 0  & 0  & \frac{k\omega_3}{m_3^2}\\
  0  & 1 & 0  &  0\\
  0 & 0 & 1  & 0 \\ 
  \frac{k\omega_3}{m_3^2} & 0  &  0 & 1+\frac{k^2}{m_3^2}
\end{array}\right)\epsilon.
$$
This leads to a graviton spectral components
$$
A(k,x^\mu)= \frac{1}{3}\bigg(-1+\frac{\omega_3^2}{m_3^2}\bigg) e^{i\omega_3 t-i k_3 z}, \qquad E_{ij}=\frac{k^2}{3m_3^2}\delta_{iz}\delta_{jz} e^{i\omega_3 t-i k_3 z}-\frac{1}{3}\delta_{ij}\frac{k^2}{3m_3^2} e^{i\omega_3 t-i k_3 z},
$$
$$
2D=\bigg(1+\frac{k^2}{3m_3^2}\bigg)e^{i\omega_3 t-i k_3 z},\qquad B_{z}=\frac{k \omega_3}{3m_3^2}e^{i\omega_3 t-i k_3 z}.
$$
For the propagation in arbitrary directions, the above components are
$$
A(k,x^\mu)= \frac{1}{3}\bigg(-1+\frac{\omega_3^2}{m_3^2}\bigg) e^{i\omega_3 t-i k\dot r}, \qquad E_{ij}=\frac{k_i k_j}{3m_3^2} e^{i\omega_3 t-i k\cdot r}-\frac{\delta_{ij}k^2}{9m_3^2} e^{i\omega_3 t-i k \cdot r},
$$
\begin{equation}\label{eva}
2D=\bigg(1+\frac{k^2}{3m_3^2}\bigg)e^{i\omega_3 t-i k \cdot r},\qquad B_{i}=\frac{k_i \omega_3}{3m_3^2}e^{i\omega_3 t-i k \cdot r}.
\end{equation}
The spin 2 perturbation $E_{ij}$ is traceless. It includes only components proportional to the momentum.
Under a gauge transformation  it changes as
$$
E^\prime_{ij}=E_{ij}-\partial_i \xi_j^s+\frac{1}{3}\delta_{ij}\nabla\cdot \xi^s+\partial_{(i}\xi^v_{j)},
$$
where one has the decomposition $\xi_i=\xi_i^s+\xi_i^v$ with $\partial_i \xi^{iv}=0$. The tensor $E_{ij}$ is itself decomposed into 
a general as a scalar and vector  part
$$
E^s_{ij}=(k_i k_j-\frac{1}{3}\delta_{ij})E(k),\qquad E_{ij}^v=\frac{E_i k_j+E_j k_i}{2},
$$
with $E(K)$ arbitrary and $k^iE_i(k)=0$, plus a tensor part $E^T_{ij}$ such that
$$
E^T_{ii}=0,\qquad k^i E^T_{ij}=0.
$$
The tensor in \eqref{eva} is purely of the form 
$$
E^s_{ij}=(k_i k_j-\frac{1}{3}\delta_{ij})E(k),
$$
with $E(k)=\frac{1}{3m_3^2}$. In other words, it does not contain a vector or tensor component.  It is also known from theory of gravitational perturbations that the tensor component $E^T_{ij}$ is invariant under gauge transformations. Since it is zero here, it will be zero under any gauge transformation. Based on this, it seems that the graviton component $h_{\mu\nu}^{3}$  can not be considered as a true spin two excitation, as a gauge transformation will not generate a tensor excitation.

Therefore, the failure of the argument in \eqref{discofever} is to consider the mode with mass $m_3^2$
as having a component which contributes to the identity.  Once this mode is separated from the spin 2 perturbation, and $c(k)$ and $c^c(k)$ are declared to be standard, then the canonical commutation relations give a $4\times 4$ system for $a(k)$, $a^c(k)$,  $b(k)$ and $b^c(k)$.  This system is not overdetermined.

In view of this,  the  reasoning given in \eqref{promi} and \eqref{discofever}, by taking into account that $B$ is not contributing, should be modified as follows. Due to the equations of motion given above for $\Psi_{\mu\nu}$, it is seen that the massless mode only contributes to $P_{1\alpha\beta}$ while the massive one contributes to $P_{2\alpha\beta}$. It is clear that the commutator structure is schematically
$$
[Q_{1\alpha\beta}, Q_{2\mu\nu}]=[\Psi_{\alpha\beta}, \dot{\Psi}_{\mu\nu}]=[\Psi^1_{\alpha\beta}, \dot{\Psi}^1_{\mu\nu}]+[\Psi^2_{\alpha\beta}, \dot{\Psi}^2_{\mu\nu}]\equiv 0,
$$
$$
[Q_{1\alpha\beta}, P_{1\mu\nu}]=\frac{1}{4\kappa}[\dot{\Psi}^1_{\alpha\beta}, \Psi^1_{\mu\nu}]=iI_{\mu\nu,\alpha\beta}\delta(x-x^\prime),
$$
$$
[Q_{2\alpha\beta}, P_{2\mu\nu}]\equiv-\frac{1}{4\kappa}[\Psi^2_{\alpha\beta}, \dot{\Psi}^2_{\mu\nu}]\equiv iI_{\mu\nu,\alpha\beta}\delta(x-x^\prime),
$$
$$
[P_{1\alpha\beta}, P_{2\mu\nu}]=0.
$$
All the above conclusions should be interpreted "up to a gauge terms" and neglecting terms which will vanish when mean values are taken. The first of the above formulas implies that \begin{equation}\label{argen}
[a_p(k), a_{q}^c(k^\prime)]=4\kappa\delta_{pq}\omega_1(k)\delta(k-k^\prime),\qquad [b_p(k), b_q^c(k^\prime)]=-4\kappa\delta_{pq}\omega_2(k)\delta(k-k^\prime).
\end{equation}
The creation annihilation algebra has a minus sign for $b_p(k)$ and $b_c^c(k)$. Otherwise, $[Q_{1\mu\nu}, Q_{2\alpha\beta}]\neq 0$. This is the standard literature prescription, since usually the massive spin 2 mode is interpreted as a ghost.

For the mode $B$ nothing is said. The commutations may be chosen as
\begin{equation}\label{argen2}
[c_p(k), c_q^c(k^\prime)]=\pm4\kappa\delta_{pq}\omega_3(k)\delta(k-k^\prime),
\end{equation}
that is, any sign may be employed.  Usually, the standard choice with the plus sign is employed. The ghost is usually atributed to the massive spin 2 excitation.

\section{The full prescription and the universality properties of the effective action}
\subsection{The first type of quantization}
The above discussion was related to free graviton fields $h_{in\mu\nu}$, which are relevant for studying scattering as they enter in the LSZ formula \eqref{lsznose}. The full formula is
$$
\hat{S}=:\exp\bigg\{\frac{2}{\kappa^2}\int h^{\mu\nu}_{in}(x)\square\bigg[\frac{1}{2}\bigg(1-\frac{\kappa^2 \alpha}{4}\square \bigg) P^2-\bigg(1-\frac{\kappa^2(\alpha-3\beta)}{2}\square\bigg) P^{0-s}
$$
\begin{equation}\label{lsznose2}
+\bigg(\frac{1}{2}-\frac{\kappa^2(4\beta-3\alpha)\square}{8}\bigg) P^{0-w}+\frac{\kappa^4}{8\lambda}\square[P^1-P^{0-w}]\bigg]_{\alpha\beta,\mu\nu} \frac{\delta}{\delta J_{\alpha\beta}(x)}d^4x.\bigg\}
:Z(J)\bigg|_{J=0},
\end{equation}
with 
$$
h_{in}^{\mu\nu}=\sum_p\int\frac{d^3k}{(2\pi)^{\frac{3}{2}}\omega(k)}[a_{\mu\nu} a_p(k)e^{i\omega(k)t-i k\cdot x}+a_{\mu\nu}^\ast a_p^c(k)e^{-i\omega(k)t+i k\cdot x}]
$$
$$
+\sum_p\int\frac{d^3k}{(2\pi)^{\frac{3}{2}}\omega_2(k)}[b_{\mu\nu}b_p(k) e^{i\omega_2(k)t-i k\cdot x}+b^\ast_{\mu\nu} b_p^c(k)e^{-i\omega_2(k)t+i k\cdot x}].
$$
$$
+\sum_p\int\frac{d^3k}{(2\pi)^{\frac{3}{2}}\omega_3(k)}[c_{\mu\nu}c_p(k) e^{i\omega_3(k)t-i k\cdot x}+c^\ast_{\mu\nu} c_p^c(k)e^{-i\omega_3(k)t+i k\cdot x}].
$$
The sum is over all the polarizations but the space of physical states $|\Psi>$ eliminates some of them. These states have to be consistent with the Stelle gauge, which is $\partial_\mu h^{\mu\nu}$. The algebra of the creation and annihilation is \eqref{argen}-\eqref{argen2}. The scattering amplitude $A_{i}^j$ corresponds to sandwich with covariant and contravariant states $<_i|$ and $|^j>$ or viceversa.

Note that, if the oscillator algebra were non standard, still the covariant/contravariant formalism for the $S$ matrix described in \eqref{nofederal} will take care of the results. It is unlikely that such non standard algebra will invalidate the model. 
 \subsection{The second approach}
In the second approach for the $S$ matrix was described in \eqref{lszvarias3}, and has to be generalized to the present case. This may be hard, specially taking into account now that the gauge symmetry is playing a role. One has to modify the propagator of the model
$$
D_{\alpha\beta, \mu\nu}=\frac{1}{\square}\bigg[\frac{2}{1-\frac{\kappa^2\alpha}{4}\square } P^2+\frac{8\lambda}{\kappa^4\square} P^1-\frac{1}{1-\frac{\kappa^2}{2}(3\beta-\alpha)\square } P^{0-s}
$$
$$
+\frac{2}{1-\frac{\kappa^4}{4\lambda}\square-\kappa^2(\beta-\frac{3}{4}\alpha)\square} P^{0-w}\bigg]_{\alpha\beta,\mu\nu}.
$$
In order to take into account that the creation and annihilation operators $b_p(k)$, $b_q^c(k^\prime)$are the ones having a wrong sign,  the last expression may be worked out by partial fractions as
$$
D_{\alpha\beta, \mu\nu}=\bigg[\frac{2P_2}{\square}+\frac{\frac{\kappa^2\alpha}{2}}{1-\frac{\kappa^2\alpha}{4}\square } P^2+\frac{8\lambda}{\kappa^4\square^2} P^1-\frac{1}{\bigg(1-\frac{\kappa^2}{2}(3\beta-\alpha)\square\bigg)\square} P^{0-s}
$$
$$
+\frac{2}{\bigg(1-\frac{\kappa^4}{4\lambda}\square-\kappa^2(\beta-\frac{3}{4}\alpha)\square\bigg)\square} P^{0-w}\bigg]_{\alpha\beta,\mu\nu}.
$$
Assume that for the masses going to infinite, that is, $\alpha\to $ and $\beta\to 0$ the result has to be  GR fixed by the Stelle gauge. Therefore it may be reasonable to impose that the terms are unchanged except the second, which is related to the ghost massive spin two mode. Since these terms are turned on when the masses are finite, and these modes have the wrong oscillator algebra, i will assume that the propagator that enters in this quantization method is
$$
D_{\alpha\beta, \mu\nu}=\bigg[\frac{2P_2}{\square}+\frac{\frac{\kappa^2\alpha}{2}}{1-\frac{\kappa^2\alpha}{4}\square } P^2-\frac{8\lambda}{\kappa^4\square^2} P^1-\frac{1}{\bigg(1-\frac{\kappa^2}{2}(3\beta-\alpha)\square\bigg)\square} P^{0-s}
$$
$$
+\frac{2}{\bigg(1-\frac{\kappa^4}{4\lambda}\square-\kappa^2(\beta-\frac{3}{4}\alpha)\square\bigg)\square} P^{0-w}\bigg]_{\alpha\beta,\mu\nu}.
$$
That is, only the second  changed the sign.  Now employ \eqref{inverso}  and find the corresponding kinetic operator
$$
O^m_{\alpha\beta, \mu\nu}=\bigg[\frac{\square\bigg(1-\frac{\kappa^2\alpha}{4}\square\bigg) }{2-k^2\alpha \square} P^2+\frac{\kappa^4\square^2}{8\lambda} P^1-(1-\frac{\kappa^2}{2}(3\beta-\alpha))\square P^{0-s}
$$
\begin{equation}\label{matieres}
+\frac{2}{1-\frac{\kappa^4}{4\lambda}\square-\kappa^2(\beta-\frac{3}{4}\alpha)\square} P^{0-w}\bigg]_{\alpha\beta,\mu\nu}.
\end{equation}
This is the analogous of the operator \eqref{inthearmy} found for the Pais-Uhlenbeck scalar field, adapted to the present situation. The gauge fixing term did not change. 

At this point, one should analyze carefully the issue of gauge invariance, since those modifications may give an issue in this aspect.  Consider a gneeric non local action of the form \cite{biswas}
\begin{equation}\label{biswas}
L=\sqrt{-g}[R+RF_1(\square)R+R_{\mu\nu}F_2(\square)R^{\mu\nu}+R_{\mu\nu\alpha\beta}F_3(\square)R^{\mu\nu\alpha\beta}].
\end{equation}
Its linearization is of the form \cite{biswas}
$$
S_s=-\frac{2}{\kappa^2}\int \bigg[\frac{1}{2}h^{\mu\nu} a(\square)\square  h_{\mu\nu}+ h^{\mu\nu}b(\square)\partial^\gamma \partial_{(\nu}h_{\mu)\gamma}
$$
\begin{equation}\label{reiname}
+h  c(\square)\partial_\alpha \partial_\beta h^{\alpha\beta}+\frac{1}{2}hd(\square)\square  h+h^{\mu\nu}\frac{f(\square)}{\square} \partial_\mu\partial_\nu\partial_\alpha \partial_\beta  h^{\alpha\beta}\bigg]d^4x.
\end{equation}
The functions $a(\square)$, .., $f(\square)$ are not linearly independent, since they are parameterized by three functions $F_i(\square)$ with $i=1,2,3$ defining the lagrangian. It was shown in \cite{biswas} 
that
$$
a=1-\frac{1}{2}F_2(\square)\square -2F_3(\square)\square, \qquad b=-a,
$$
$$
c=1+2F_1(\square)\square +\frac{1}{2}F_2(\square)\square, \qquad d=-c,
$$
$$
f=-2F_1(\square)\square -F_2(\square)\square -2F_3(\square)\square.
$$
These expressions shows a linear dependence between the coefficients, since it is obvious that
\begin{equation}\label{fortu3}
a+b=0, \qquad c+d=0,\qquad b+c+f=0.
\end{equation}
This is valid without the gauge fixing terms. Taking these generic models into account, note that with the help of \eqref{fortu}  and taking into account \eqref{matieres}, it is found after some calculation that
$$
a=\frac{\bigg(1-\frac{\kappa^2\alpha}{4}\square\bigg) }{1-\frac{k^2\alpha}{2} \square},\qquad a=-b,
$$
$$
c=\frac{1}{3}a+\frac{1}{3} \bigg(1-\frac{\kappa^2}{2}(3\beta-\alpha)\square \bigg),\qquad d=-c.
$$
The first two \eqref{fortu3} are satisfied. 
However
\begin{equation}\label{matieres2}
f=-\frac{2}{3}\bigg(1-\frac{\kappa^2\alpha}{4}\square\bigg)+\frac{1}{3} \bigg(1-\frac{\kappa^2}{2}(3\beta-\alpha)\square \bigg)
\neq -b-c.
\end{equation}
This does not correspond to any of these non local models and it look that it may violate gauge invariance.

The above conclusion however, may be premature. The coefficients should satisfy these relations without the gauge fixing term. That is the reason for which the above coefficients were not calculated taking into account this term. The kinetic term \eqref{matieres} is gauge fixed. Therefore, one may declare that 
$$
f=-b-c,
$$
and identify the unwanted part as a gauge fixing term. This choice leads to
\begin{equation}\label{matieres3}
f=-\frac{1}{3}\frac{\bigg(1-\frac{\kappa^2\alpha}{4}\square\bigg) }{1-k^2\alpha\square}+\frac{1}{3} \bigg(1-\frac{\kappa^2}{2}(3\beta-\alpha)\square \bigg).
\end{equation}
The difference between \eqref{matieres2} and \eqref{matieres3} is 
$$
\Delta f=-\frac{\kappa^2\alpha}{3}\frac{\bigg(1-\frac{\kappa^2\alpha}{4}\square\bigg)}{1-\frac{\kappa^2\alpha}{2}\square}.
$$
If this corresponds to a gauge fixing term, the full gauge fixing is the original Stelle one plus this contribution, namely
$$
L_{gf}=-\frac{\kappa^2\alpha}{3}h^{\mu\nu}\frac{\bigg(1-\frac{\kappa^2\alpha}{4}\square\bigg)}{\square\bigg(1-\frac{\kappa^2\alpha}{2}\square\bigg)} \partial_\mu\partial_\nu\partial_\alpha \partial_\beta  h^{\alpha\beta}-\frac{\kappa^2}{2\lambda}  h^{\mu\nu}\square  \partial_\mu\partial^\alpha h_{\alpha\nu}.
$$
These terms contribute to the classical equations of motion as
$$
H_{\mu\nu}-\frac{\kappa^2\alpha}{3}\frac{\bigg(1-\frac{\kappa^2\alpha}{4}\square\bigg)}{\square\bigg(1-\frac{\kappa^2\alpha}{2}\square\bigg)} \partial_\mu\partial_\nu\partial_\alpha \partial_\beta  h^{\alpha\beta}-\frac{\kappa^2}{2\lambda} \square  \partial_{(\mu}\partial^\alpha h_{\nu)\alpha}=0,
$$
with $H_{\mu\nu}$ the contribution from the lagrangian of the modified model without gauge fixing.
This term satisfies $\partial^\mu H_{\mu\nu}=0$, as the coefficients were selected for this part to be gauge invariant. Therefore
$$
-\frac{\kappa^2\alpha}{3}\frac{\bigg(1-\frac{\kappa^2\alpha}{4}\square\bigg)}{1-\frac{\kappa^2\alpha}{2}\square} \partial_\nu\partial_\alpha \partial_\beta  h^{\alpha\beta}-\frac{\kappa^2}{4\lambda} \square^2\partial^\alpha h_{\nu\alpha}-\frac{\kappa^2}{4\lambda} \square  \partial_{\nu}\partial^\alpha\partial^\mu h_{\mu\alpha}=0.
$$
The minimal way to satisfy this identity is to postulate that $\partial^\mu h_{\mu\alpha}=0$, which is again the gauge employed along the text. In other words, the above corresponds to non local gauge invariant gravity action, fixed with in the harmnic gauge with horribly chosen gauge fixing term.

Due to \eqref{fortu3}, there are several possible choices for the $F_i(\square)$ functions. The minimal choice is $F_3(\square)=0$, leading to
$$
F_1(\square)=\frac{1}{2\square}\bigg[\frac{4}{3}\frac{\bigg(1-\frac{\kappa^2\alpha}{4}\square\bigg) }{1-\frac{k^2\alpha}{2} \square}+\frac{1}{3} \bigg(1-\frac{\kappa^2}{2}(3\beta-\alpha)\square \bigg)-2\bigg],\qquad F_2(\square)=\frac{1}{\square}\bigg[1-\frac{\bigg(1-\frac{\kappa^2\alpha}{4}\square\bigg) }{1-\frac{k^2\alpha}{2} \square}\bigg].
$$
The lagrangian \eqref{biswas} plays the analogous role to \eqref{inthearmy} for the Pais-Uhlenbeck field, in this quantization scheme.

The LSZ rules are now
$$
\hat{S}^m=:\exp\bigg\{\frac{2}{\kappa^2}\int h^{\mu\nu}_{in}(x)\bigg[\frac{\square\bigg(1-\frac{\kappa^2\alpha}{4}\square\bigg) }{2-k^2\alpha \square} P^2+\frac{\kappa^4\square^2}{8\lambda} P^1-(1-\frac{\kappa^2}{2}(3\beta-\alpha))\square P^{0-s}
$$
\begin{equation}\label{masem}
+\frac{2}{1-\frac{\kappa^4}{4\lambda}\square-\kappa^2(\beta-\frac{3}{4}\alpha)\square} P^{0-w}\bigg]_{\alpha\beta,\mu\nu} \frac{\delta}{\delta J_{\alpha\beta}(x)}d^4x.\bigg\}
:Z(J)\bigg|_{J=0}.
\end{equation}
For a generic observable $O(P, X)$ constructed in powers of $\hat{X}$ and $\hat{P}$, calculate the mean value $<O(P,X)\eta >$ and analitically continue the ghost variables to values $P=\to-iP$ and $ X\to-iX$. This is in harmony with the all above findings. 

Even though the exponential and $Z(J)$ correspond to gauge invariant models fixed with the gauge $\partial_\mu h^{\mu\nu}=0$, this is for a fixed gauge. It is of interest to prove that the same happens for any gauge. This is an interesting lead for the future. Another approach, perhaps rusty, is to calculate the S-matrix and impose gauge invariance in axiomatic form.

\subsection{General form of the renormalized effective action}
The LSZ formulas derived in the text are either \eqref{lsznose2}  or \eqref{masem}, depending on the chosen approach. These formulas are based on the knowledge of the path integral $Z(J)$.  The advantage of the gauge fixing term that Stelle employs in \cite{stelle1}  is that several universality properties  are known. The partition function in this gauge is
$$
Z( \xi, \overline{\xi}, K, L, J)=e^{i G( \xi, \overline{\xi}, K, L, J)}=\int Dh^{\mu\nu}D\eta^a D\overline{\eta}^b \exp\{i\int d^4x[\overline{\xi}_\alpha\eta^\alpha+\xi_\alpha\overline{\eta}^\alpha+\kappa J_{\mu\nu}h^{\mu\nu}]\}
$$
\begin{equation}\label{eliana}
\exp\{i\int d^4x[L_s-\frac{\kappa^2}{2\lambda}F_\tau\square F^\tau+\overline{\eta}_\tau F_{\mu\nu}^\tau D^{\mu\nu}_{\alpha}\eta^\alpha+\kappa K_{\mu\nu}D^{\mu\nu}_\alpha \eta^\alpha +\kappa^2 L_\alpha \partial_\beta \eta^\alpha \eta^\beta]\}.
\end{equation}
Here the current coupling is  $J_{\mu\nu} h^{\mu\nu}$ and $J_{\mu\nu}$ may be identified as the energy momentum tensor $J_{\mu\nu}=T_{\mu\nu}$ for an external source. The quantity $\lambda$ is simply a constant parameter in the gauge fixing term, as discussed in the previous section.  The quantities $K_{\mu\nu}$ and $L_\alpha$ are sources. The $\eta_a$ and $\overline{\eta}_a$ fields are the Faddeev-Popov ghosts and anti-ghost respectively. The standard transition amplitude is 
$$
Z( \xi, \overline{\xi}, J)=Z( \xi, \overline{\xi}, J, K,L )|_{K=L=0}.
$$
From the  expression of the above functional,  it follows that the gauge ghost action for the gravitational perturbation can be expressed as follows
$$
S_{gf}=\int d^4x[L_s-\frac{\kappa^2}{2\lambda}F_\tau\square F^\tau+\overline{\eta}_\tau F_{\mu\nu}^\tau D^{\mu\nu}_{\alpha}\eta^\alpha+\kappa K_{\mu\nu}D^{\mu\nu}_\alpha \eta^\alpha +\kappa^2 L_\alpha \partial_\beta \eta^\alpha \eta^\beta].
$$
The operator related to the Stelle gauge choice is $F^\tau= \partial_\nu h^{\tau\nu}$. Furthermore $F^\tau_{\mu\nu}=\delta^\tau_\mu \partial_\nu$. The mass dimensions of the fields are directly seen from this action 
\begin{equation}\label{masaso}
[h^{\mu\nu}]=0,\qquad [\eta^a]=[\overline{\eta}^a]=1, \qquad [K_{\mu\nu}]=[L_\alpha]=3.
\end{equation}
Here, as before, the values for the ghost and anti-ghost are non uniquely defined. The choice above is however valid, and is the one that is usually employed. The operator $D_\alpha^{\beta\gamma}$ is defined by
\begin{equation}\label{infdif}
D_\alpha^{\mu\nu}\xi^\alpha= \partial^\mu \xi^\nu +\partial^\nu \xi^\mu-\eta^{\mu\nu}\partial_\alpha \xi^\alpha
+ h^{\alpha\mu} \partial_\alpha \xi^\nu+ h^{\alpha\nu} \partial_\alpha \xi^\mu-\xi^\alpha\partial_\alpha h^{\mu\nu}-h^{\nu\nu}\partial_\alpha \xi^\alpha.
\end{equation}
As is well known, there is a deep relation between the functional $G$ given above and the effective action $\Gamma$. The relation is 
$$
e^{i G(J, \xi, \overline{\xi}, K, L)}=\int_{\text{C.T.D}} Dh^{\mu \nu} D\eta^a D\overline{\eta}^a \exp\{i\int \;d^4x\;  [\Gamma( h, \eta, \overline{\eta}, K, L)+
$$
\begin{equation}\label{conexa2}
\overline{\xi}_\alpha\eta^\alpha+\xi_\alpha\overline{\eta}^\alpha+\kappa J_{\mu\nu}h^{\mu\nu}]\},
\end{equation}
where the notation C.T.D. stands for connected three diagrams. The physical meaning of the last formula is the following.  If functional form of the effective action $\Gamma( h, \eta, \overline{\eta}, K, L)$  has been obtained exactly, the functional $G(J, \xi, \overline{\xi}, K, L)$ can be found by using the effective action in place of the original action and by calculating Feynman diagrams only at tree level, without loops. 

An important point is that, with the gauge described in the above sections, the Stelle effective action has the universal form
$$
\Gamma_n=\int d^4x\bigg[A(h_{\mu\nu})+\frac{\delta S_{0
}}{\delta h^{\mu\nu}}P^{\mu \nu}+(\kappa K_{\rho\sigma}-\overline{\eta}_\tau \overleftarrow{F}^\tau_{\rho\sigma})\bigg(\frac{\delta  D^{\rho\sigma}_\alpha}{\delta h^{\mu\nu}}\eta^\alpha P^{\mu\nu}
$$
$$
-\frac{\partial P^{\rho\sigma}}{\delta h^{\mu\nu}} D^{\mu\nu}_\alpha\eta^\alpha - D^{\rho\sigma}_\alpha(Q^\alpha_\tau \eta^\tau)\bigg)
$$
\begin{equation}\label{diverge}
+\kappa L_\sigma \bigg(\kappa Q^\sigma_\tau\partial_\beta \eta^\tau-\kappa \partial_\beta(Q^\sigma_\tau \eta^\tau)-\kappa(\partial_\tau \eta^\sigma)Q^\tau_\beta -\frac{\delta Q^\sigma_\tau}{\delta h^{\mu\nu}}\eta^\tau D_\beta^{\mu\nu}\bigg) \eta^\beta\bigg].
\end{equation}
This is exactly the formula (6.17) of the original reference \cite{stelle2}. Here the following quantity 
$$
A(h_{\mu\nu})=\sqrt{-g}\bigg[-\frac{2}{\kappa^2+\delta\kappa^{ 2}}R+(\beta+\delta\beta) R^2-(\alpha+\delta\alpha) R_{\mu\nu}R^{\mu\nu}\bigg],
$$
has been introduced and $S_{0}$ is the classical lagrangian of Stelle gravity, in other words, the limit of $A(h_{\mu\nu})$ with the $\delta$ quantities taken to zero.   The quantities $P^{\mu\nu}(h^{\alpha\beta})$ and $Q_{\nu}^{\mu}(h^{\alpha\beta})$ are functions of $h^{\alpha\beta}$ solely, and are not determined without a proper loop analysis for the effective action. 

It should be recalled that other gauges can be employed for quantization of the Stelle model,  however the above universal form is not ensured in those cases. 

\section{Final comments and possible future research programs}

 In the present work, two quantization procedures were presented.  The first is \eqref{lsznose2}, which looks simply as the standard textbook formula for calculating amputated Green functions,, but with modified oscillator algebra. The quantity $Z(J)$ is essentially the Stelle path integral, written in the euclidean setting. The results are then continued to imaginary times to obtain the Lorenzian scattering matrix. The operators in the exponential in \eqref{lsznose2} are written in terms of creation annihilation operator with the correct commutation algebra. Observables such as the wave form $<h_{\mu\nu}>$ have to be calculated by employing these corrected creation annihilation operators.

If the algebra appearing is a non standard one, still the formalism \eqref{nofederal} will take care of negative norm states.

The above prescription may rise skepticism, since it is very similar to the one neglected during the last 80 years due to the apparent problem of unitarity. Note however, that the oscillator algebra of the exponential \eqref{lsznose2} is corrected. In author´s opinion, there is no need for this to be wrong. Since the present text is extensive, the following extremely rough heuristic argument in favor of the above quantization may be useful. The unitarity problem is usually invoked by arguing that, with the oscillator algebra of canonical quantization, there appear states with negative norm $<n|m>\leq 0$. These states may be divided  by their norm, and the result will be positive definite. Of course, this argument is usually neglected since the standard norm in Quantum Mechanics is
$$
<\psi|\psi>=\int \psi^\ast \psi d^4x,
 $$
and the multiplication $\psi\to i \psi$ then $\psi^\ast\to -i\psi^\ast$, this does not change the norm since $-i^2=1$. However, this reasoning may be a trap. The above norm is  positive definite right at the beginning, so there is no need to do anything. Negative norms arise from other prescriptions such as
$$
<\psi|\psi>=\int \psi^\ast(-x)\psi(x) d^4x,
 $$
discussed in the text. It may make sense in this context  to "normalize by dividing by the norm" thus making everything positive. The 80 years of investigations about this topic may be interpreted as investigations in favor of this extremely primitive procedure. The results of \cite{salvio}, and this is a very personal opinion, may be interpreted as correcting the norm by the covariant/contravariant procedure described in the text.

The second type of quantization is given in \eqref{masem}.  The expression is more complicated, as the action of the exponential involves a non standard operator. It was conjectured along the text that in this form, the results will be gauge invariant. However, this is to be studied further. The exponential operator acting on $Z(J)$, is non local.

Personally, i would apply the first type of quantization as a first attempt. It looks simpler and i believe that it may give interesting results in the physics of gravitational waves. However, the study of this second quantization is of interest as well.

Recently, a reference \cite{kleefeld} appeared which apparently presents a consistent quantization of the Pais-Uhlenbeck oscillator. A natural question is  how the methods developed there apply for Quadratic Gravity. A work that seems to have started this program is \cite{kleefeld2}, and this research may have some overlap with the present one.

The main motivation of the above prescriptions is that the Stelle theory is renormalizable and this property is not spoiled by the formalism developed in  \cite{salvio}. The path integral can be given in terms of an effective action that can be renormalized with a finite number of counter terms \cite{stelle1} which, in some specific gauge, leads to the general form  \eqref{diverge}. The  tree diagrams approximation for calculating correlations can then be employed. The Slavnov-Taylor identities \cite{Slavnov}-\cite{Taylor} ensure gauge invariance order by order. A review of this renormalization procedure may be found in  \cite{yo}, further details can be found in the textbooks \cite{buchbinder}-\cite{hamber}. The prescriptions with covariant and contravariant Hilbert spaces insure positive norm states, even if some oscillator has a non standard algebra.

There are several potential applications related to the present formalism, in particular related to high precision gravitational waves. A first may be the study of the spectrum of gravitational waves, by starting with the quantum model and by taking the classical limit. This quantum/classical procedure was initiated in \cite{duff}, this formalism is not only restricted to problems of gravity. The applications of these methods to gravity were developed further, however they are usually employed in the context of GR. Examples of works in this area are \cite{mogulo}-\cite{mogulo5},  \cite{eng}-\cite{eng2}, some of these works are focused on  gravitational bremsstrahlung, the radiation of gravitational waves due to the dynamical evolution of a system of particles.  The calculation of this effect requires to take the mean value of the waveform $<h_{\mu\nu}>$. This calculation, for Quadratic Gravity, involves the prescription presented here, in particular the mapping of the oscillator structure constants to standard ones or, equivalently, the mapping $P_2\to i P_2$ and $Q_2\to i Q_2$ described in \eqref{aprescribir}. There are several works related to this topic. The work \cite{duff5} for instance, considers this type of problems with a worldline formalism, and it may be a good task to generalize it to the present context. The works \cite{anton1}-\cite{anton4} present several results about wave forms, bound states and Hawking radiation which are worthy to be reconsidered  as well with the present quantization procedure. 

There exist several old attempts for describing the nuclear potential, beyond the well established Yukawa interaction, which can be generalized to study gravity. The works of \cite{fubini1}-\cite{fubini3} and \cite{chew} are devoted  to the definition of a nuclear potential in terms of scattering amplitudes.   These findings motivated a large amount of work devoted to the definition of such potentials and there appeared attempts to introduce an effective description of the relativistic scattering in terms of an effective potential and a Lippmann-Schwinger type of equation \cite{logu1}-\cite{logu3}. These approaches were employed in order to study the two body problem in atomic and nuclear physics for instance in \cite{kang}-\cite{sugar}. Relations to the eikonal procedure \cite{levisucher}-\cite{lipatov} were found in \cite{todoroveikonal}. Later on, attempts to study the two body problem in gravity was considered in \cite{todorovgravity}. These authors are able to obtain some approximations for the effective problem, but they can not go to higher orders since they employ GR, which is non renormalizable.  These results may be reconsidered in the present formalism, as renormalizability is not spoiled. There exist certain two body formalisms in the literature, a well known
one is \cite{damour}, that have better convergence properties than \cite{todorovgravity}. However, these approaches are purely classical.

The use of coherent states in order to study the classical limit of gravity and scattering process, which was employed in \cite{ochirov}-\cite{ochirov2} may be useful for studying bremstrahlung as well, together with the results of \cite{weinberg}-\cite{kulish}. The issue of collinear divergences in gravity may be also a relevant task, In author´s opinion, the addition of massive modes in gravity may not significantly alter the results in \cite{collinear1} or \cite{collinear2}-\cite{collinear4}, however a checkup is desirable.

Not less important, the reformulation of the eikonal limit of gravity, in view of the quantization presented here, is a possible  relevant research. An extensive but perhaps incomplete list of works dealing with such topics are \cite{levisucher}-\cite{vietnam}. There are  computational codes \cite{dubna2}-\cite{dubna4} and new formalisms \cite{prinz} that may facilitate the calculations required when studying these subjects. 

The renormalization of Quadratic Gravity presented here is related to perturbations along flat spaces. It looks  that applications related to curved backgrounds such as \cite{gaddam1}-\cite{gaddam2} are outside of the present formalism. However, already in 1973 it was found in reference \cite{pionero0} a connection between quantum Feynman diagrams and the perturbation series of the classical Schwarzschild background. This topic was explored further in \cite{damga}-\cite{solon},  related work is \cite{review3}-\cite{porto6}. The study of these connections between classical and quantum aspects of gravity with the present quantization schemes is also  a possible lead. In addition, the renormalization group of the model, studied with the present techniques may be of clear interest \cite{buccio1}-\cite{buccio3}. Further applications may be found in \cite{salvioad1}.

Finally, the Lyra version of Quadratic Gravity \cite{lyra1}-\cite{lyra7} and the study of processes which involve the interactions between graviton, matter and the scale fields, and in particular, its renormalization properties, is another possible interesting cosmological application. There are also apparent super renormalizable generalizations of Quadratic Gravity \cite{modesto1}-\cite{modesto5}, and a treatment of their ghost structure is a desirable work for the future.

In any case, the possibility of obtain these predictions with a consistent version of Quadratic Gravity may be a fascinating opportunity at present times, where the technology of high precision gravitational waves is fully under construction.

\section*{Acknowledgements}
The author is supported by CONICET, Argentina and by the Grant
PICT 2020-02181.  I am indebted to Diego Buccio, by some very useful discussion.

\end{document}